\documentclass[aps,eprint,showpacs,showkeys,amsmath,amssymb,superscriptaddress]{revtex4-2}
\usepackage{natbib}
\usepackage{float}
\usepackage{amsmath}
\usepackage{amssymb}
\usepackage{graphicx}
\usepackage{bm}
\usepackage{color}
\usepackage{epic}
\usepackage{eepic}
\usepackage[utf8]{inputenc}
\usepackage{pifont}
\usepackage{color}
\usepackage{nicefrac}
\usepackage[hidelinks]{hyperref}
\usepackage[caption=false]{subfig}
\usepackage{array}
\usepackage{siunitx} % for the units
\usepackage{tikz}
\usepackage{verbatim} % for block comments
\usepackage[export]{adjustbox}
\usepackage{hyperref}
\hypersetup{colorlinks=True,
citecolor=red,
linkcolor=blue, 
urlcolor=black}

\newcommand{\rev}[1]{{\color{black} #1}}

\newlength\imageheight % To determine the height of the tallest subfigure (used to center subfigures vertically)

\begin{document}
\preprint{APS/123-QED}
%
%\title{Rigid fibers interacting with a triangular obstacle in a microchannel}
\title{Dynamics of rigid fibers interacting with triangular obstacles in microchannel flows.}
\author{Zhibo Li}\
\thanks{These authors contributed equally to this work.}
\affiliation{Laboratoire de Physique et Mécanique des Milieux Hétérogènes (PMMH), UMR7636 CNRS, ESPCI Paris, PSL Research University, Sorbonne Université, Université Paris Cité, 75005 Paris, France}

\author{Clément Bielinski}
\thanks{These authors contributed equally to this work.}
\affiliation{Laboratoire d'Hydrodynamique (LadHyX), CNRS, École polytechnique, Institut Polytechnique de Paris, Palaiseau, 91120, France}

\author{Anke Lindner}
\email{anke.lindner@espci.fr}
\affiliation{Laboratoire de Physique et Mécanique des Milieux Hétérogènes (PMMH), UMR7636 CNRS, ESPCI Paris, PSL Research University, Sorbonne Université, Université Paris Cité, 75005 Paris, France}

\author{Olivia du Roure}
\email{olivia.duroure@espci.fr}
\affiliation{Laboratoire de Physique et Mécanique des Milieux Hétérogènes (PMMH), UMR7636 CNRS, ESPCI Paris, PSL Research University, Sorbonne Université, Université Paris Cité, 75005 Paris, France}

\author{Blaise Delmotte}
\email{blaise.delmotte@cnrs.fr}
\affiliation{Laboratoire d'Hydrodynamique (LadHyX), CNRS, École polytechnique, Institut Polytechnique de Paris, Palaiseau, 91120, France}

\date{\today}
\begin{abstract}
    Fiber suspensions flowing in structured media are encountered in many biological and industrial systems. Interactions between fibers and the transporting flow as well as fiber contact with obstacles can lead to complex dynamics.
    In this work, we combine microfluidic experiments and numerical simulations to study the interactions of a rigid fiber  with an individual equilateral triangular pillar in a 
    microfluidic channel.
    Four dominant fiber dynamics are identified: transport above or below the obstacle, pole vaulting and trapping, in excellent agreement between experiments and modeling. The dynamics are classified as a function of the length, angle and lateral position of the fibers at the channel entry. We show that the orientation and lateral position close to the obstacle are responsible for the fiber dynamics and we link those to the initial conditions of the fibers at the channel entrance. Direct contact between the fibers and the pillar is required to obtain strong modification of the fiber trajectories, which is associated to irreversible dynamics.
    Longer fibers are found to be more laterally shifted by the pillar than shorter fibers that rather tend to remain on their initial streamline.
  Our findings could in the future be used to design and optimize microfluidic sorting devices to sort rigid fibers by length. % Length permitted 500 words.
\end{abstract}
\maketitle

%%%%%%%%%%%%%%%%%%%%%%%%%%%%%%%%%%%%%%%%%

\section{Introduction}
Fluids containing small elongated particles play a crucial role in many fields of modern technology, such as paper manufacturing \cite{Lundell2011}, drag reduction \cite{PASCHKEWITZ2004}, composite materials fabrication \cite{Wang2022Polymers}, and pollution control problems \cite{Benedini2020, Chandrappa2021}. In many instances, small fibers must navigate through crowded environments embedded with obstacles: micro-plastic fibers can propagate in soils and cause pollution of groundwater \cite{re_shedding_2019,engdahl_simulating_2018}, wood-pulp fibers interact with the fabric mesh underneath during the formation of paper sheets \cite{Lundell2011}, and pathogenic filaments made of parasites, such as bacterial biofilm streamers, can clog tortuous capillaries or complex structures such as stents \cite{rusconi_laminar_2010,drescher_biofilm_2013}. 
The motion of elongated particles is much more complex than spherical ones due to their asymmetric shape. Their dynamics results from the interplay between the surrounding background flow, internal elastic forces, hydrodynamic interactions and eventually interactions with solid walls and embedded obstacles \cite{duRoure2019,Chakrabarti2020,Makanga2023}. 
Interactions with obstacles are used in microfluidic particle sorting devices based on the so-called deterministic lateral displacement (DLD) technique, which was initially developed for spherical particles \cite{Huang2004}. DLD uses successive collisions with pillars in a background flow to sort particles based on their size or mechanical properties. It has been successfully extended to separate  DNA fragments \cite{Wang2015, Chen2015}, pathogenic bacteria \cite{Beech2018}, cells \cite{Kabacaoglu2018, Loutherback2012}, and blood parasites \cite{Holm2011}. However, the sorting of rigid elongated objects, such as micro-plastics, with obstacles \rev{has not yet been reported}.

Indeed, while the motion of elongated particles in  unbounded and  confined viscous flows has been widely investigated \cite{duRoure2019}, research on the dynamic interactions between a rigid fiber and an obstacle is still scarce. When freely transported in a viscous flow, the trajectory of the center of mass of a rigid fiber will follow the streamlines. Hydrodynamic interactions with bounding channel walls can additionally lead to rotation, reorientation and transverse oscillations of the fiber \cite{Nagel2017,cappello_transport_2019}. 
Combining experiments, theory and numerical simulations, Makanga \textit{et al.} \cite{Makanga2023} recently showed that, in the absence of a background flow, the interactions between a sedimenting fiber and an obstacle can either induce a large lateral displacement or permanent trapping depending on the obstacle shape, fiber length and/or deformability. In the presence of an ambient flow field, the 2D simulations of a semi-flexible polymer in a periodic array of circular obstacles by Chakrabarti \textit{et al.} \cite{Chakrabarti2020} show that various modes of transport occur depending on the incidence of the incoming flow in the lattice. While instructive, this work does not explore the effect of rigidity and lacks experimental validations.

 Studying fluid-structure interactions of fibers in the presence of obstacles is a particularly rich topic as fiber transport does not only result from the interaction of the finite size slender object with a complex flow field (created by the presence of the obstacle), but also by possible contact between the fiber and the obstacle. \rev{In theory, such direct contact is prevented by lubrication forces in vanishing Reynolds number flows. However, direct contact is possible in practice due to surface roughness or small attractive forces between fibers and surfaces.} %Despite considering flows at vanishing Reynolds numbers where direct contact between objects is prevented by lubrication forces, surface roughness or the presence of small attractive forces between fibers and objects might promote such direct contact.

Migration of fibers between streamlines can thus result from reversible interactions, induced by streamline curvature, but also by irreversible interactions induced by direct fiber/obstacle contact. Modeling direct contact between fibers and obstacles is very challenging as the details of such interactions are often unknown and not fully controlled from experiments. In this paper we overcome these difficulties by performing a combined experimental and modeling study where a simplified approach is used in the model to simulate fiber and obstacle contact. Direct comparison between very well controlled model experiments and simulations allows to adjust the contact parameters in the simulation and to obtain excellent agreement between experiments and simulations, as far as fiber trajectories, orientations and time scales are concerned.  In this way we effectively capture the fiber dynamics and the combined role of fluid-structure interactions and fiber-obstacle contact. This allows us to perform a systematic study of cross-stream migration of rigid fibers interacting with a triangular pillar in a confined microchannel as a function of their initial orientation and position at the channel entry and their length, and to evaluate the sorting potential.

This paper is organized as follows: in Sec.~\ref{sec:experimental_method}, we briefly introduce the fabrication method of the rigid fibers and the experimental setup.
Section \ref{sec:numerical_method} presents the numerical methods used to compute the flow field and the motion of the fibers.
Section~\ref{sec:results_flow_fiber_dynamics} describes the flow field and the fiber dynamics, and Sec.~\ref{sec:results_lateral_displacement} focuses on the sorting potential. The main conclusions of this study are summarized in Sec.~\ref{sec:conclusions}.

\section{Experimental method}
\label{sec:experimental_method}

\subsection{Rigid fiber fabrication and characterization}
The elongated rigid fibers used in this study are prepared by shearing an emulsion of SU-8 polymer droplets in a glycerol-ethanol mixture and exposing the stretched droplets to ultraviolet (UV) light (see Fig.~\ref{subfig:SU8_fabri}).
The UV radiations photo-crosslink the SU-8 and yield chemically highly stable colloidal SU-8 fibers \cite{fernandez2019synthesis}, as shown on the right panel in Fig.~\ref{subfig:SU8_fabri}. The fibers are mostly straight and have high \rev{length-to-diameter} aspect ratios.
The length and radius of the fibers are controlled by tuning the viscosity of the solvent and the shear stress.
In practice, this corresponds to adjusting the ratio between glycerol and ethanol and the stirring speed.
In our experiments, the solvent consists of \qty{70}{\percent} glycerol and \qty{30}{\percent} ethanol by weight, and it is stirred at 300 rpm.
\rev{As shown in Fig.~\ref{subfig:SU8_L}, in the experiments, most of the fibers have a length ranging from \numrange{40}{180} \unit{\um} and a radius around $2\,\unit{\um}$, which corresponds to a length-to-diameter aspect ratio of \numrange{10}{40}.} The Young's modulus of crosslinked SU-8 is found to be $E=\numrange{0.9}{7.4} \,\unit{GPa}$ \cite{xu2016characterization}.
The fibers can be considered undeformable in our experiments due to their high flexural rigidity, which is on the order of $10^{-14}\,\unit{Nm^2}$. 
%Considering a fiber with a length of \qty{100}{\um} and diameter of \qty{4}{\um} as a cantilever beam, the maximum deflection under a uniform distributed load is $\delta_{\rm B} = ql^4/8EI$, where $q$ is the load per unit length, $l$ the length of a beam, $E$ Young's modulus, and $I$ the second moment of area. Based on the mechanical property of the SU-8 \cite{xu2016characterization}, to have an experimentally detectable deflection, which corresponds to the physical size of 1 pixel in our experiment, the load is estimated to be approximately $q_{\rm re} \sim 2.5 \times 10^{-3}$N/m. The Stokes drag per unit length exerted on the fiber perpendicular to the long axis direction is about $q_{\rm st} \sim 8 \pi \mu \left[ \ln(2 \varepsilon) \right]^{-1} U \sim  1 \times 10^{-3}$N/m, which is smaller than $q_{\rm re}$, where $\varepsilon$ is the fiber aspect ratio and $U$ the applied velocity. As a result, the deformation of fibers in our experiment is negligible.
%Based on the mechanical property of the SU-8 \cite{xu2016characterization}, the persistence length is estimated to be $l_{\rm p} = B/\kappa T \sim 10^7\,\unit{\m}$ where $\kappa$ is the Boltzmann constant, $T$ the absolute temperature, and $B$ the bending stiffness, $B=EI$, which is a function of Young's modulus $E$ and the second moment of area $I$ of the beam cross-section about the axis of interest, length of the beam and beam boundary condition. The persistence length is much larger than the contour length $L$ indicating that the fibers are rigid.

\subsection{Experimental setup}
\begin{figure}[H]
    \centering
    \subfloat[\label{subfig:SU8_fabri}]{\includegraphics[width=0.35\textwidth]{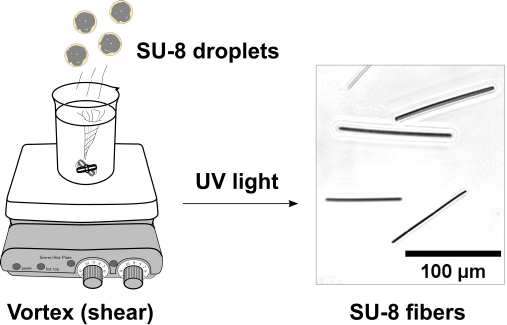}}
    \hfill
    \subfloat[\label{subfig:SU8_L}]{\includegraphics[width=0.26\textwidth]{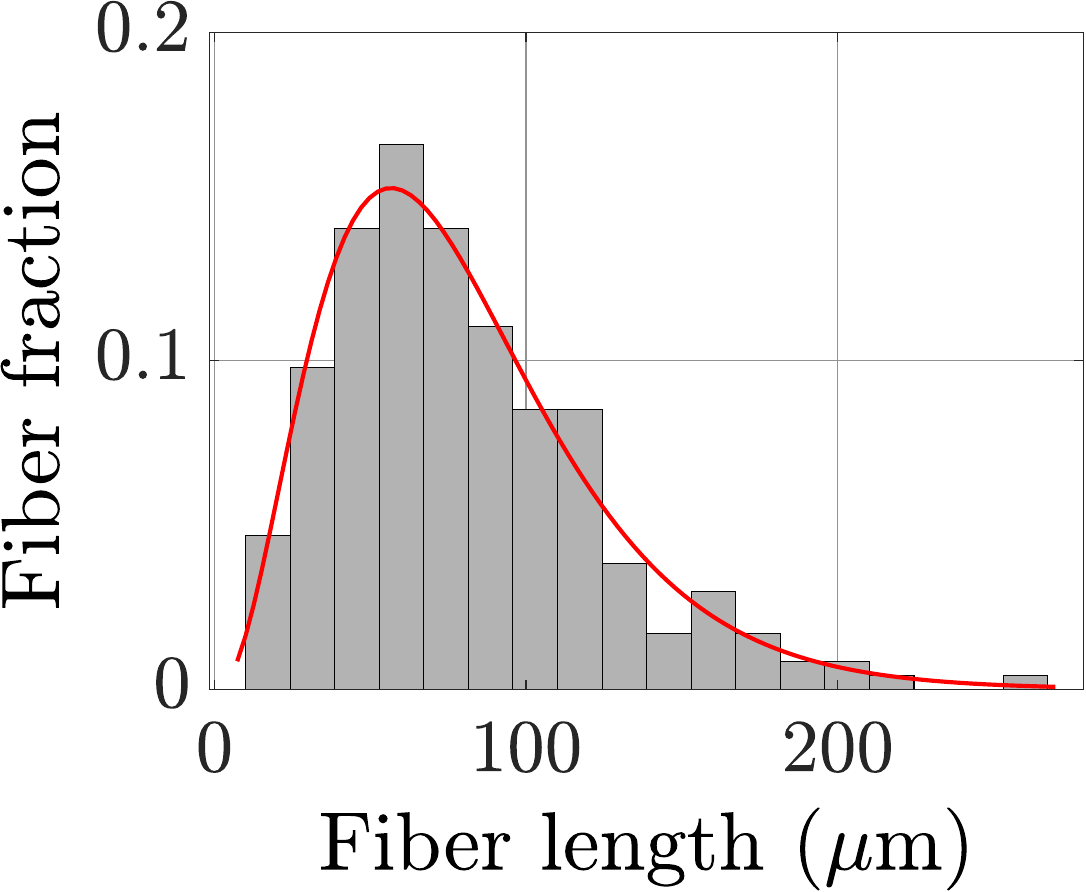} \hspace{0.5cm}\includegraphics[width=0.269\textwidth]{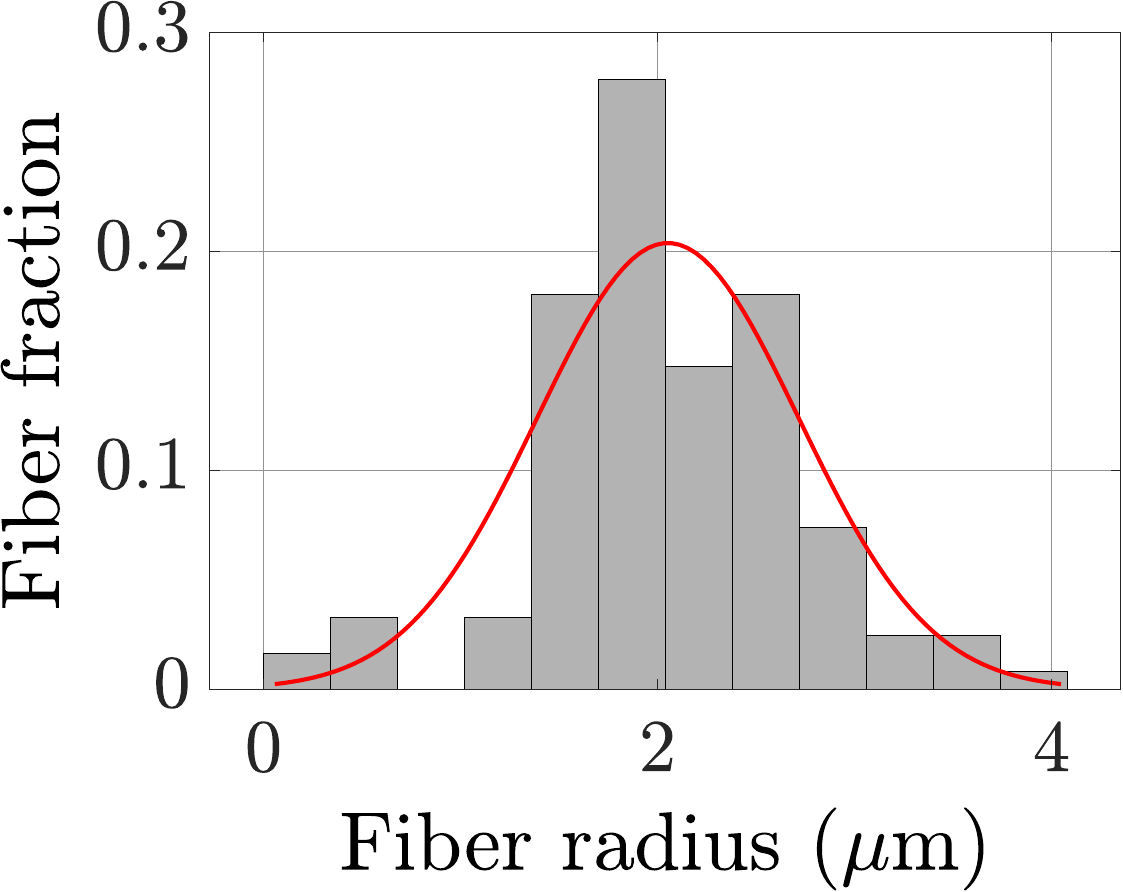}}
    \\
    \subfloat[\label{subfig:exp_setup}]{\includegraphics[width=0.48\textwidth]{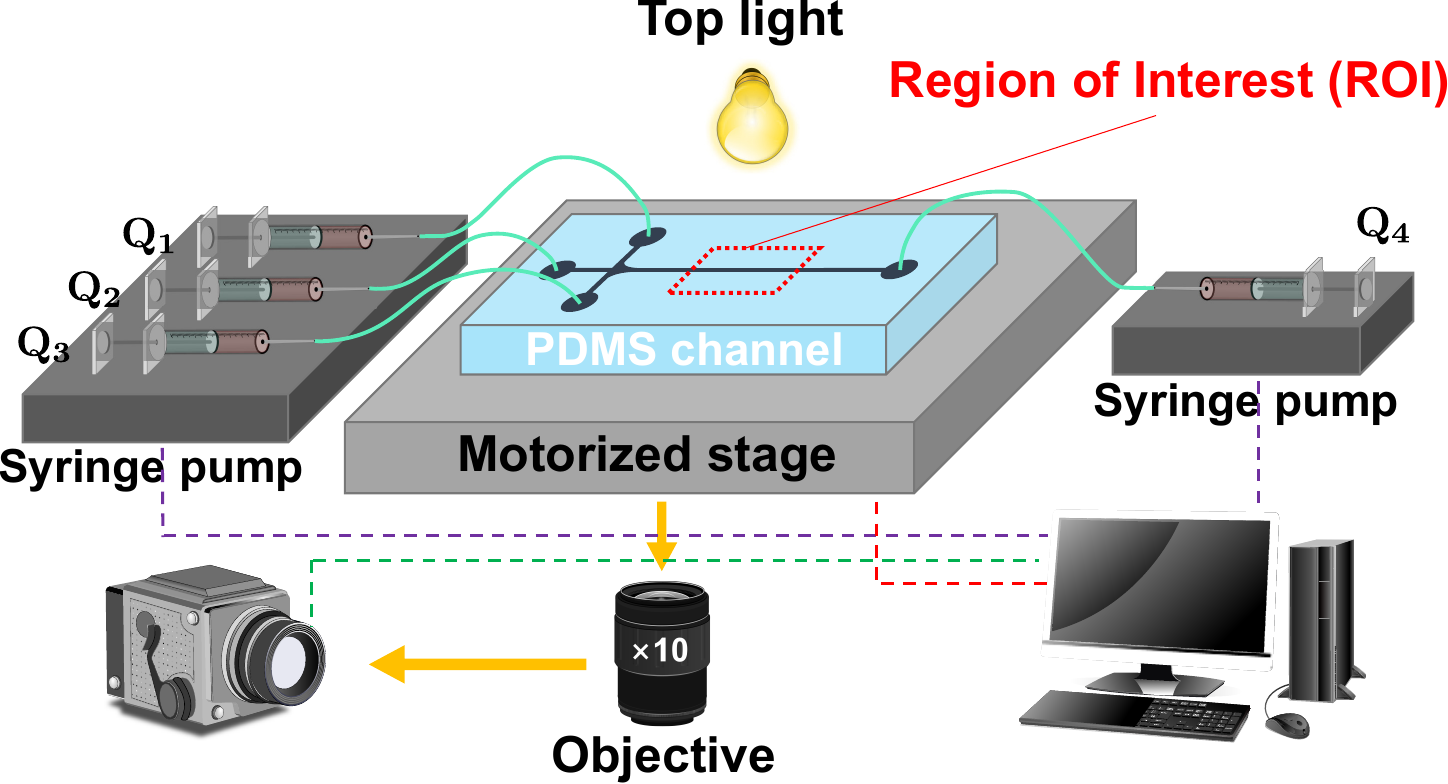}}
    \hfill
    \subfloat[\label{subfig:num_setup}]{\includegraphics[width=0.45\textwidth]{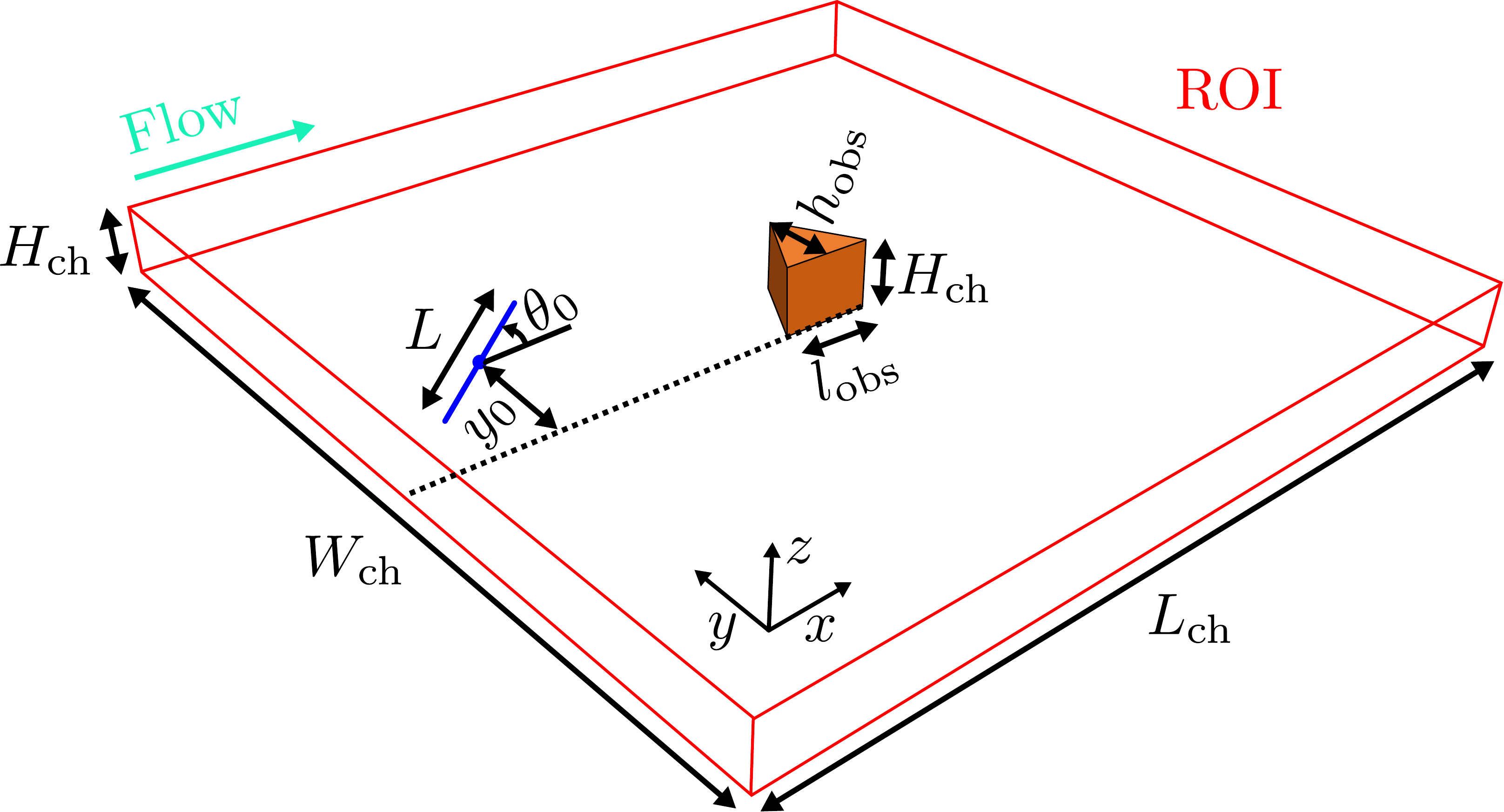}}
    \caption{\rev{Experimental method and geometrical parameters.} (a) The rigid SU-8 fiber fabrication process. The right panel is a real image of the SU-8 fibers under the microscope. (b) The length and radius distributions of the SU-8 fibers. (c) A detailed sketch of the experimental setup showing the microfluidic channel, flow control, and data acquisition.  (d) Zoom in on the region of interest (ROI) showing the main geometrical parameters of the study (figure not to scale).}
    \label{fig:setup}
\end{figure}

We conduct the experiments in a polydimethylsiloxane (PDMS) microchannel of width $W_{\rm ch}=800\,\unit{\um}$, height $H_{\rm ch}=40\,\unit{\um}$, and length $L_{\rm ch}=20\,\unit{\mm}$ with three inlets and one outlet (see Fig.~\ref{subfig:exp_setup}).
A triangular pillar with the same depth $H_{\rm ch}$ as the channel is placed in the middle of the microchannel (see Fig.~\ref{subfig:num_setup}). Its base is aligned with the flow direction and the triangle is of height $h_{\rm obs}=75\,\unit{\um}$ and base $l_{\rm obs}=2 h_{\rm obs}/\sqrt{3}$. 
The experiments are performed on an inverted microscope (Zeiss Axio Observer A1).
The PDMS channel is placed on a motorized stage (ASI MS-2000 XY automated stage) to precisely control its position in the $x$ and $y$ directions (horizontal plane).
An insert is moved in the $z$-axis via a piezo element with a range of 150 µm with nanometer accuracy. A syringe pump (CETONI GmbH, neMESYS Low Pressure module 290N) drives the experimental fluids with controlled flow rates.
The suspension containing rigid fibers in a glycerol and ethanol mixture is delivered from the middle inlet with a flow rate $Q_2=1\,\unit{\nano\litre}$/s. Fiber concentration is very low to assure that fibers enter the channel and interact with the obstacle one by one.
Two lateral inlets inject the same glycerol and ethanol mixture with the flow rate $Q_1=Q_3=5\,\unit{\nano\litre}$/s.
These lateral flows focus the fibers into a narrow band in the center of the channel width, increasing the probability of their interaction with the obstacle. 
% We cannot focus fibers in the midplane with respect to the channel height, but we only observe and consider fibers in a narrow band around the midplane to prevent fibers from feeling the velocity gradient in the direction of the channel height \Anke{Does this last sentence make sense? And if yes, do we want to give a range of disctances around the midplane where we consider fibers? Also do we need to explain the Poiseuille flow in the direction of the channel height? May be not and also may be not here...}
Flow-focusing also aligns the fibers parallel to the main flow direction, reducing the range of initial fiber orientations.
At the outlet, the flow rate is set to $Q_4=-(Q_1+Q_2+Q_3)=-11\,\unit{\nano\litre}$/s to stabilize the flow. The flow can be reversed to release fibers that remain trapped on the obstacle.
In the experiments, the density $\rho$ and dynamic viscosity $\mu$ of the suspending fluid are respectively $1190\,\unit[per-mode = symbol]{\kilogram\per\meter^3}$ and $340\,\unit{\milli\pascal\second}$ \cite{Alkindi2008PhysicalGlycerol}. Notice that the density of the raw SU-8 resin is reported to be about $1200\,\unit[per-mode = symbol]{\kilogram\per\meter^3}$, which is very close to the density of the solvent. Hence, sedimentation of fibers is negligible in the experiments.

SU-8 fibers are observed under bright-field microscopy, and a sCMOS camera (ORCA-Flash4.0 V3 Digital CMOS camera, Hamamatsu) records images of their evolution in the channel through a 10\texttimes\,objective (N-Achroplan 10x/0.25 Ph1 M27, Zeiss). The camera's exposure time is \qty{10}{\milli\second} to avoid image  blurring, and the sampling frequency is \qty{100}{\hertz} at best performance with an image resolution of \numproduct{2048x2048} pixels. The images are processed using a homemade MATLAB code, which includes background removal, tubular structure enhancement, noise reduction using Gaussian blurring, binarization, skeletonization, and B-spline reconstruction.

Before the experiments, we perform micro-particle image velocimetry (µPIV) (LaVision GmbH) at different depths in the microchannel with the same suspending fluid and flow rates as in the experiments to characterize the flow field.
The fluid velocity at the channel centerline $U_{\rm center}$ is in the order of 600\,µm/s, which gives a Reynolds number ${\rm Re} = \rho U_{\rm center}h_{\rm obs}/\mu$ in the order of $10^{-4}$.

\section{Numerical method}
\label{sec:numerical_method}
\subsection{Numerical setup}

The microfluidic channel used in the numerical simulations has the same width $W_{\rm ch}=800\,\unit{\um}$ and height $H_{\rm ch}=40\,\unit{\um}$ as the one used in the experiments (see Fig.~\ref{subfig:num_setup}), and it has a length $L_{\rm ch}=2400\,\unit{\um}$.
An equilateral triangular pillar of height $h_{\rm obs}=75\,\unit{\um}$ and base $l_{\rm obs}=2 h_{\rm obs}/\sqrt{3}$ is placed in the middle of the channel with the baseline aligned with the flow direction, again matching the experimental conditions.
The channel is long enough so that the flow at the inlet and outlet of the domain is not significantly disturbed by the presence of the pillar.

A rigid fiber of length $L$ and cross-section radius $a$ is positioned in the midplane of the channel far away from the pillar.
Its center of mass is initially located at ${\bf r_0}=(x_0=L_{\rm ch}/4,y_0,z_0=H_{\rm ch}/2)$, which is far enough from the pillar so that the fiber does not feel any flow disturbance.
In the experiments, most of the fibers enter the microfluidic channel close to the center with respect to the channel width and with a small initial angle $\theta_0$ due to the flow-focusing.
To explore similar initial conditions the initial lateral position $y_0$ is varied in the simulations between the base and the apex of the pillar and the initial angle is varied in the range $-10^{\circ} \leq \theta_0 \leq 10^{\circ}$.
The fiber radius is set to $a=2\,\unit{\um}$ and the fiber length spans from $0.5l_{\rm obs}$ to $1.4l_{\rm obs}$ to match the geometrical properties of the fibers used in the experiments.

Owing to the low Reynolds number reported in the experiments, the fluid flow in the microfluidic channel is governed by the Stokes equation
\begin{equation}
\nabla p - \rho{\bf f}_{\rm b} = \mu \nabla ^2 {\bf u}
\label{eq:Stokes_equation}
\end{equation}
where $p$, $\rho$, $\mu$ and ${\bf u}$ are respectively the pressure, the density, the dynamic viscosity and the velocity of the fluid, and ${\bf f}_{\rm b} = (f_{{\rm b}x},0,0)$ is the constant force parallel to the channel walls that generates the flow.
The value of $f_{{\rm b}x}$ is adjusted to have the same velocity $U_{\rm center}$ at the channel centerline as in the experiments.
No-slip condition (${\bf u}={\bf 0}$) is set on the channel walls and on the surface of the obstacle, and periodic boundary conditions are set at the channel inlet and outlet.

\subsection{Computation of the flow field}
\label{sec:flow_LBM}
The flow field is computed in three dimensions using the lattice Boltzmann method (LBM) \cite{Succi2001,Krueger2016}.
The LBM is based on the lattice Boltzmann equation
\begin{equation}
\lambda_i({\bf r} + {\bf e}_i\Delta t, t + \Delta t) - \lambda_i({\bf r},t) = -\left(\lambda_i - \lambda_i^{\rm{eq}}\right)\Delta t/\tau + f_{{\rm b}_i}
\label{eq:LBE}
\end{equation}
where $\lambda_i({\bf r},t)$ is a distribution function that gives the probability of finding a fluid particle at position ${\bf r}$ and time $t$ flowing at the discrete velocity ${\bf e}_i$, and $\Delta t$ is the time step.
Here, the D3Q19 lattice is used and thus $i=0-18$, and the collision term is approximated by the Bhatnagar-Gross-Krook collision operator \cite{Bhatnagar1954}.
The relaxation time $\tau$ is related to the kinematic viscosity of the fluid $\nu$ by $\tau = 3\nu + \frac{1}{2}$.
The equilibrium distribution function $\lambda_i^{\rm eq}$ is defined as
\begin{equation}
\lambda^{\rm eq} _{i}({\bf r},t)= \omega _i \rho \left[1 + \frac{{\bf u}\cdot {\bf e}_i}{c_{\rm s}^2} + \frac{({\bf u}\cdot{\bf e}_i)^2}{2c_{\rm s}^4} -\frac{{\bf u}\cdot{\bf u}}{2c_{\rm s}^2} \right]
\end{equation}
where $c_{\rm s}=1/\sqrt{3}$ is the lattice speed of sound and the $\omega_i$'s are weight factors with $\omega_0=1/3$, $\omega_{1-6}=1/18$ and $\omega_{7-18}=1/36$.
\rev{The source term $f_{{\rm b}_i} = \frac{\omega_i}{c_{\rm s}^2}{\bf f}_{\rm b}\cdot{\bf e}_i$ is added to account for the body force that triggers the flow field.}
The fluid density $\rho$ and velocity ${\bf u}$ are respectively computed as the zeroth and first order moments of the distribution function $\lambda_i$
\begin{align}
&\rho({\bf r},t) = \sum _{i=0}^{18} \lambda_i ({\bf r},t)\rev{,}& &
{\bf u}({\bf r},t) = \frac{1}{\rho}\sum _{i=0}^{18} \lambda_i ({\bf r},t){\bf e}_i.&
\end{align}
The no-slip boundary conditions applied on the channel walls and on the surface of the pillar are achieved using a standard bounce-back scheme.
\rev{The background flow field is \rev{steady} and is \rev{therefore} computed only once.}
\subsection{Computation of the fiber dynamics}

The viscous flow transports the fiber within the channel and thus exerts mechanical stress on it. In addition, the fiber can experience contact forces when it meets the obstacle surface. Internal bending and tensile forces are imposed on the rigid fiber in order to keep it straight.  The interplay between the viscous and contact forces and the internal tensile and bending forces determines the fiber dynamics and trajectories. Below we briefly introduce the \rev{bead-spring model} used to account for these elastohydrodynamic couplings and to handle contact of the fiber with the obstacle in the simulations.  \rev{The bead-spring model has been widely used, and validated, in the literature to study the motion of rigid and flexible fibers in viscous flows \cite{yamamoto1993method,schlagberger2005orientation,wada2009hydrodynamics,manghi2006propulsion,delmotte2015general,marchetti2018deformation,schoeller2021methods,slowicka2022buckling,Makanga2023}.}
\subsubsection{\rev{Internal elastic forces}}
% The fiber is modeled by a chain of $n$ rigid spherical beads of radius $a$ that are connected together by internal elastic forces (see Fig.~\ref{subfig:sketch_fiber}) to fulfill the inextensibility condition and then keep the fiber length $L$ constant over time.
The fiber is modeled as a chain of $n$ rigid spherical beads of radius $a$ that are linked together by internal elastic forces ${\bf F}^{\rm E}$ (see Fig.~\ref{subfig:sketch_fiber}) to keep the fiber shape unchanged over time.
These forces are derived from an elastic potential $H$  \cite{Gauger2006,marchetti2018deformation,Makanga2023}
\begin{align}
    &
    {\bf F}^{\rm E} = -\nabla H
    &
    &
    \text{with}
    &
    &
    H = \sum_{i=2}^n \left[ \frac{S}{4a} \left( \left|{\bf t}_i\right| - 2a\right)^2 \right] + \sum_{i=2}^{n-1}\left[ \frac{B}{2a}\left(1 - {\bf \hat t}_{i+1} \cdot {\bf \hat t}_i \right) \right]
    &
\end{align}
where the first sum accounts for stretching (i.e.\ tensile) forces and the second sum accounts for bending forces.
Here, $S = E \pi a^2$ is the stretching coefficient and $B = E \pi a^4 /4$ is the bending coefficient, where $E$ is the fiber Young's modulus. \rev{This model can be used to simulate both flexible and rigid fibers within the same framework.}
In the simulations we chose $E=26$\,MPa, which is smaller than the Young's modulus of the fibers used in the experiments, but it is large enough to simulate rigid fibers and prevents making the problem too stiff to be solved numerically. 
${\bf t}_i$ is the vector linking the center of mass of beads $i$ and $i-1$ and ${\bf \hat t}_i = {\bf t}_i/\left|{\bf t}_i\right|$. 

\subsubsection{\rev{Contact  with the obstacle}}
\begin{figure}[H]
\settoheight{\imageheight}{\includegraphics[width=0.3\textwidth]{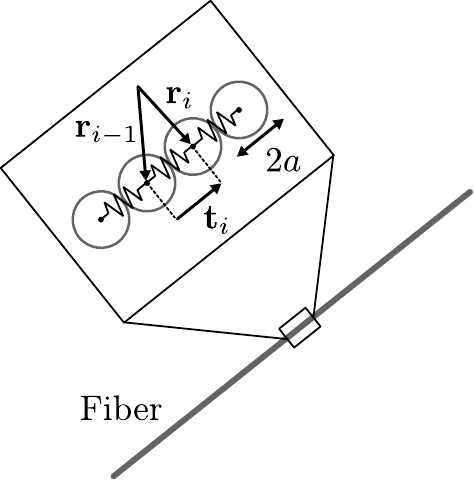}}
\centering
    \subfloat[\label{subfig:sketch_fiber}]{\includegraphics[width=0.3\textwidth]{sketch_fiber-eps-converted-to.pdf}}
    \hfill
    \subfloat[\label{subfig:sketch_fiber_obstacle}]{\tikz\node[minimum height=\imageheight]{\includegraphics[width=0.33\textwidth]{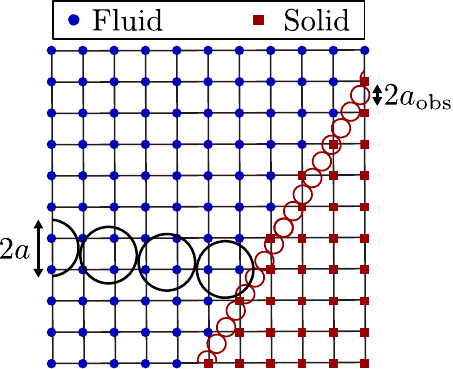}};}
    \hfill
    \subfloat[\label{subfig:sketch_fiber_obstacle_steric_force}]{\tikz\node[minimum height=\imageheight]{\includegraphics[width=0.28\textwidth]{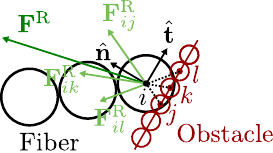}}; }
    \caption{%Discretization of the fiber and the obstacle by spherical beads.
    \rev{Numerical method.} (a) Sketch showing the discretization of the fiber by rigid spherical beads connected by springs. (b) Representation of the Eulerian grid where the fiber and the obstacle are immersed. Fluid nodes are shown as blue circles and solid nodes as red squares. (c) Zoom on the contact between the fiber and the obstacle surface. The resulting repulsive force ${\bf F}^{\rm R}={\bf F}_{ij}^{\rm R}+{\bf F}_{ik}^{\rm R}+{\bf F}_{il}^{\rm R}$ is not strictly normal to the obstacle surface (the red line) because of the discretization using spherical beads.}
    \label{fig:sketches_fiber_obstacle}
\end{figure}

The fiber is immersed in the Eulerian grid used to compute the flow field by LBM, as depicted in Fig.~\ref{subfig:sketch_fiber_obstacle}, where fluid nodes are represented as blue circles and solid nodes where no-slip boundary conditions apply are shown as red squares.
When the fiber comes close to the obstacle,
%it will slow down due to the no-slip boundary condition on the surface but also because of
it will experience short range contact forces.
To model this contact,  the surface of the pillar is discretized with small beads in order to \rev{smooth} the stepped shape of the Eulerian grid and to tune the effective pillar roughness through the bead radius $a_{\rm obs}$.
A repulsive force ${\bf F}^{\rm R}$ \cite{Dance2004CollisionBE} is added to the fiber beads that are closer than a given cutoff $R_{\rm ref}$ to the pillar beads to prevent the fiber from penetrating inside the obstacle.
The external repulsive force between the fiber bead $i$ and the obstacle bead $j$ is defined as
\begin{align}
{\bf F}^{\rm R}_{ij} = \begin{cases}
    -\frac{F_{\rm ref}}{a+a_{\rm obs}}\left[ \frac{R^2_{\rm ref} - \left|{\bf r}_{ij}\right|^2}{R^2_{\rm ref} - \left(a + a_{\rm obs}\right)^2} \right]^{4}{\bf r}_{ij} & \text{if}~ \left| {\bf r}_{ij} \right| < R_{\rm ref} \\
    {\bf 0} & \text{otherwise}
  \end{cases}
\end{align}
where $F_{\rm ref}=6\pi\mu a U_{\rm center}$ and ${\bf r}_{ij}$ is the vector between the fiber bead $i$ and the obstacle bead $j$.
The total repulsive force ${\bf F}^{\rm R}$ acting on the $i^{\rm th}$ fiber bead is the sum of all the repulsive forces ${\bf F}^{\rm R}_{ij}$ (see Fig.~\ref{subfig:sketch_fiber_obstacle_steric_force}).
The discretization of the fiber and the pillar using beads leads to a repulsive force that is not strictly normal to the pillar surface.
It can be decomposed as
\begin{equation}
    {\bf F}^{\rm R} = {\rm F^R_n} \hat{\bf n} + {\rm F^R_t} \hat{\bf t}
\end{equation}
where $\hat{\bf n}$ and $\hat{\bf t}$ are respectively the unit vectors normal and tangent to the pillar surface.
The normal component is a repulsive force and the tangential component can be described as a ``friction force".

% An obstacle bead radius $a_{\rm obs}=0.6\,\unit{\um}$ was found to reproduce the intrinsic roughness of the pillar used in the experiments.
% The cutoff distance $R_{\rm ref}$ is chosen in such a way the fiber beads are allowed to overlap with the obstacle beads but they are not allowed to penetrate inside the pillar.
% The fiber can thus get close to the no-slip condition without remaining \Clement{stuck inside the numerical obstacle (red squares in Fig.~\ref{subfig:sketch_fiber_obstacle})} where the velocity is zero.
\subsubsection{\rev{Hydrodynamic interactions and equations of motion}}
Once the internal and external forces are obtained, the velocity of the fiber \rev{elements} is computed from the mobility relation \cite{guazzelli2011physical}

\begin{equation}
    {\bf U}_i = {\bf u}_i +  \sum_{j=1}^n {\bf M}_{ij} {\bf F}_j 
    \label{eq:bead_vel}
\end{equation}
\rev{where ${\bf U}_i$ is the total velocity of bead $i$, ${\bf u}_i$ is the velocity induced by the background flow ${\bf u}^{\infty}$ (computed with LBM, see Sec.\ \ref{sec:flow_LBM}) and $\sum_{j=1}^n {\bf M}_{ij} {\bf F}_j $ is the velocity induced by the non-hydrodynamic forces ${\bf F} = {\bf F}^{\rm E} + {\bf F}^{\rm R}$ acting on the $n$ fiber beads. ${\bf M}$ is the so-called mobility matrix that contains all hydrodynamic interactions between the fiber beads, i.e.\ their velocity induced by the flow disturbances generated by the non-hydrodynamic forces.} 
\rev{The computation of ${\bf u}_i$ is similar to the approach used in the immersed boundary method to impose no-slip boundary conditions on the particle surface \cite{Krueger2016,Peskin1977}:
\begin{equation}
  {\bf u}_i(t)  = \sum_{\bf x}\Delta x^3 {\bf u^{\infty}}({\bf x},t)d({\bf r}_i(t)-{\bf x}),
\end{equation}
where $\Delta x$ is the spacing of the Eulerian grid used to compute the flow by LBM at each node ${\bf x}$, ${\bf r}_i(t)$ is the position of the $i$th fiber bead at time $t$, and $d$ is the kernel function.
The kernel function is factorized in 3D as:
\begin{equation}
  d({\bf x}) = \phi(x)\phi(y)\phi(z)/\Delta x^3,
\end{equation}
with
\begin{equation}
  \phi\left(x\right) = \begin{cases} \frac{1}{4}\left(1+ \cos \frac{\pi x}{2}\right)~~~\text{if}~\left|x\right| \leq 2\Delta x \\
    0 ~~~\text{else}
    \end{cases}
    \text{.}
\end{equation}
We simulated the sedimentation of a single bead in an unbounded domain, and, by equating the gravity force and the Stokes drag, we verified this stencil function approximates well the no-slip condition for a bead of radius $a = \Delta x$.}

In this work we use the Rotne-Prager-Yamakawa mobility matrix defined as \cite{Wajnryb2013}
\begin{align}
    {\bf M}_{ij} = \begin{cases}
        \left( \mathbb{I} + \frac{a^2}{3}\nabla^2 \right) \mathbb{T}({\bf r}_{ij}) & i\neq j \\
        \frac{1}{6 \pi \mu a} \mathbb{I} & i=j
      \end{cases}
    \end{align}
where $\mathbb{I}$ is the $3\times3$ identity matrix, $\mathbb{T}$ is the Oseen tensor and ${\bf r}_{ij}$ the vector between fiber beads $i$ and $j$.

The new position of the fiber beads ${\bf r}$ is then computed by integrating ${\rm d}{\bf r}/{\rm d}t = {\bf U}$ with an implicit second order backward differentiation formula (BDF2) method to handle the stiffness of the system. 
\rev{The time step is set to 10$^{-7}$~s, which is about 3 times smaller than the characteristic bending time $t_{\rm b}=\mu(2a)^4/B\approx 2.7\times10^{-7}$s and 1.6 times smaller than the characteristic stretching time $t_{\rm s} = 6\pi\mu a/(S/2a) = 12\mu/E \approx 1.6\times10^{-7}$s, and thus ensures numerical stability.}

The mobility matrix ${\bf M}$ \rev{accounts for the disturbances induced by the fiber on the flow, and for its drag anisotropy, through the hydrodynamic interactions (HI) between the fiber segments; but it neglects the corrections of these HI due to the channel walls and obstacle surface. However, as discussed in Appendix \ref{sec:fiber-wall HI} these corrections are small compared to the fiber velocity induced by the ambient flow.} The fiber nevertheless slows down at the approach of the obstacle due to the no-slip boundary condition leading to a vanishing velocity at the obstacle surface. However, lubrication forces are not taken into account and the obstacle-fiber interaction is solely modelled by repulsive forces between the fiber and the object corresponding to direct fiber/obstacle contact. \rev{This is a strong simplification since for perfectly smooth surfaces, lubrication forces prevent direct contact at vanishing Reynolds number.} In the experiments, of course, surfaces are never perfectly smooth allowing for direct contact even at low Reynolds numbers, but the exact contact conditions are difficult to quantify or to control.

We thus chose here to model the fiber-obstacle interactions using the simplified approach of a short range repulsive force. Careful comparison between simulations and experimental results obtained with very well controlled model experiments allows us to adjust the parameters of the simulations. 
%to correctly capture the effective role of particle-obstacle interactions.  
\rev{By conducting a sensitivity analysis we} found that a cutoff distance $R_{\rm ref} = 1.54\,\unit{\um}$ and an obstacle bead radius $a_{\rm obs} = 0.6\,\unit{\um}$ were optimal to prevent artificial overlapping between the fiber and the obstacle surface
%and to mimic its effective roughness.
and also provide excellent quantitative agreement for all comparisons between experiments and simulations (see Secs.~\ref{sec:results_flow_fiber_dynamics} and \ref{sec:results_lateral_displacement}). This proves that our ``effective" approach captures correctly the role of the complicated fiber-obstacle interactions on the fiber dynamics. 
We would like to stress that such agreement is surprisingly good given the fact that our model neglects near-field hydrodynamic interactions between the fiber and the obstacle. But as mentioned above, solving the exact flow in the thin gap separating the two objects is out of reach due to their unknown surface topography, and our adjustable model offers a robust  alternative. 

\section{Flow disturbance and fiber dynamics}
\label{sec:results_flow_fiber_dynamics}

\subsection{The disturbance flow field}
\begin{figure}[H]
\centering
    \subfloat[\label{subfig:velocity_fields_LBM_PIV}]{\includegraphics[width=0.65\textwidth]{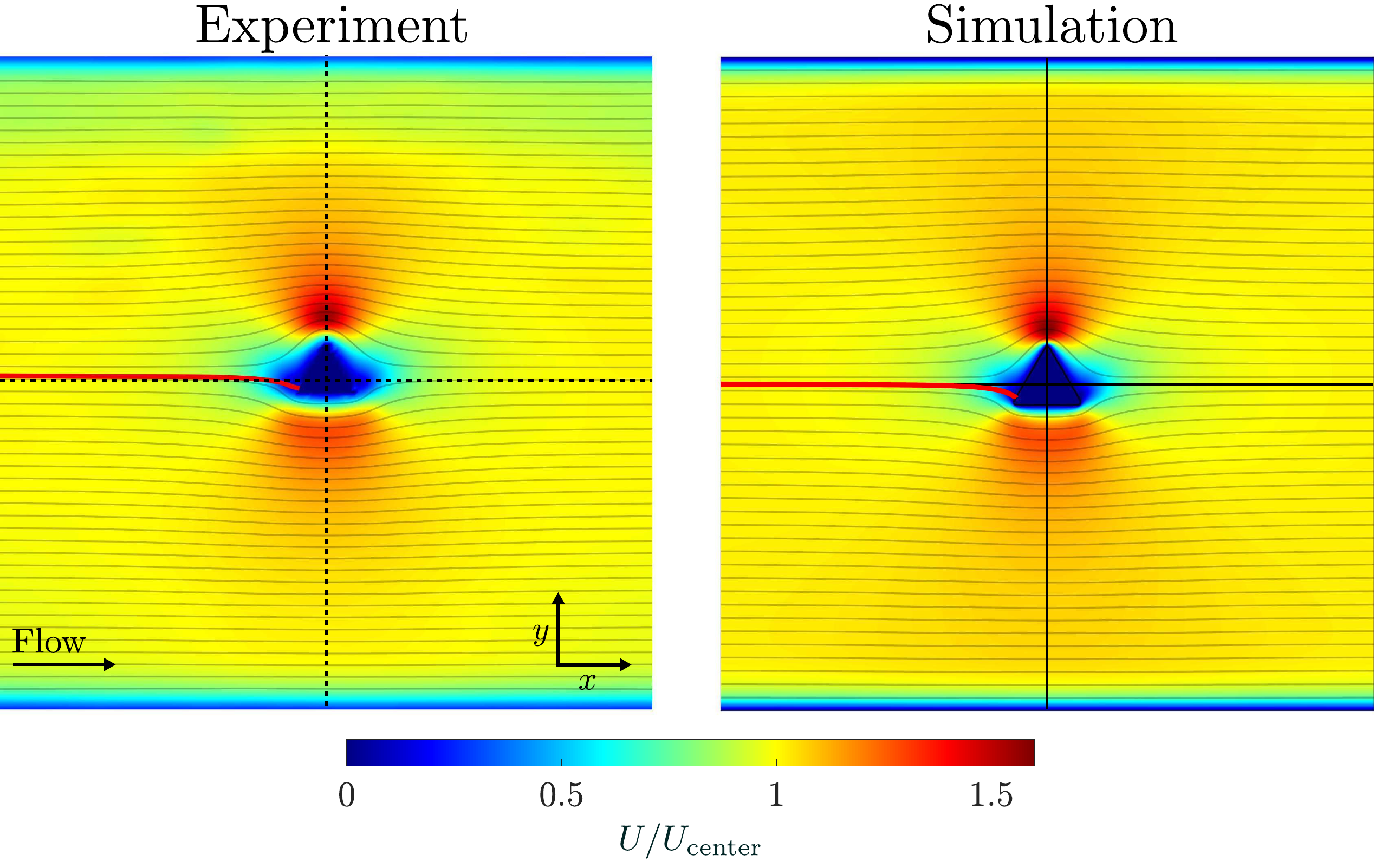}}
    \\
    \subfloat[\label{subfig:velocity_profiles_along_x_LBM_PIV}]{\includegraphics[width=0.4\textwidth]{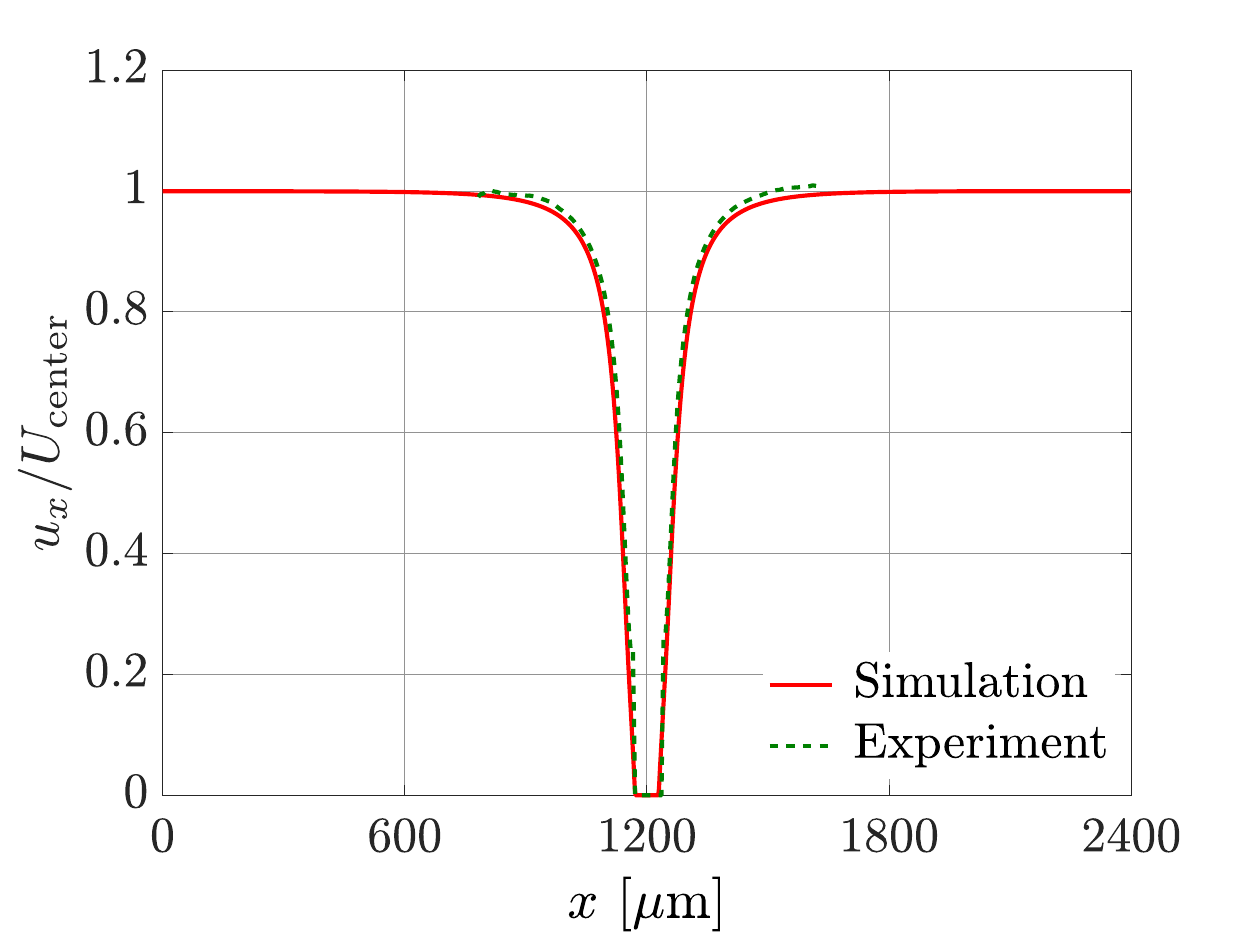}}
    \hspace{1cm}
    \subfloat[\label{subfig:velocity_profiles_along_y_LBM_PIV}]{\includegraphics[width=0.4\textwidth]{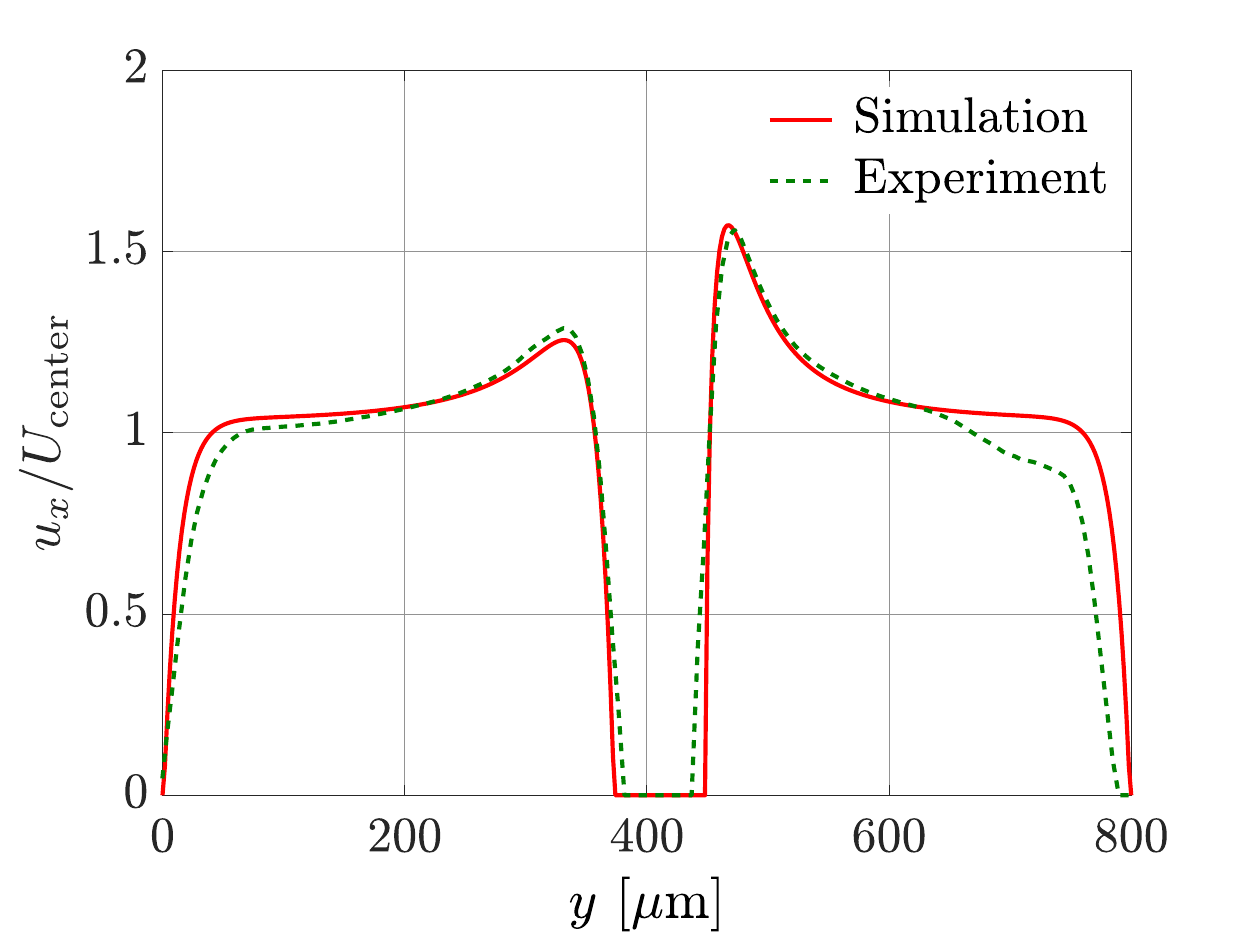}}
    \caption{Velocity fields in the vicinity of the pillar in the experiments and simulations. (a) Normalized velocity magnitude and streamlines obtained experimentally by µPIV (left) and computed by LBM simulation (right). The thick red line is the flow separatrix. (b) Velocity profiles along the $x$-axis (horizontal lines in (a)). (c) Velocity profiles along the $y$-axis (vertical lines in (a)). $U_{\rm center}$ is the velocity magnitude at the channel centerline.}
    \label{fig:velocity_fields}
\end{figure}

We first characterize the flow disturbance created by the presence of the obstacle. The flow field computed by the LBM is shown in Fig.~\ref{fig:velocity_fields} and compared to the µPIV measurements in the midplane of the experimental channel.  Note that in the Hele-Shaw configuration of our channel a Poiseuille flow develops in the $z$ direction. The flow velocity is maximal at the midplane where the velocity gradient in the $z$ direction is zero. We thus concentrate both in the experiments and in the simulations on the fiber dynamics at the midplane, where they are solely given by the flow properties in the $x$ and $y$ directions and where the out of plane shear can be neglected.
Panel \subref{subfig:velocity_fields_LBM_PIV} gives the streamlines and the normalized velocity magnitude in the neighborhood of the pillar (only a portion of the channel is represented here).
The main features of the experimental flow are well captured by the simulation. The velocity is zero along the bounding walls ($y=0$ and 800\,µm) and on the pillar surface.
The fluid is sharply accelerated right above and below the pillar and it is almost uniform far away from the pillar.  The confinement of the flow in the shallow Hele-Shaw like channel leads to a localization of the flow disturbance close to the obstacle and to a strong acceleration zone at the apex of the triangle. The lateral extent of the flow disturbance scales with the channel height \cite{Liron1976} as shown in Appendix \ref{sec:supplementary_material_flow} and gets more and more localized with decreasing channel height. At the same time the velocity gradients close to the obstacle increase. 
The red thick line is the flow separatrix which separates the streamlines going above and below the obstacle.
The velocity profiles along the $x$ and $y$ axes (horizontal and vertical lines in panel \subref{subfig:velocity_fields_LBM_PIV}) are respectively reported in Figs.~\ref{subfig:velocity_profiles_along_x_LBM_PIV} and \ref{subfig:velocity_profiles_along_y_LBM_PIV}.
The flow fields in the experiment and the simulation are in excellent quantitative agreement. Note that all streamlines are symmetric due to the symmetry of the obstacle and the fact that the Reynolds number is small.

\subsection{Fiber dynamics}
\begin{figure}[H]
    \centering
    \subfloat[\label{subfig:comparison_exp_num_below}]{\includegraphics[width=0.49\textwidth]{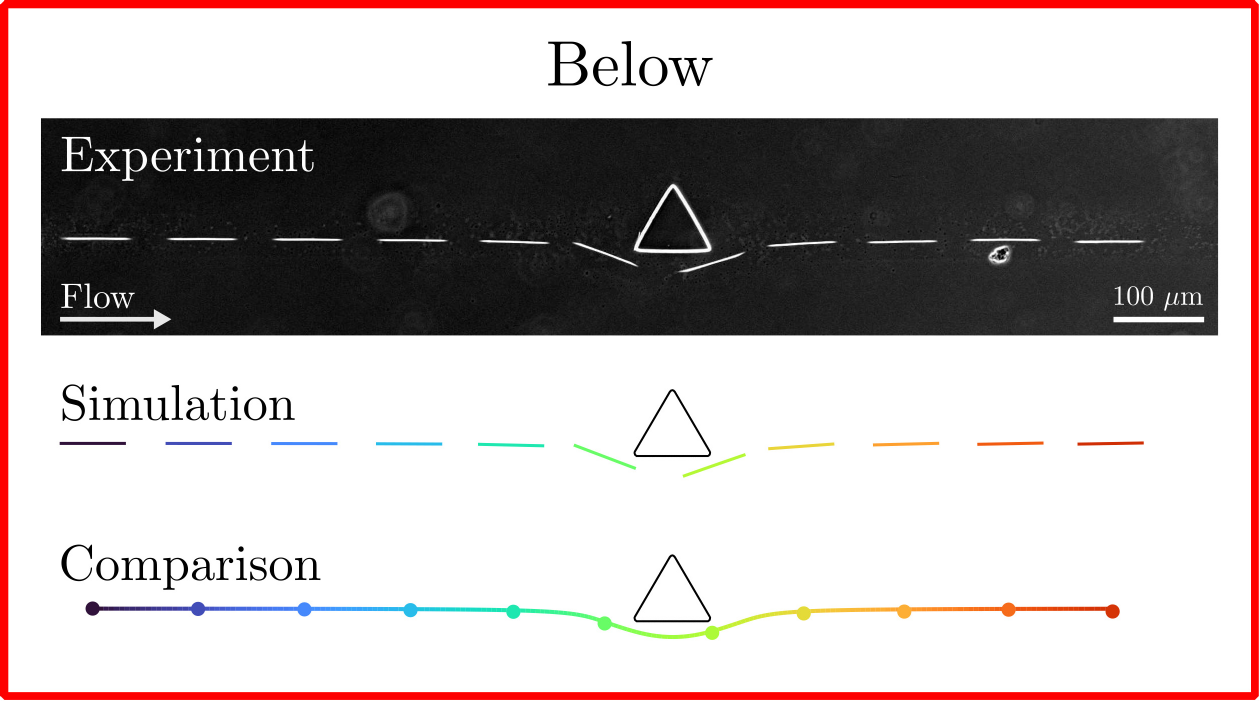}}
    \hfill
    \subfloat[\label{subfig:comparison_exp_num_above}]{\includegraphics[width=0.49\textwidth]{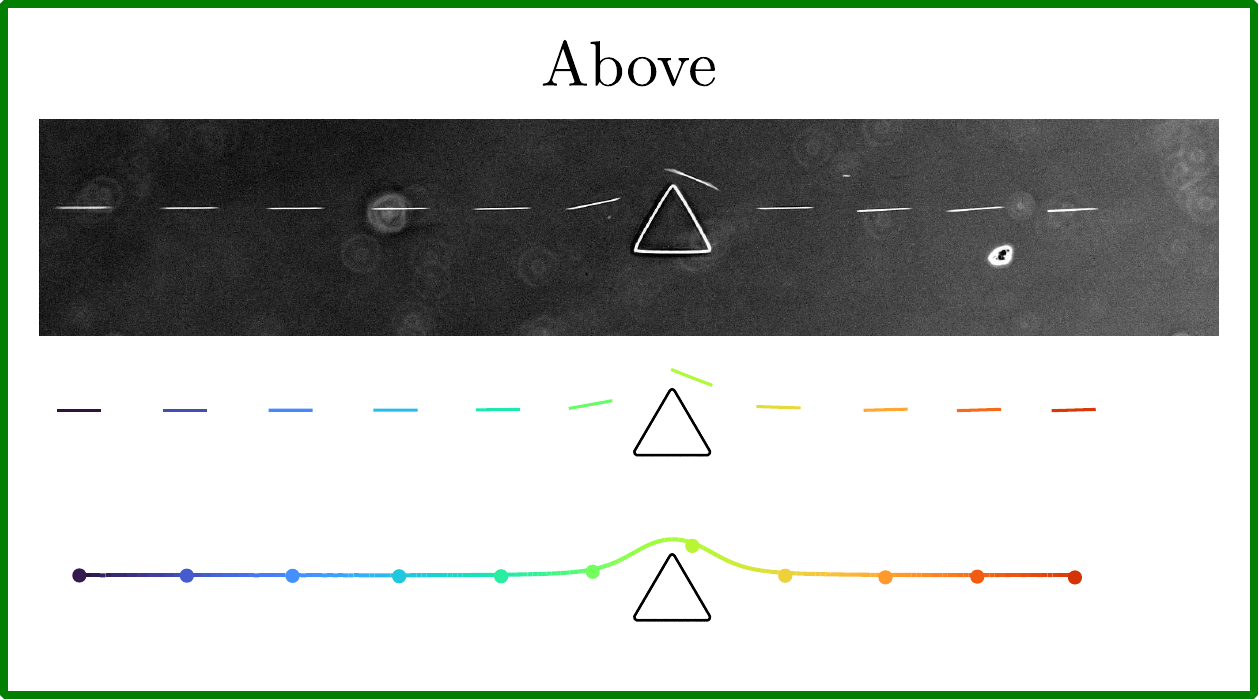}}
    \\[0.1cm]
    \subfloat[\label{subfig:comparison_exp_num_pole_vaulting}]{\includegraphics[width=0.49\textwidth]{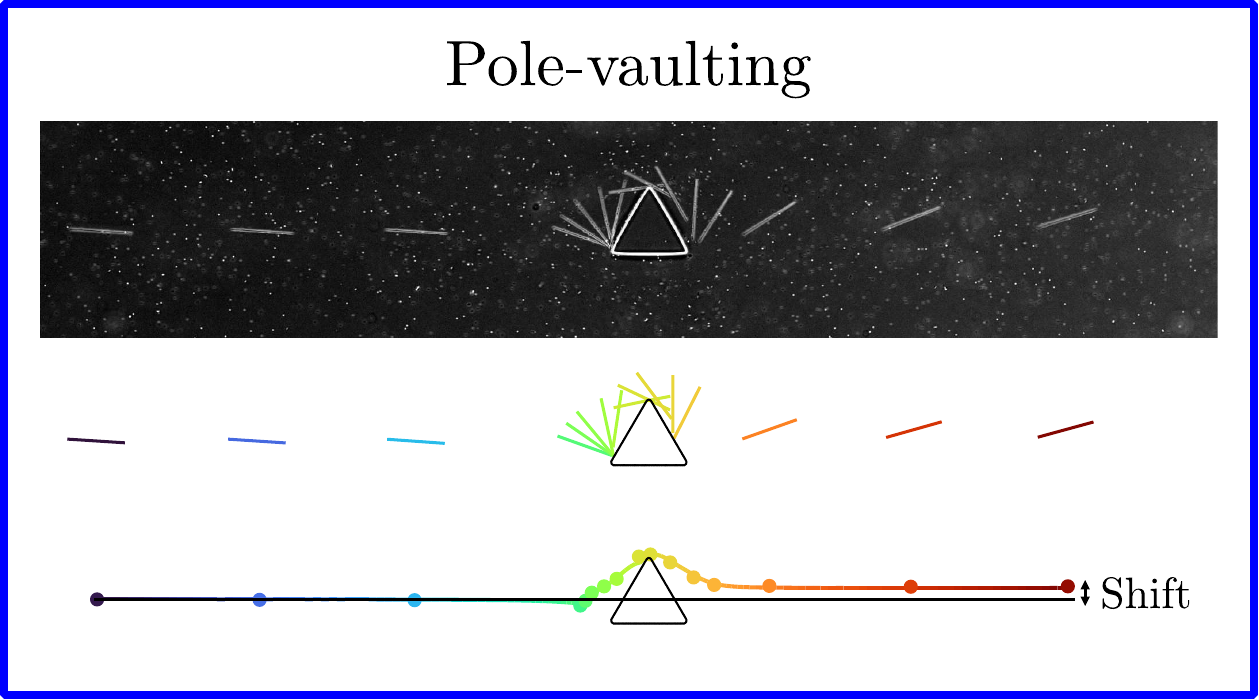}}
    \hfill
    \subfloat[\label{subfig:comparison_exp_num_trapping}]{\includegraphics[width=0.49\textwidth]{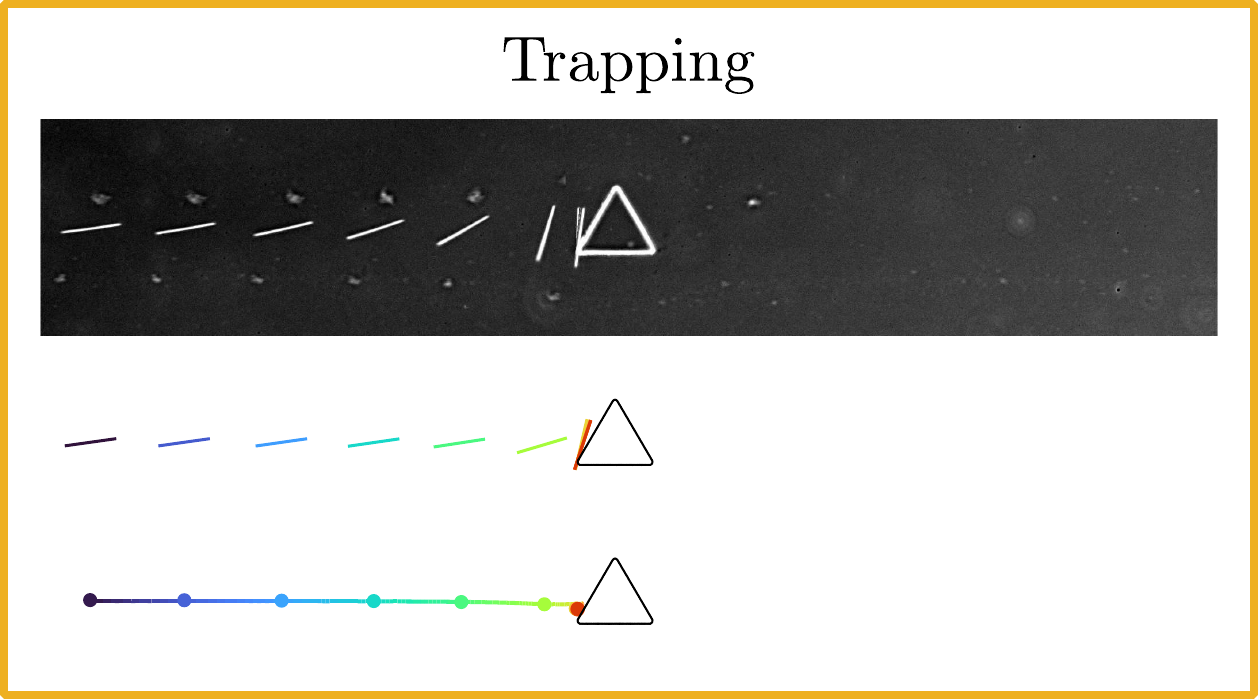}}
    \\[0.3cm]
     \includegraphics[width=0.55\textwidth]{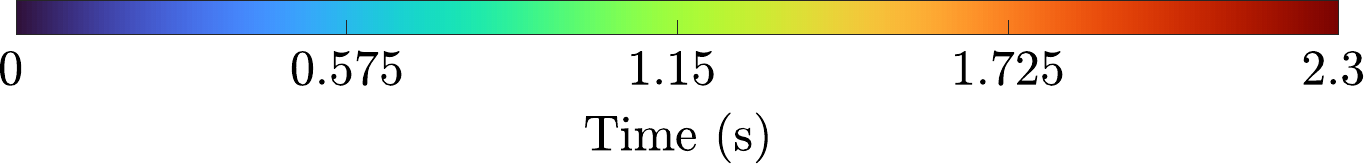}
    \caption{The four different fiber dynamics observed experimentally and accurately reproduced by simulations. The first line of each panel shows images from the experiments, the second line represents the chronophotograph of the simulation and the third line compares the trajectories of the fiber center of mass in the experiment (dots) and in the simulation (solid line). (a) The fiber goes below the pillar. \rev{Initial conditions are: $\theta_{\rm 0} = 0^{\circ}$, $y_0/h_{\rm obs} = 0.197$, and $L/l_{\rm obs} = 0.94$ both in the experiment and simulation.} (b) The fiber goes above the pillar. \rev{Initial conditions are: $\theta_{\rm 0} = 0^{\circ}$, $y_0/h_{\rm obs} = 0.678$, and $L/l_{\rm obs} = 0.64$ both in the experiment and simulation.} (c) Pole-vaulting. Vertical black double-arrow indicates the lateral shift that results from the interaction with the obstacle. \rev{Initial conditions are: $\theta_{\rm 0} = -3.7^{\circ}$, $y_0/h_{\rm obs} = 0.373$, and $L/l_{\rm obs} = 0.82$ both in the experiment and simulation.}(d) Permanent trapping. \rev{Initial conditions are: $\theta_{\rm 0} = 8^{\circ}$, $y_0/h_{\rm obs} = 0.351$, and $L/l_{\rm obs} = 0.75$ both in the experiment and simulation.}. The colors in the two last rows of each panel indicate the time with the color-code indicated at the bottom of the figure. All figures share the same scale bar (100\,µm).}
    \label{fig:comparisons_exp_num}
\end{figure}

The dynamics of rigid fibers is investigated as a function of their length $L$, initial angle $\theta_0$ and lateral position $y_0$ at the channel entry. A lateral position of $y_0=0$ corresponds to the position of the base of the triangle. We thus refer to increasing $y_0$ as positioning the fiber ``higher" whereas decreasing $y_0$ corresponds to positioning the fiber ``lower" along the triangle.
More than 200 experiments and 1300 numerical simulations have been performed with a single isolated fiber for various initial conditions.
Four different fiber dynamics have been observed both experimentally and numerically depending on $\theta_0$, $y_0$ and $L$.
Typical examples from experiments and simulations are shown in Fig.~\ref{fig:comparisons_exp_num}. For these examples the exact initial conditions from the experiments have been used as a starting point for the simulations.
In panel \subref{subfig:comparison_exp_num_below}, the fiber is initially parallel to the flow ($\theta_0=0$) and almost aligned with the base of the pillar ($y_0 \approx 0$).
It simply follows a  mostly symmetric trajectory and goes below the pillar.
This dynamics is referred to as ``Below" in the following.
For panel \subref{subfig:comparison_exp_num_above}, the fiber has initially a larger $y_0$ positioning it ``higher" with respect to the base of the triangle.
It also follows a mostly symmetric trajectory which, here, goes above the pillar.
This dynamics is referred to as ``Above".
For these two cases, the fiber does not approach the obstacle closely and nearly goes back to its initial lateral position and angle far away downstream. %because the flow and the vorticity are symmetric.
However, for intermediate initial lateral positions
%\Anke{Would it be ok to discuss things with respect to the flow separatrix and not the obstacle center of mass which is not really meaninful? We have introduced the separatrix right before that that could work? Or we can just say intermediate $y_0$ then we do not give the idea of the separatrix separating above and below away yet! But I would delete the term "center of mass of the triangle".})  
the fiber seems to establish contact with the pillar before passing either below or above it.
The fiber in panel \subref{subfig:comparison_exp_num_pole_vaulting} has such an intermediate initial position and a slightly negative angle.
In this case, the fiber strongly interacts with the pillar.
Its front remains blocked at the left edge of the obstacle for a short period of time, resulting in the rotation of the fiber around the front followed by a switch of its front and rear. This dynamics is referred to as ``Pole-vaulting" motion. Here, the fiber does not go back to its initial configuration ($\theta_0,y_0$) far away downstream.
It migrates across streamlines and remains laterally shifted as indicated in panel \subref{subfig:comparison_exp_num_pole_vaulting}.
Finally for panel \subref{subfig:comparison_exp_num_trapping} the fiber has a positive initial angle and an intermediate initial lateral position.
It gets trapped at the left tip of the pillar and finds an equilibrium position there, which is referred to as ``Trapping" in the following.
%Trapping events result from a balance of the hydrodynamic forces exerted on the fiber on both sides of the contact point.  Since the flow is stronger below the point of contact than above, the fiber chooses an asymmetric configuration toward the top to balance these two sides. Trapping events are thus very sensitive to the fiber length and to the position of this contact point.

%In the experiment shown in Fig.~\ref{subfig:comparison_exp_num_trapping}, the fiber starts rotating far upstream from the pillar, which is not the case in the simulation. This may be due to the presence of debris that might locally disturb the flow field. This could also explain why the angle of the fiber at equilibrium is a bit higher in the experiment.
Despite slight differences observed, that may be attributed to experimental imperfections \rev{and to the fact that fiber-wall hydrodynamic interactions are neglected},  \rev{we always obtain a very good agreement in both space and time between experiments and  simulations whenever initial conditions are identical}.
%and the characteristics of the four reported dynamics are reproduced in experiments and simulations. 
Movies of the four cases presented in Fig.~\ref{fig:comparisons_exp_num} are provided in the Supplementary Materials.

\subsection{Effect of the initial conditions on the fiber dynamics}
\begin{figure}[H]
    \centering
    \includegraphics[width=0.8\textwidth]{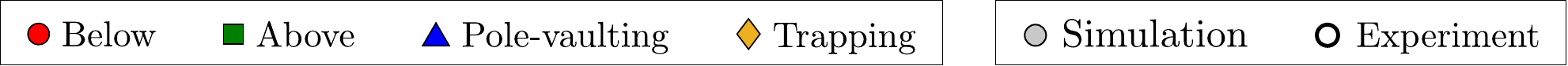}
    \\[-0.2cm]
    \subfloat[\label{subfig:dynamics_exp_num_L0o8}]{\includegraphics[width=0.48\textwidth]{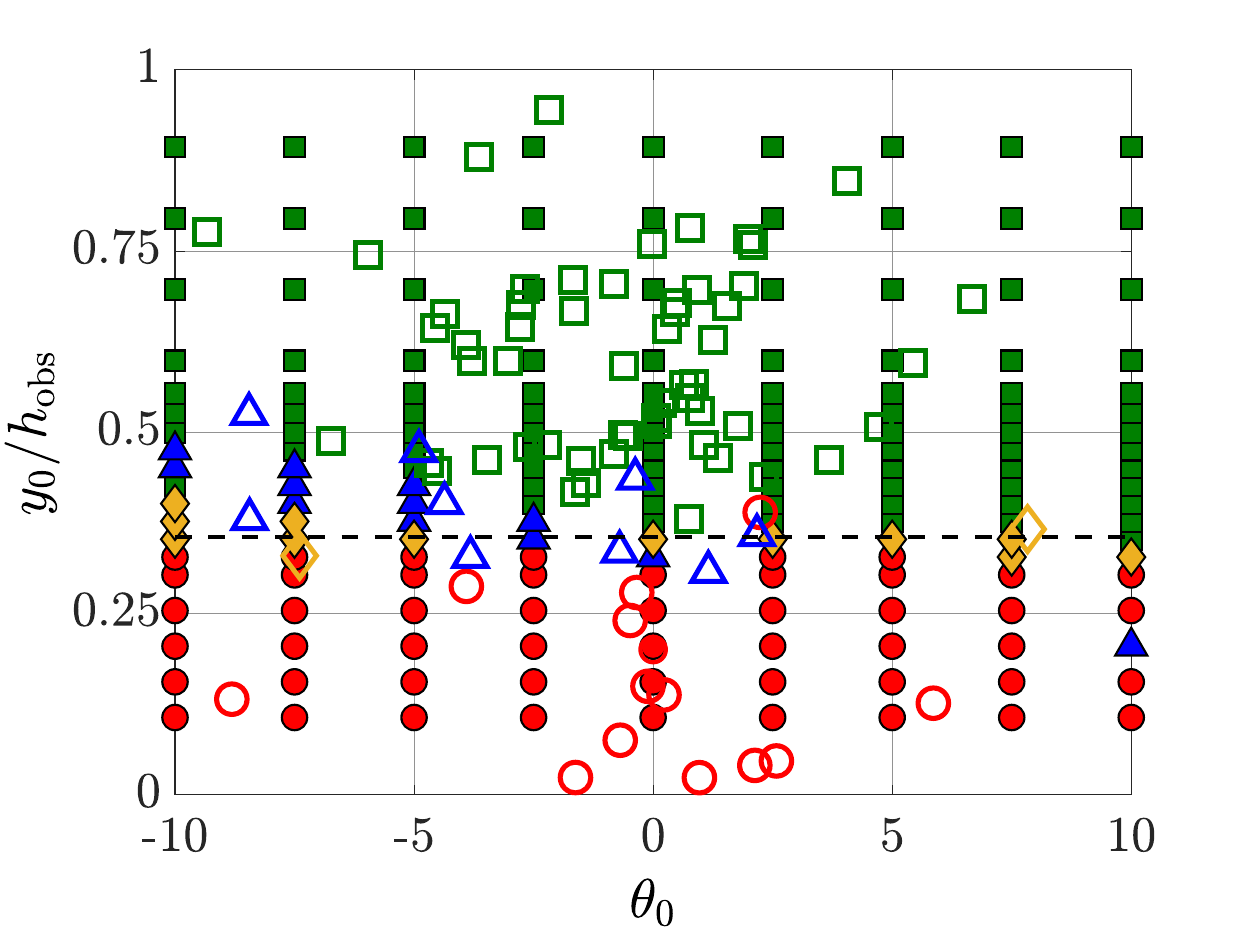}}
    \hfill
     \subfloat[\label{subfig:dynamics_3D}]{\includegraphics[width=0.48\textwidth]{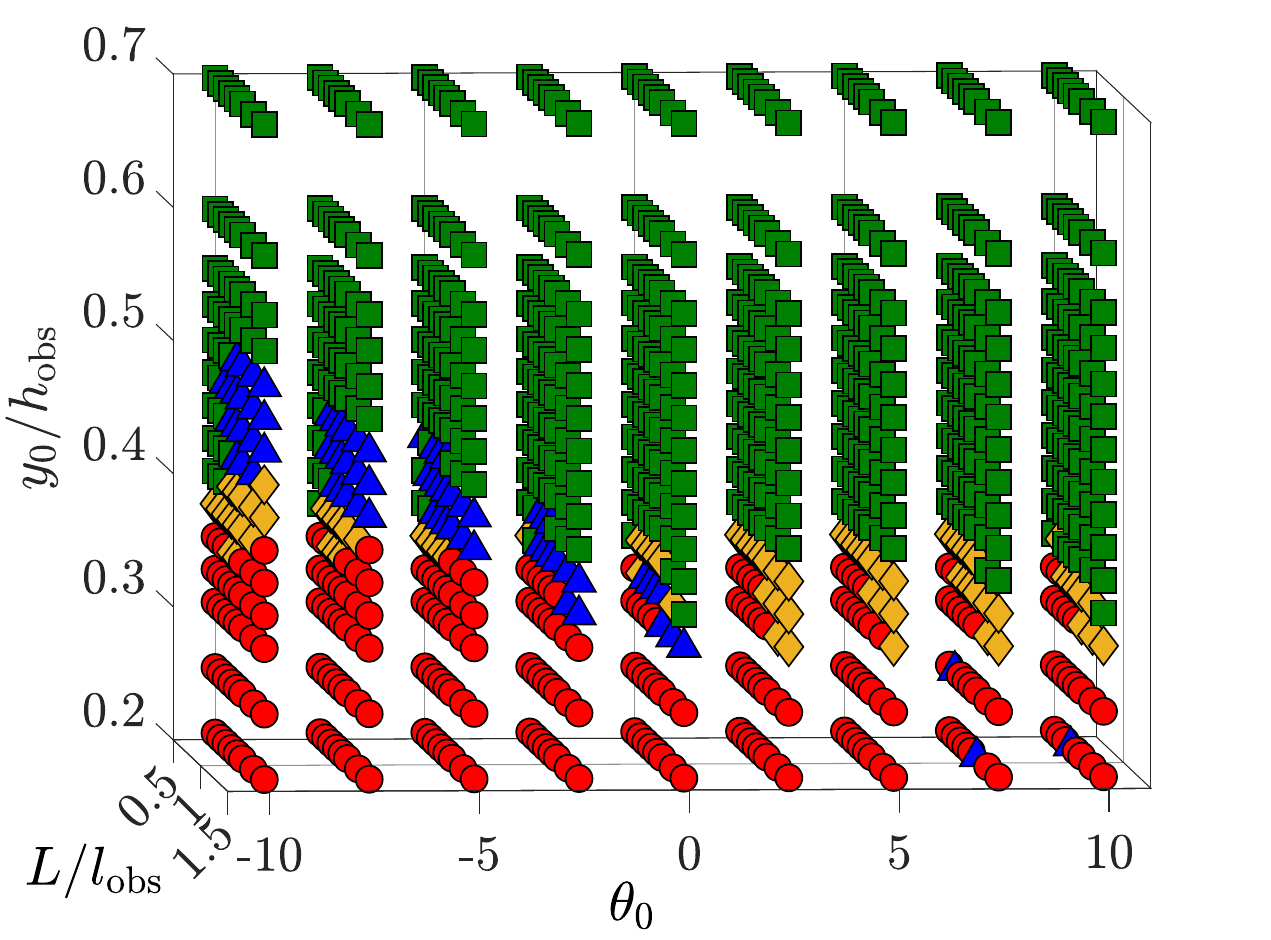}}
    \caption{Effect of the fiber initial configuration on the resulting dynamics. (a) Comparison between experiments (open symbols) and simulations (closed symbols). The fiber lengths are $0.5l_{\rm obs} \leq L \leq 1.5l_{\rm obs}$ in the experiments and $L=0.8l_{\rm obs}$ in the simulations, which is the average length of the fibers in the experiments. The dashed line represents the position of the flow separatrix at the entrance of the channel.
    (b) Simulated dynamics for different fiber lengths ranging in $0.5l_{\rm obs} < L < 1.4l_{\rm obs}$.
    Red circles: The fiber goes below the pillar; Green squares: The fiber goes above the pillar; Blue triangles: Pole-vaulting; Yellow diamonds: Permanent trapping.}
    \label{fig:effect_initial_configuration_dynamics}
\end{figure}

 As illustrated in Fig.~\ref{fig:comparisons_exp_num}, the dynamics of the fibers is  dependent on their initial angle $\theta_0$ and lateral position $y_0$ when they enter the channel.
This is shown more broadly in Fig.~\ref{fig:effect_initial_configuration_dynamics} where
panel \subref{subfig:dynamics_exp_num_L0o8} is a comparison of the fiber dynamics observed experimentally (open symbols) and obtained numerically (closed symbols) for a large number of initial configurations $(\theta_0,y_0)$, and for fiber length $0.5l_{\rm obs} \leq L \leq 1.5l_{\rm obs}$ in the experiments and $L=0.8l_{\rm obs}$ in the simulations, which is the average length of the fibers in the experiments. The different dynamics are determined visually from the trajectories, with the exception of the ``Pole-vaulting" dynamics. As the velocity of the fiber head is not always strictly zero during the rotation phase both in the experiments and simulations we consider the head of the fiber to be ``blocked" if $U_{\rm head} < 0.01 U_{\rm center}$ in the simulations, and if it has no perceptible motion in the experiments.
%\Olivia{Olivia: for the experiments I guess it is a range of Lengths. Zhibo, could you give the range here (and also in the caption of fig5)? \Zhibo{Done!} $Y_O$ is normalized by $h_{obs}$ while L is normalized by $L_{obs}$. is there a reason not to normalize by the same number?} \Zhibo{For the normalization, I don't think there is a specific reason. Maybe it's just more natural to normalize the $y$ using verticle $h_{obs}$ and $L$ using horizontal $l_{obs}$?}.
In the experiments, most of the fibers enter the channel with a small initial angle ($\theta_0 \approx 0$) due to the flow-focusing, and very few of them have an initial angle $|\theta_0| > 5$.
The black dashed line indicates the position of the flow separatrix far away from the pillar.
It separates the ``Below'' and ``Above'' dynamics well.
The fibers that are initially located at a certain distance below the separatrix go below the obstacle (red circles), and those that are initially located above the separatrix pass above the obstacle (green squares) regardless of their initial angle.
Most of these fibers exhibit a  mostly symmetric trajectory and do not approach the pillar very closely.
The situation is more complex close to the separatrix, where the fibers may interact directly with the pillar.
Here, the four dynamics coexist and the behavior of the fibers is strongly dependent on their initial angle.
For negative initial angles, fibers above the separatrix are more likely to have a pole-vaulting motion (blue triangles), while for positive angles they generally slide over the pillar and pass above it.
However, some pole-vaulting events also happen for positive angles.
For these rare cases, the fiber front reaches the left vertex of the pillar, rotates counter-clockwise and passes below the pillar.
The fibers that are initially located very close to the flow separatrix may also remain trapped for both positive and negative initial angles (yellow diamonds). These events occur slightly above the separatrix for negative angles and slightly below the separatrix for positive angles.  Trapping events result from a balance of the hydrodynamic forces exerted on the fiber on both sides of the contact point \rev{(see Appendix \ref{sec:appendix_hydrodynamic_forces_trapping})}.
Since the flow is stronger below the point of contact than above, the fiber chooses an asymmetric configuration toward the top to balance the forces on the two sides (see Fig.~\ref{subfig:comparison_exp_num_trapping}). Trapping events are thus very sensitive to the position of this contact point and can only occur in a very narrow range of initial conditions.

The dynamics observed for the experiments and the simulations are in excellent agreement.
In both cases, all the trapping events are located very close to the separatrix and the pole-vaulting events are observed for the same range of values of ($\theta_0,y_0$).
As trapping results from a very fine balance of the hydrodynamic forces acting on the fiber, it is very sensitive to experimental noise such as disturbances of the flow field
%and to the roughness of the obstacle,
which could explain why there are fewer trapping events observed in the experiments than in the simulations. 
% \Anke{I would move thise sentence later.} Note that, according to our simulations, friction is not \textit{necessary} for trapping to occur, but it increases the range of positions leading to trapping. \Anke{end of sentence to be moved.}

The length of the fibers also influences the resulting dynamics in some cases.
Figure \ref{subfig:dynamics_3D} shows the fiber dynamics obtained from simulations while varying $(\theta_0,y_0)$ as well as the fiber length $L$.
We recover the same regions as in panel \subref{subfig:dynamics_exp_num_L0o8}, with the ``Below" and ``Above" dynamics respectively for low and high $y_0$, the diagonal of pole-vaulting for $\theta_0 <0 $, and all the trapping events around the flow separatrix.
The fiber length does not affect the resulting dynamics when $y_0/h_{\rm obs} < 0.2$ or $y_0/h_{\rm obs} > 0.5$ because for these initial lateral positions the fiber only weakly interacts with the pillar.
It passes either below or above the pillar depending only on the value of $y_0$.
On the contrary, for $y_0/h_{\rm obs} \in [0.2;0.5]$ the fiber interacts closely with the pillar and the resulting dynamics also depends on the fiber length.
At a given $(\theta_0,y_0)$, long fibers are more likely to pole-vault or get trapped than short fibers, that generally follow the ``Above" or ``Below" dynamics.

\subsection{Fiber-obstacle interactions}

So far, we have shown how the configuration at the channel entry governs the dynamics of the fiber when passing the obstacle.   At the inlet, the streamlines are almost straight because the flow disturbances from the obstacle are weak. Closer to the obstacle the flow becomes more intricate and curved (see Fig.\ \ref{fig:velocity_fields}). Due to its finite size and elongated shape, the fiber will not necessarily follow these streamlines or maintain its initial orientation as it approaches the obstacle and it is the fiber configuration near the obstacle that ultimately determines its subsequent dynamics. For instance, direct contact with obstacle is evidently involved in the ``Trapping" and ``Pole-vaulting" dynamics, but the effect of close interactions on the ``Below" and ``Above" dynamics is less clear. This is why it is useful to analyze the fiber conditions close to the obstacle, correlate them with the different dynamics and link them to the initial conditions at the inlet.
%, whether this interaction is close or not. However, we have not yet described how the initial conditions, in turn, define the contact conditions and if they are unique.
%This is explained in this section.

\begin{figure}[H]
    \settoheight{\imageheight}{\includegraphics[width=0.4\textwidth]{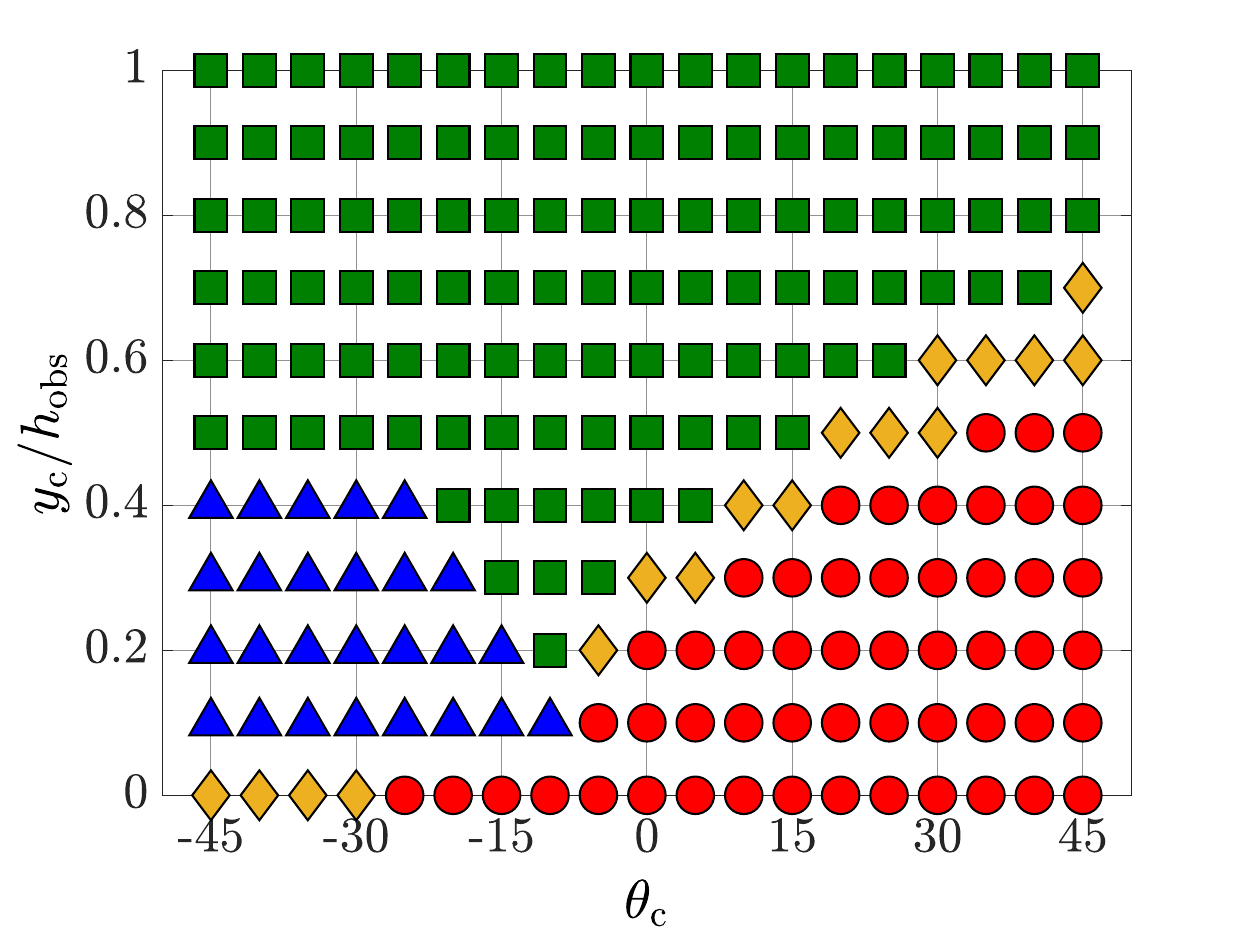}}
    \centering
    \includegraphics[width=0.45\textwidth]{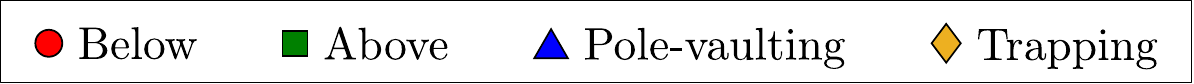}
    \\[-0.2cm]
    \subfloat[\label{subfig:contact_position_schematics}]{\tikz\node[minimum height=\imageheight]{\includegraphics[width=0.25\textwidth]{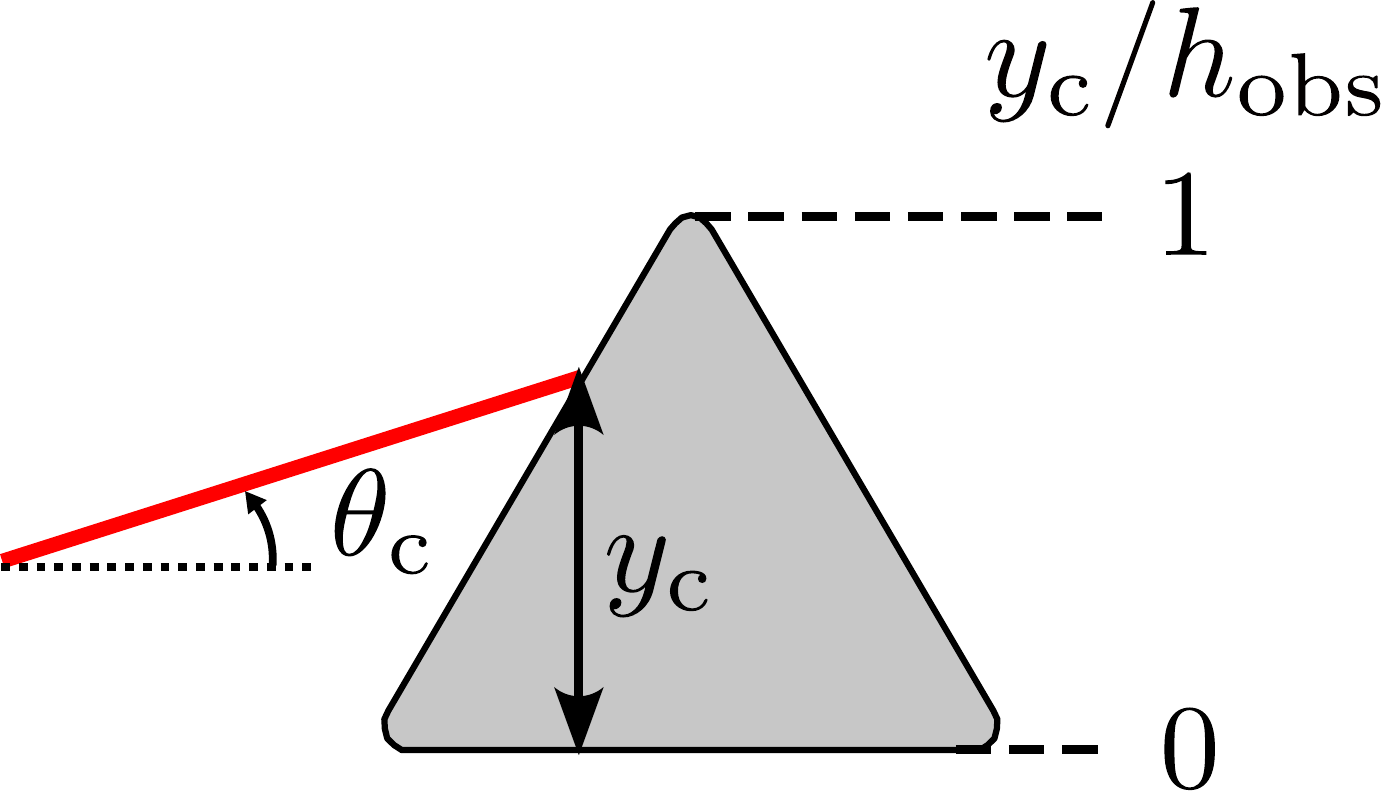}};}
    \hfill
    \subfloat[\label{subfig:num_thetaC_yC_simufromcontact}]{\includegraphics[width=0.4\textwidth]{thetaC-yC_dyn_simu_from_contact-eps-converted-to.pdf}}
    \hfill
    \subfloat[\label{subfig:sketch_competition_flow_contact_t2}]{\tikz\node[minimum height=\imageheight]{\includegraphics[width=0.28\textwidth]{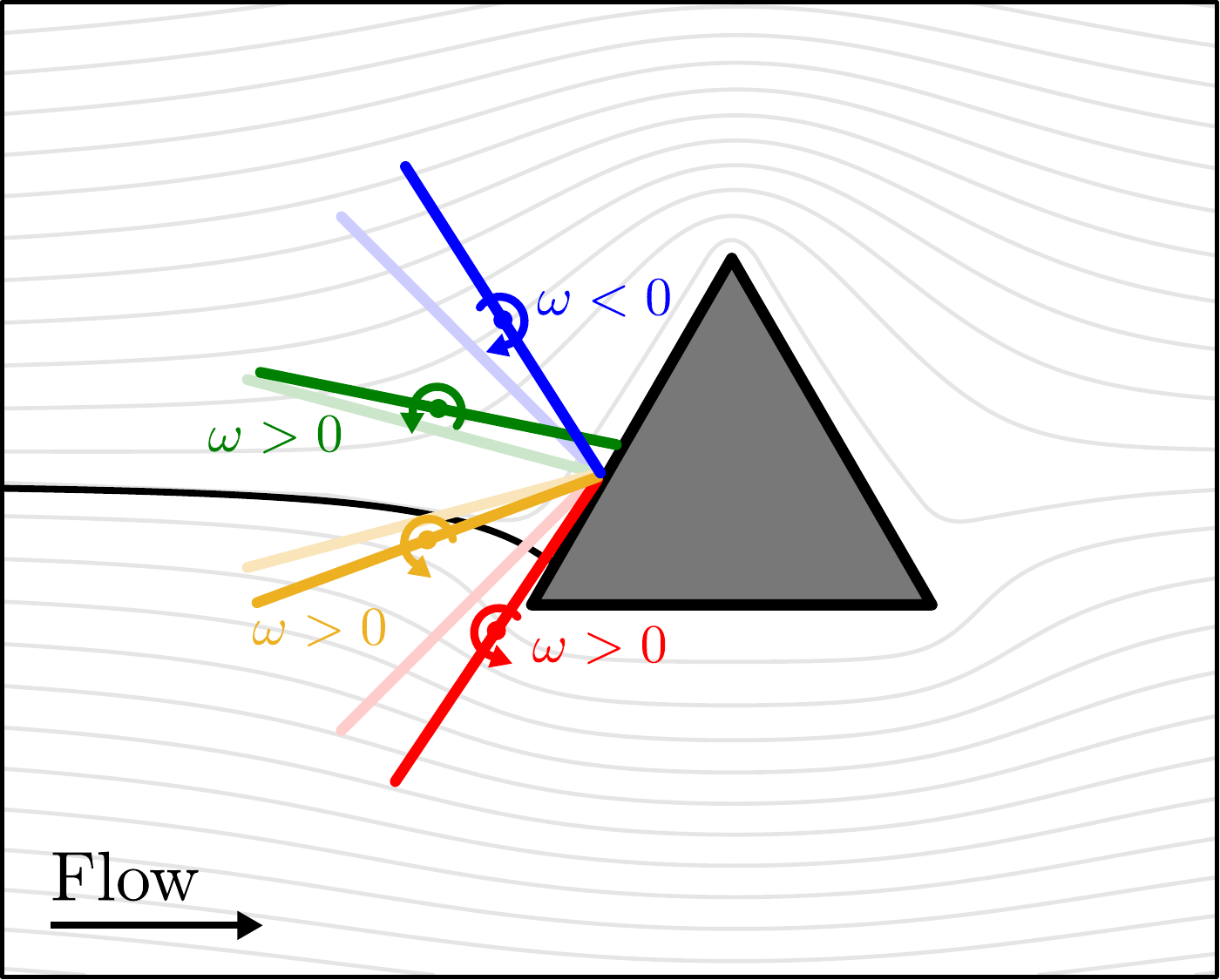}}; }
    \caption{Effect of the fiber contact configuration on the dynamics. (a) Definitions of the contact angle $\theta_{\rm c}$ and the contact position $y_{\rm c}$. (b) Phase diagram showing the dynamics of the fiber as a function of $\theta_{\rm c}$ and $y_{\rm c}$ in simulations where the fiber is initially in contact with the pillar at a well-controlled position $(\theta_{\rm c}=\theta_0, y_{\rm c}=y_0)$. The fiber length is $L/l_{\rm obs}=1$. Red circles: The fiber goes below the pillar; Green squares: The fiber goes above the pillar; Blue triangles: Pole-vaulting; Yellow diamonds: Permanent trapping. (c) Sketch showing the motion of  fibers starting at a contact position $y_{\rm c}/h_{\rm obs} = 0.4$ and $\theta_{\rm c}=-45^{\circ},-15^{\circ},15^{\circ}$ and $45^{\circ}$. Same color code as in panel (b) is used to represent the fiber dynamics, light colors correspond to the contact position while darker colors to subsequent moment. \rev{Black streamline: flow separatrix}.}
    \label{fig:contact_dyn}
\end{figure}

We will specifically analyze fibers in direct contact with the obstacle in this section. We define direct contact to take place when the repulsive force ${\bf F}^{\rm R}$ becomes non zero in the simulations. In the experiments we use visual observations and define contact when no visible gap between fiber and pillar can be observed.  
The contact conditions between the fiber and the pillar are characterized by the angle $\theta_{\rm c}$ and the lateral position $y_{\rm c}$ of the fiber when it first touches the pillar (see Fig.~\ref{subfig:contact_position_schematics}).
Their influence on the motion of the fiber is discussed in what follows.

Figure \ref{subfig:num_thetaC_yC_simufromcontact} gives the resulting dynamics in simulations for which the fiber is initially placed directly in contact with the pillar at a well controlled contact configuration $\theta_0 = \theta_{\rm c}$ and $y_0 = y_{\rm c}$.
In these simulations, the fiber length is set to $L/l_{\rm obs} = 1$.
The four fiber dynamics occupy well distinct regions in the $(\theta_{\rm c},y_{\rm c})$ space.
Pole-vaulting events occur when $\theta_{\rm c} < 0$ and $0.1 \leq y_{\rm c}/h_{\rm obs} \leq 0.4$.
The ``Below" dynamics is obtained for both positive and negative contact angles, and up to $y_{\rm c}/h_{\rm obs} = 0.5$.
The ``Above" and ``Below" domains are well separated by a thin region of trapping events, which reveals the existence of an equilibrium contact configuration between these two dynamics.
There are also some trapping events for $y_{\rm c}/h_{\rm obs} = 0$ and $\theta_{\rm c} \leq -30^\circ$.
For those cases, the fiber rotates and slides around the left apex of the pillar, and it finds an equilibrium position where it remains trapped.

The motion of the fiber just after the contact, and thus its direction of rotation, results from the complex interplay between the flow and the repulsive force.
The angular velocity of the fiber $\omega$ can be decomposed as the sum of the rotation induced by the contact and the rotation induced by the flow.
It depends on both the orientation of the fiber and its position along the edge of the pillar, which, themselves, vary over time.
This is illustrated in Fig.~\ref{subfig:sketch_competition_flow_contact_t2} on an example where the same lateral contact position leads to the four different dynamics when the contact angle is varied.
This sketch shows the evolution of four fibers starting at $y_{\rm c}/h_{\rm obs} = 0.4$ at $t = t_0$ (light fibers), but with different contact angles $\theta_{\rm c}$.
The blue and red fibers have a high $\left| \theta_{\rm c} \right|$ and therefore sample many streamlines and feel a high velocity gradient.
The rotation of these fibers is mainly due to the flow which strongly pushes the blue one clockwise ($\omega < 0$) and the red one counter-clockwise ($\omega > 0$).
As a result, the blue fiber will have a pole-vaulting motion while the red one will pass below the pillar at a later time.
For intermediate contact angles, the fibers feel a lower velocity gradient and so the contact force also plays a role in their
rotation.
The green and yellow fibers thus have a lower angular velocity than the blue and red ones.
They both rotate counter-clockwise, but the green one is pushed up by the flow and slides over the edge of the pillar, and the yellow one rotates around its head and will find an equilibrium position and remain trapped on the left apex of the pillar. 

\begin{figure}[H]
    \settoheight{\imageheight}{\includegraphics[height=0.41\textwidth]{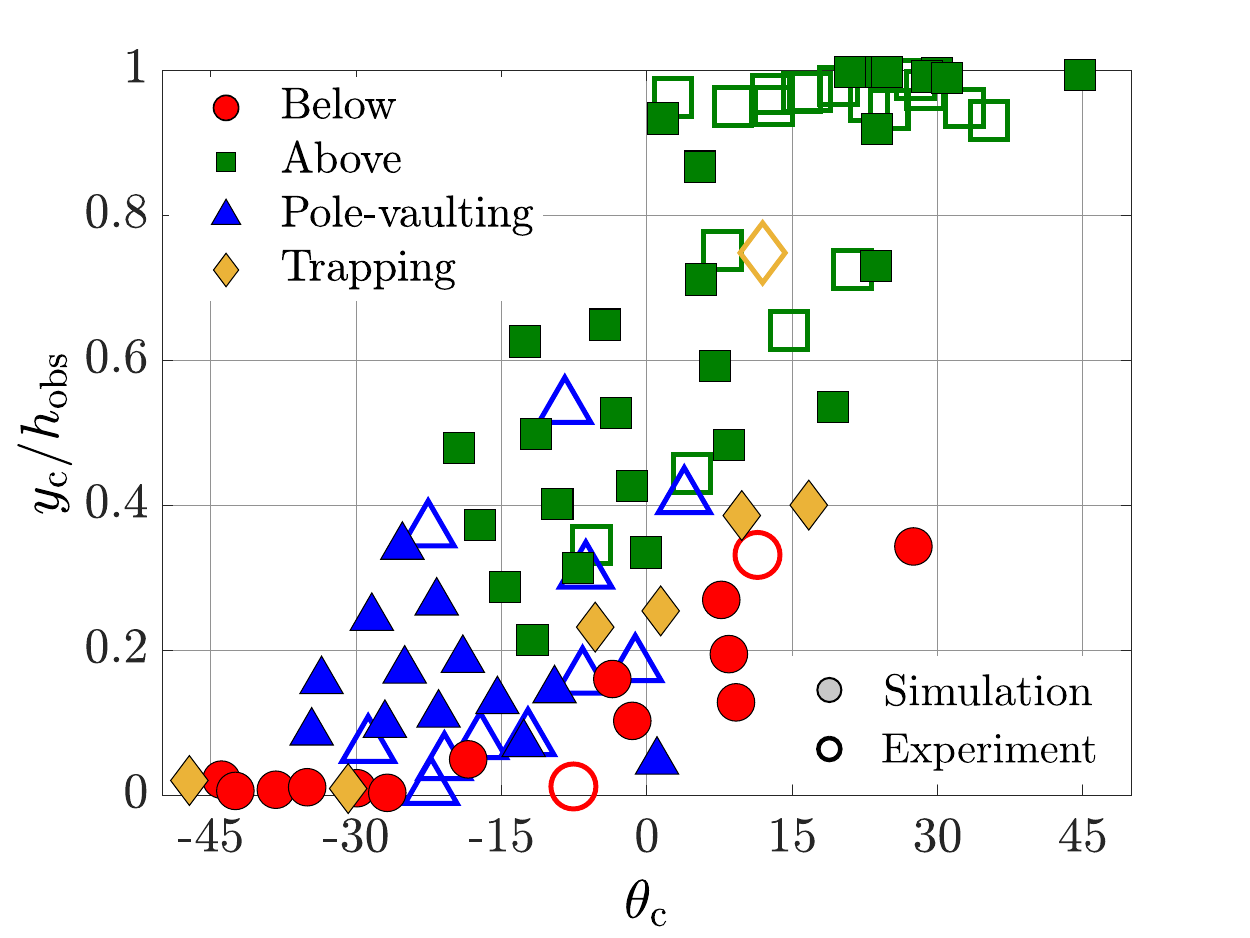}}
    \centering
    \subfloat[\label{subfig:exp_num_thetaC_yC}]{\includegraphics[height=0.38\textwidth]{thetaC-yC_dyn_simu_from_far_theta_0m10to10_L1_and_exp-eps-converted-to.pdf}}
    \hfill
    \subfloat[\label{subfig:chronophotographs_contact}]{\tikz\node[minimum height=\imageheight]{\includegraphics[height=0.3\textwidth]{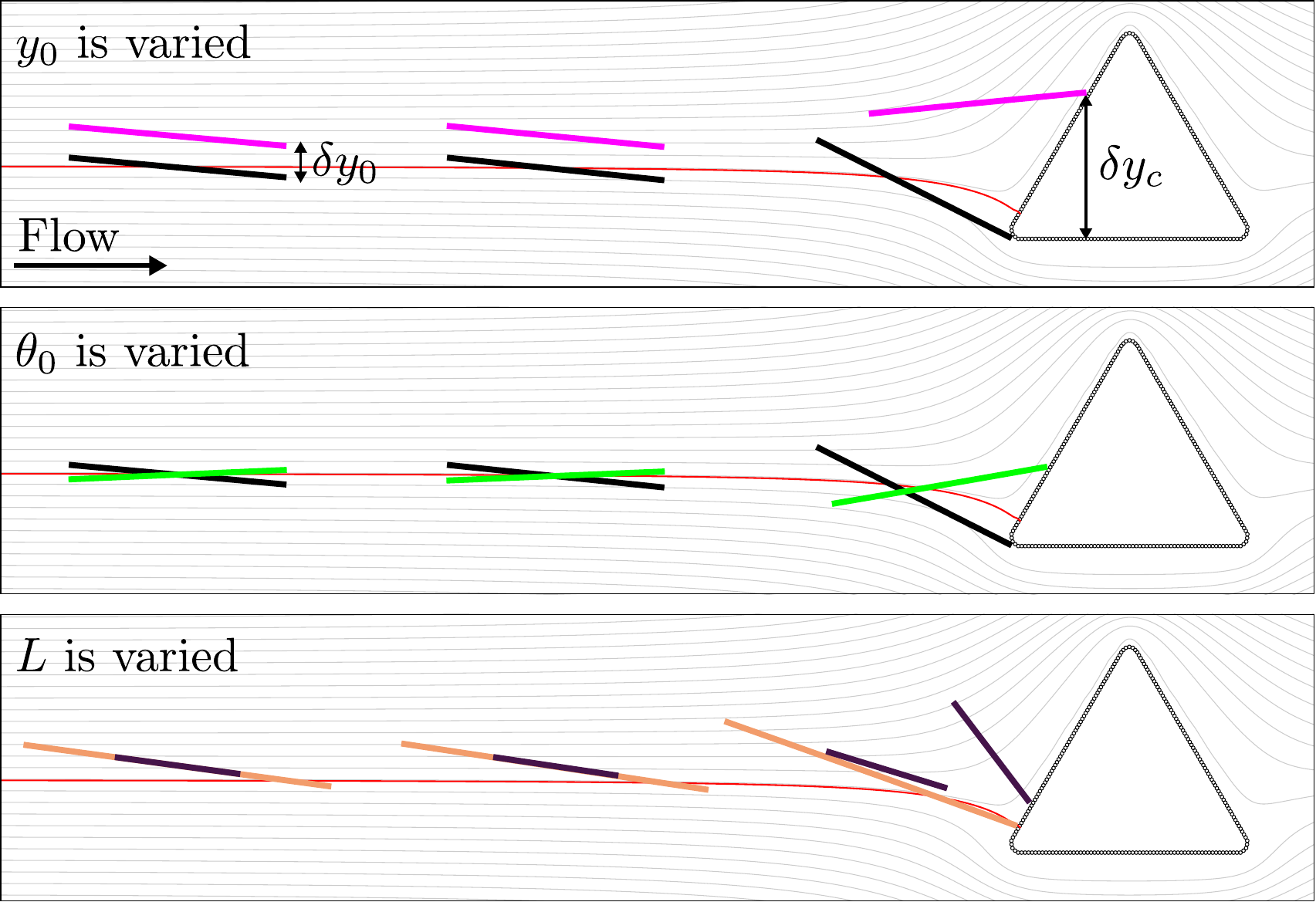}}; }
    \caption{%Relation between the initial conditions and the contact configurations.
    (a) Phase diagram representing the dynamics of fibers initially located far away from the pillar as a function of their contact configuration $\theta_{\rm c}$ and $y_{\rm c}$. Open symbols: experiments; closed symbols: simulations at $L/l_{\rm obs}=1$.
    (b) Typical examples of chronophotographs illustrating the high sensitivity of the contact configuration on $y_0$ (top), $\theta_0$ (middle) and $L$ (bottom). \rev{Red line: flow separatrix.}}
    \label{fig:vectorMaps}
\end{figure}

To investigate whether all contact conditions shown in Fig.~\ref{subfig:num_thetaC_yC_simufromcontact} can be reached when the fiber transported by the flow approaches the obstacle,  
%, the fiber is initially placed in contact with the pillar, which is of course not the case in real experiments.
we report in Fig.~\ref{subfig:exp_num_thetaC_yC} the range of contact conditions obtained for experiments (open symbols) and simulations (closed symbols) when fibers are released at the channel entry within the range of initial conditions:  $-10^{\circ} \leq \theta_0 \leq 10^{\circ}$ and $0 \leq y_0/h_{\rm obs} \leq 1$. There is an overall good agreement between the dynamics obtained experimentally and numerically in Fig.~\ref{subfig:exp_num_thetaC_yC}.
Some pole-vaulting events and one trapping event are observed experimentally at higher $y_{\rm c}/h_{\rm obs}$ compared to the simulations, which can be attributed to roughness effects and local disturbances of the flow field. The fiber dynamics at a given ($\theta_{\rm c},y_{\rm c}$) is the same in both Figs.~\ref{subfig:num_thetaC_yC_simufromcontact} and \ref{subfig:exp_num_thetaC_yC}.  
This means the fiber dynamics is uniquely determined by the configuration at contact $\theta_{\rm c}$ and $y_{\rm c}$ at a given fiber length. 
 Interestingly, several contact configurations cannot be reached and no data is observed in the top left and bottom right corners of the plot.

It is therefore interesting to relate the contact configurations to the initial conditions.
The mapping between the contact configurations and the initial conditions
\begin{equation}
    (\theta_{\rm c}, y_{\rm c}) = f(\theta_0, y_0, L)
\end{equation}
is complex due to the triangular obstacle that disturbs the flow field in its vicinity and the coupling of the fiber orientation and position due to its elongated shape. A thorough sensitivity analysis of the mapping is carried out in Appendix \ref{sec:appendix_effect_fiber_initial_position_length_contact} whereas we here briefly showcase the sensitivity of the function $f$ with respect to its parameters.
We perturb each of these parameters independently in Fig.~\ref{subfig:chronophotographs_contact}.
In the first panel, a small lateral shift above the separatrix ($\delta y_0 = 0.15 h_{\rm obs}$) leads to a significantly higher contact point ($\delta y_{\rm c}  =  0.7 h_{\rm obs}$) and opposite orientations at contact. This is due to the opposite curvature of the streamlines above and below the separatrix. Similarly, a small change in the initial orientation leads to contact configurations with opposite orientations (see second panel). Finally, the fiber length also affects the contact configuration: the longest fiber reaches the obstacle earlier than the shortest one, and has therefore less time to rotate before contact (see third panel).

% Finally, Fig.~\ref{subfig:exp_num_thetaC_yC} also 
% reports the dynamics of the fibers released at the channel entry as a function of their contact conditions.
% \Clement{I think we can remove the previous sentence since Fig.~\ref{subfig:exp_num_thetaC_yC} has already been introduced above}

%obtained numerically (closed symbols) and experimentally (open symbols) as a function of $\theta_{\rm c}$ and $y_{\rm c}$, but here, the fiber is initially located far away from the pillar (only the dynamics for which contact occurs are reported).
%In this situation the contact configuration depends on the trajectory followed by the fiber before it touches the pillar, and thus $\theta_{\rm c} \neq \theta_0$ and $y_{\rm c} \neq y_0$ in contrast to Fig.~\ref{subfig:num_thetaC_yC_simufromcontact}.

In this section we have analyzed the dynamics of fibers getting very close to the obstacle, corresponding to fibers released close to the separatrix. In this range, small differences in the initial conditions lead to strong differences in the fiber trajectories and orientation close to the obstacle due to the finite size effects in the complex disturbance flow field around the obstacle.  This explains  why in the region close to the flow separatrix very different fiber dynamics can be observed. Trajectories that do not enter into contact with the pillar just pass above or below the obstacle. Only a limited range of initial conditions leads to fiber contact with the obstacle where all four dynamics are observed. ``Trapping" separates the ``Above" and ``Below" dynamics and ``Pole-vaulting" is observed for specific contact conditions. 

To conclude this section, we briefly investigate the effect of obstacle roughness on the different dynamics, and more particularly on the ``Pole-vaulting" and ``Trapping" cases.  In Appendix \ref{sec:appendix_role_friction} we \rev{use numerical simulations} to compare the fiber dynamics obtained in Fig.~\ref{subfig:dynamics_exp_num_L0o8} for a rough obstacle, i.e.\
 exerting tangential ``friction" forces on the fiber at contact, with a perfectly smooth obstacle, for which contact forces are strictly normal to the obstacle surface. Our simulations show that trapping and pole-vaulting  still occur in the absence of roughness, but over a smaller range of initial conditions.  These results confirm that roughness is not needed but promotes these two dynamics by increasing their likelihood at contact. 
 
The last aspect we want to investigate in this study is the influence of fiber dynamics and contact with obstacles on their lateral drift.

% We have specifically analyzed the role of lateral and tangential forces during contact. \Anke{Discuss Clements findings on the role of friction here? Or later? Could also be higher up in this paragraph when we discuss the different dynamics as a function of contact condition.}  

  % \Anke{Do we really want this sentence? It is a bit trivial, as in any case things should be deterministic? I would suggest to delete it.}Each of these initial conditions leads to a unique and well defined contact condition, and one can thus determine the contact conditions and the fiber dynamics from the initial conditions. 

\section{Lateral displacement}
\label{sec:results_lateral_displacement}

The trajectories of the fibers are significantly affected by the presence of the pillar and many of them do not remain on their initial streamline. They thus do not return to their initial lateral position far away downstream, after the obstacle (\textit{e.g.} Fig.~\ref{subfig:comparison_exp_num_pole_vaulting}). Such lateral displacement could be used for fiber sorting applications and needs to be understood in order to be leveraged.
In this section, we first quantify the lateral displacement both in simulations and experiments, then investigate the mechanisms at play and show that contact with the pillar enhances this effect.

\subsection{Cross-stream migration}
\label{subsec:cross-stream migration}

The lateral displacement is quantified by
\begin{equation}
    \delta = \frac{y_{\rm f} - y_0}{h_{\rm obs}}
\end{equation}
where $y_0$ and $y_{\rm f}$ are respectively the initial (upstream) and final (downstream) lateral positions of the fiber center of mass at equilibrium far away from the pillar, and $h_{\rm obs}$ is the pillar height.
Note that due to the symmetry of the streamlines, lateral displacement is only observed in the case of \textit{cross-stream migration}.

\begin{figure}[H]
    \centering
    \settoheight{\imageheight}{\includegraphics[width=0.25\textwidth]{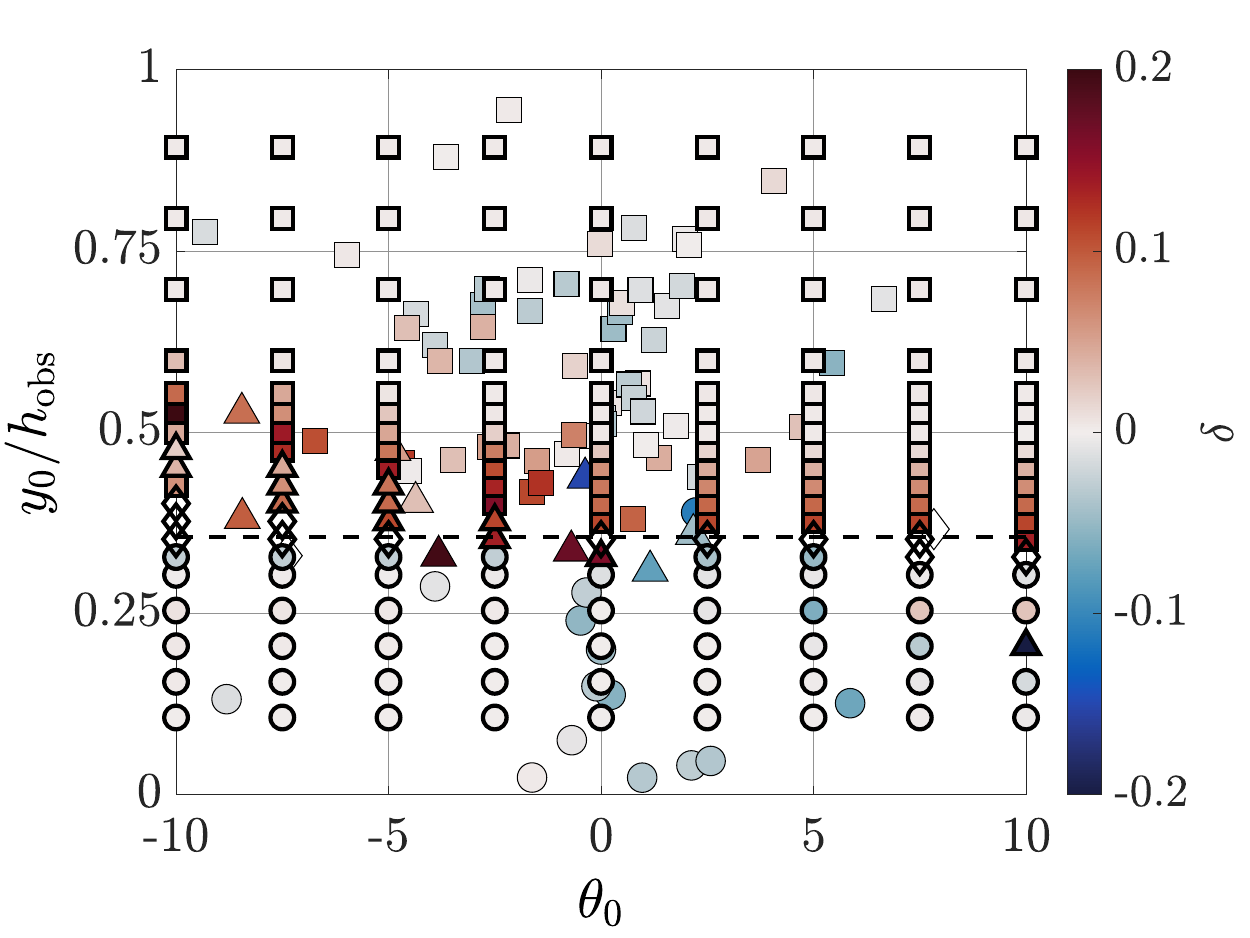}}
    \hspace*{1.2cm}\includegraphics[width=0.3\textwidth,left]{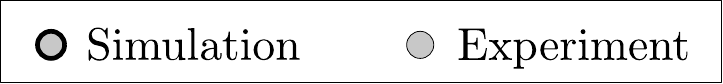}
    \\[-0.41cm]
    \subfloat[\label{subfig:deviation_exp_num_L0o8}]{\includegraphics[width=0.45\textwidth]{deviation_exp_num_L0o8_dynamics_symbols_balance_colorbar.pdf}}
    \hfill
     \subfloat[\label{subfig:deviation_3D_balance_colorbar}]{\includegraphics[width=0.5\textwidth]{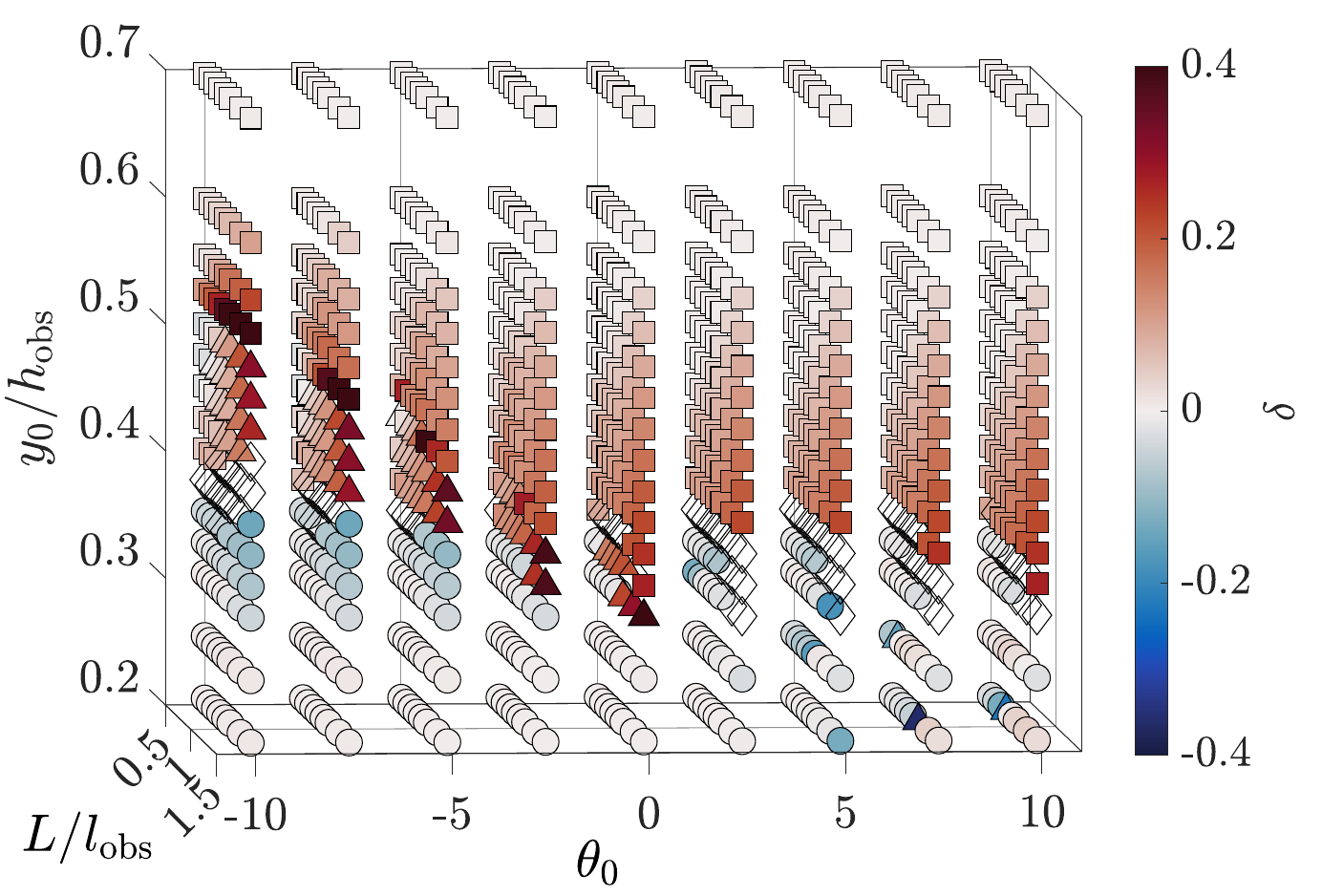}}
     \\[0.3cm]
     \includegraphics[width=0.5\textwidth]{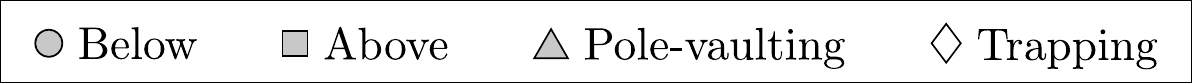}
     \\[0.5cm]
    \subfloat[\label{subfig:chronophotograph_short_long_fibers_pole_vaulting}]{\includegraphics[width=0.48\textwidth]{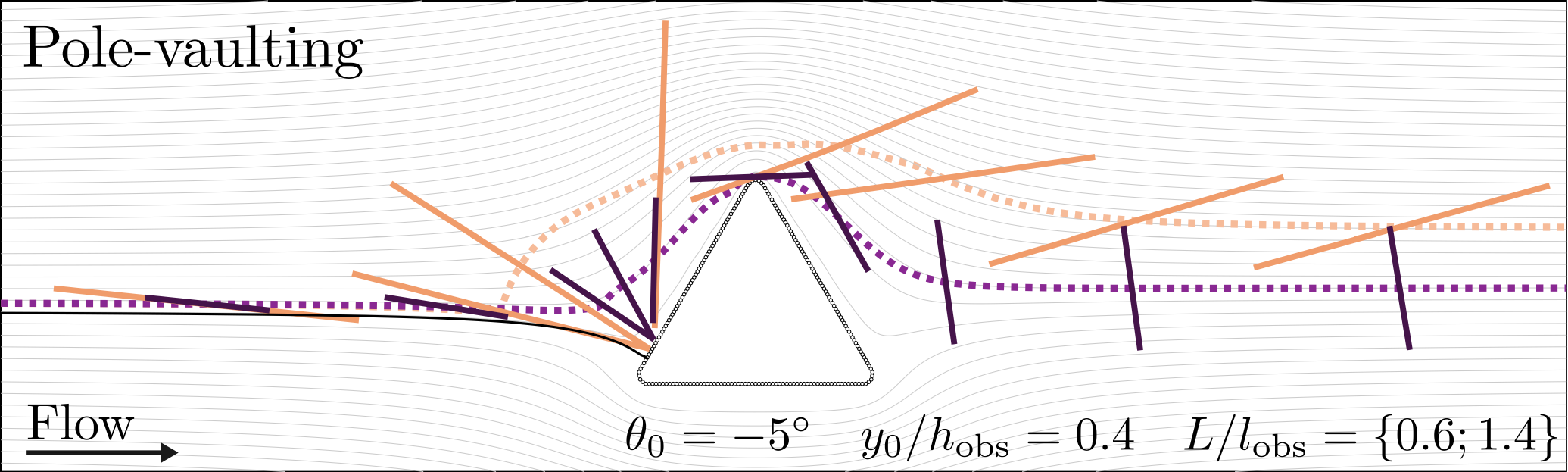}}
    \hfill
    \subfloat[\label{subfig:chronophotograph_same_length_different_initial_configuration}]{\includegraphics[width=0.48\textwidth]{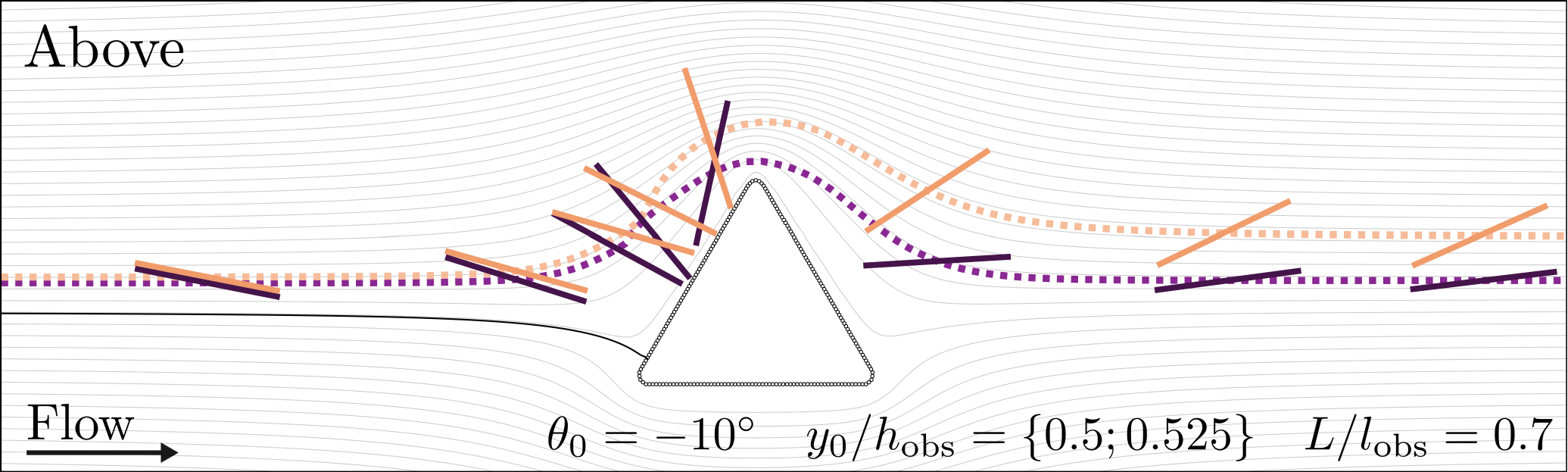}}
    \caption{Effect of the fiber initial configuration and length on the lateral displacement $\delta$.
    (a) Comparison between experiments (thin edges) and simulations (thick edges). The fiber lengths are $0.5l_{\rm obs} \leq L \leq 1.5l_{\rm obs}$ in the experiments and $L=0.8l_{\rm obs}$ in the simulations, which is the average fiber length in the experiments.
    The dashed line represents the position of the flow separatrix at the entrance of the channel.
    (b) Lateral displacement computed by numerical simulations while varying the initial angle $\theta_0$, the initial lateral position $y_0$ and the fiber length $L$.
    (c) and (d) Typical examples of chronophotographs and trajectories showing the influence of the fiber length and initial configuration on the lateral displacement.\rev{The black streamline is the flow separatrix.}}
    \label{fig:deviation_exp_num}
\end{figure}

%\begin{figure}[ht]
%\centering
%\includegraphics[width=16cm]{CrossStreamMigration_Test.png}
%\caption{\Anke{Dummy figure cross stream migration. (a) Show symbols without color for the different dynamics and color with $\delta$. Make simulations more visible than experiments (results are clearer from simulations). This panel should show the influence of the different dynamics on lateral displacement, mainly observed for pole vault, but not only. I would not show trapping (per definition no lateral displacement possible and already discussed before) (b) Same but with length. This panel is supposed to show that $\delta$ varies with $L$.} \Blaise{Ok.} \Anke{(c) trajectory with two different initial conditions showing that this leads to different $\delta$, for example.} \Blaise{Wouldn't it look too similar to Fig. 9c-d ?} \Clement{Or maybe we could show a figure similar to (d), but with the ``Above" dynamics. It will also show that the deviation is generally higher for pole-vaulting motion} \Anke{ (d) as is, showing different length lead to different $\delta$. }}
%\label{fig:CrossStreamMigration}
%\end{figure}

 Figure \ref{fig:deviation_exp_num} shows the influence of the initial condition and the fiber length on the lateral displacement.
Panel \subref{subfig:deviation_exp_num_L0o8} is a comparison of experimental data (thin edges) and simulated data (thick edges) at $L=0.8l_{\rm obs}$ (the mean fiber length in the experiments), varying $\theta_0$ and $y_0$, and panel \subref{subfig:deviation_3D_balance_colorbar} represents data extracted from the simulations while varying $\theta_0$, $y_0$ and $L$.
The fiber dynamics are coded by the same symbols as in Fig.~\ref{fig:effect_initial_configuration_dynamics}, with trapping states represented as hollow diamonds as in this case no lateral displacement can be observed.
In both panels, the darker the color, the larger the deviation.
This figure provides a link between the fiber dynamics and the lateral displacement.
The ``Pole-vaulting" dynamics leads to strong lateral deviations, while the ``Above" and ``Below" dynamics result in very small deviations, with the exception of initial conditions close to the flow separatrix (indicated by the dashed line in panel \subref{subfig:deviation_exp_num_L0o8}).
Most lateral displacements are positive meaning that the fiber will be deviated towards larger $y$ positions, but some negative (and rather small) lateral deviations are observed in particular for negative initial angles $\theta_0$.

 The lateral displacements obtained in the simulations are in rather good agreement with those observed experimentally.
In the experiments, $\delta$ is also larger close to the flow separatrix, i.e.\ in the range $0.3 \leq y_0/h_{\rm obs} \leq 0.55$ where the fibers strongly interact with the pillar, and it is rather small outside this range, where the fibers do not or weakly interact with the pillar.

Figure \ref{subfig:deviation_3D_balance_colorbar} shows the additional effect of the fiber length on $\delta$.  The lateral displacement increases with the fiber length within the window $0.3 \leq y_0/h_{\rm obs} \leq 0.55$, which means that long fibers are more laterally shifted than short fibers.
This is illustrated in Fig.~\ref{subfig:chronophotograph_short_long_fibers_pole_vaulting} which represents a typical example of chronophotographs and trajectories of a short and a long fiber following the ``Pole-vaulting" dynamics.
Both fibers have the same initial configuration $(\theta_0,y_0)$ and thus they follow the same trajectory until they approach the pillar.
Close to the pillar, the flow is highly disturbed and because the fibers have different lengths they sample different streamlines and their trajectories start to separate. Indeed, during the rotation around its tip, the short fiber's center of mass is closer to the obstacle than the center of mass of the long one.
The two fibers thus follow different streamlines after rotating around their tip during ``Pole-vaulting". At the apex of the pillar, the short fiber is horizontal and feels streamlines that are close to the obstacle and descend abruptly behind the pillar, while the long fiber is oblique and samples streamlines with a smaller vertical speed further downstream.
As a result, the long fiber has a lateral displacement $\delta=0.36$ and the short fiber $\delta=0.07$, and the two fibers end up well separated by a gap of about $0.29h_{\rm obs}$ far away downstream. This indicates that under certain conditions ($0.3 \leq y_0/h_{\rm obs} \leq 0.55$) it is therefore possible to sort fibers by length.
Such a sorting effect can occur regardless of the fiber dynamics, but is strongest for ``Pole-vaulting" (see Fig.~\ref{subfig:deviation_3D_balance_colorbar}).

The lateral displacement is also highly sensitive on the initial configuration of the fiber ($\theta_0,y_0$).
Slight variations of the initial configuration can lead to very different lateral displacements, as illustrated in Fig.~\ref{subfig:chronophotograph_same_length_different_initial_configuration}.
Both fibers have the same length and the same initial angle, but they have a slightly different initial lateral position $y_0$.
The uppermost fiber (orange, $y_0 = 0.525 h_{\rm obs}$) slides perpendicularly along the pillar and reorients only once it reaches the apex, so that it remains in a higher lateral position, while the lowest one (purple, $y_0 = 0.5 h_{\rm obs}$) also slides but reorients earlier, thus reaching the apex with a horizontal orientation and diving back down with the streamlines downstream.
This results in two very different lateral displacements ($\delta=0.01$ for the purple fiber, and $\delta=0.19$ for the orange one), and thus a large gap between both fibers downstream. 
%This finding indicates that very precise control of the initial conditions is necessary to be able to separate fibers by their length, limiting the sorting potential of the triangular pillar.

%\Blaise{I am not sure the following paragraph is useful...maybe we should choose two trajectories with different dynamics (and different $\delta$ for panel c) ? } In the simulations, $\delta$  increases suddenly near $y_0\approx 0.5 h_{\rm obs}$ at $\theta_0=-10^{\circ}$, \textit{i.e.}\ close to the transition between ``Pole-vaulting" and ``Above". Two trajectories sampled from this region are shown on Fig.~\ref{fig:deviation_exp_num}(c). The uppermost fiber (orange, $y_0 = 0.525 h_{\rm obs}$) slides perpendicularly along the pillar and reorients only once it reaches the apex, so that it remains in a higher lateral position, while the lowest one (purple, $y_0 = 0.5 h_{\rm obs}$) also slides but reorients earlier, thus reaching the apex with a horizontal orientation and diving back down with the streamlines downstream.
%\Clement{This results in two very different lateral displacements ($\delta=0.01$ for the purple fiber, and $\delta=0.19$ for the orange one), and thus a large gap between both fibers.}
%This example shows the extreme sensitivity of the lateral displacement to the initial conditions. 

\subsection{Contact enhances lateral displacement}
\label{subsec:lateral_displacement}

We have seen that significant lateral displacement is only observed for fibers released  close to the separatrix that thus pass very close to the obstacle. We now analyze the nature of the interactions between the fiber and the pillar under these conditions. From the simulations we can identify fiber trajectories where direct contact between the fibers and the pillar occurs. We recall that, in the simulations, contact is defined as the repulsive force between the fiber and the pillar surface, ${\bf F}^{\rm R}$ becoming non zero.
In the experiments, contact is assumed when no visible gap between fiber and pillar can be observed.

%\begin{figure}[ht]
%\centering
%\includegraphics[width=16cm]{MigrationAndContact.png}
%\caption{\Anke{Dummy figure migration and contact. (a) I would only show contact no contact with symbols and the color code for $\delta$. May be we can used the panel that also shows $L$ and only use the simulation data? May be the symbols for no contact could be a bit smaller? This panel should show the influence of contact on lateral displacement. I would not show trapping (per definition no lateral displacement possible and already discussed before) This figure is basically the original panel (b). (b) The contact hystogram. May be plot the blue bars in front of the orange bars? CAn we make both colored colums more visible? Plat them next to each other? (c) reversible trajectory (d) non reversible trajectory }}
%\label{fig:MigrationAndConact}
%\end{figure}

\begin{figure}[H]
\settoheight{\imageheight}{\includegraphics[width=0.48\textwidth]{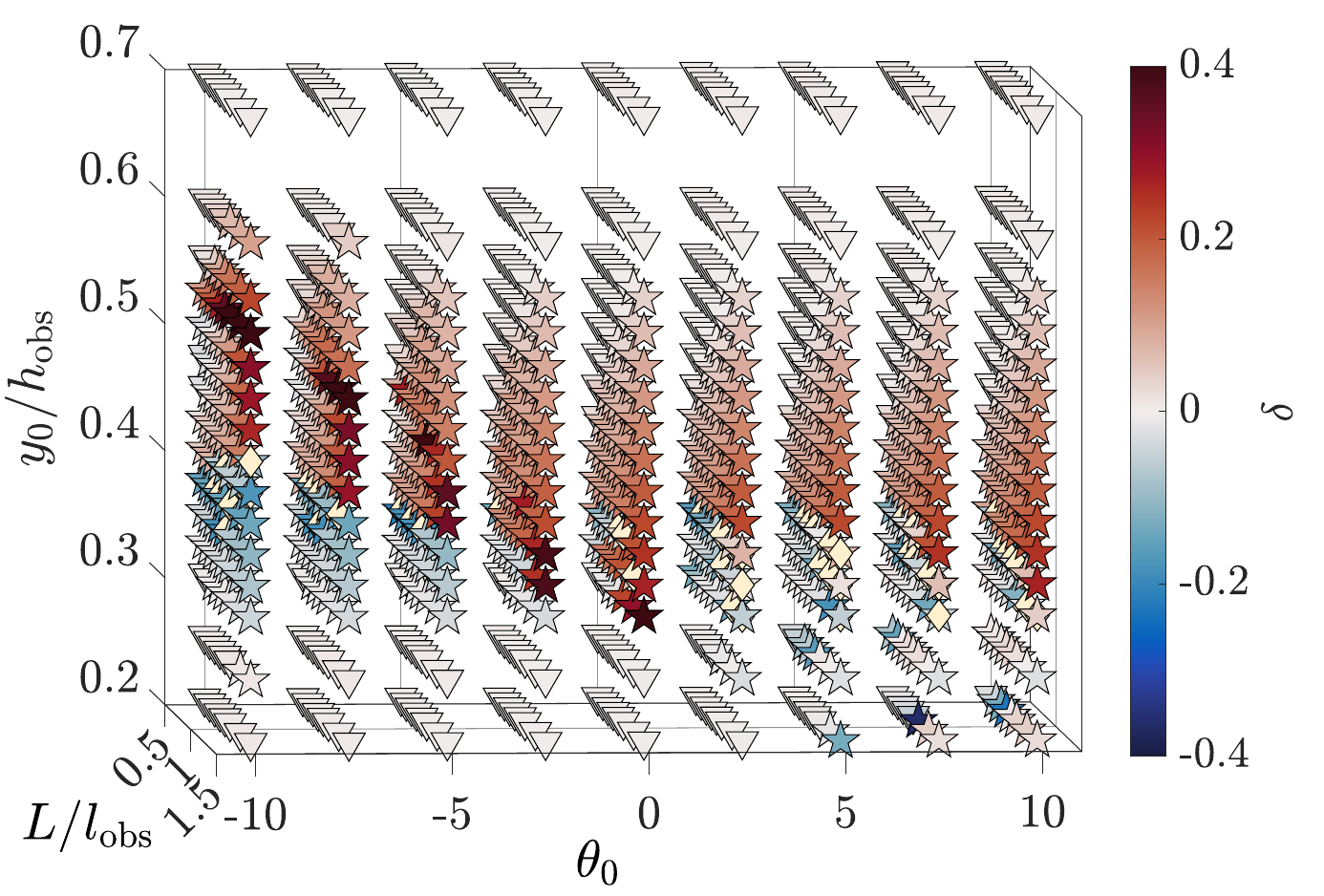}}
    \hspace*{1.3cm}\includegraphics[width=0.34\textwidth,left]{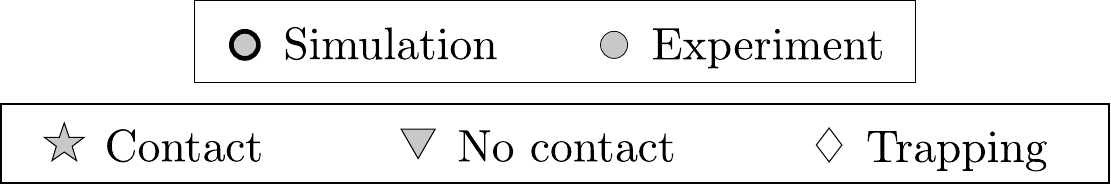}
    \\[-0.4cm]
    \centering
    \subfloat[\label{subfig:deviation_contact_combined_num_exp}]{\includegraphics[width=0.48\textwidth]{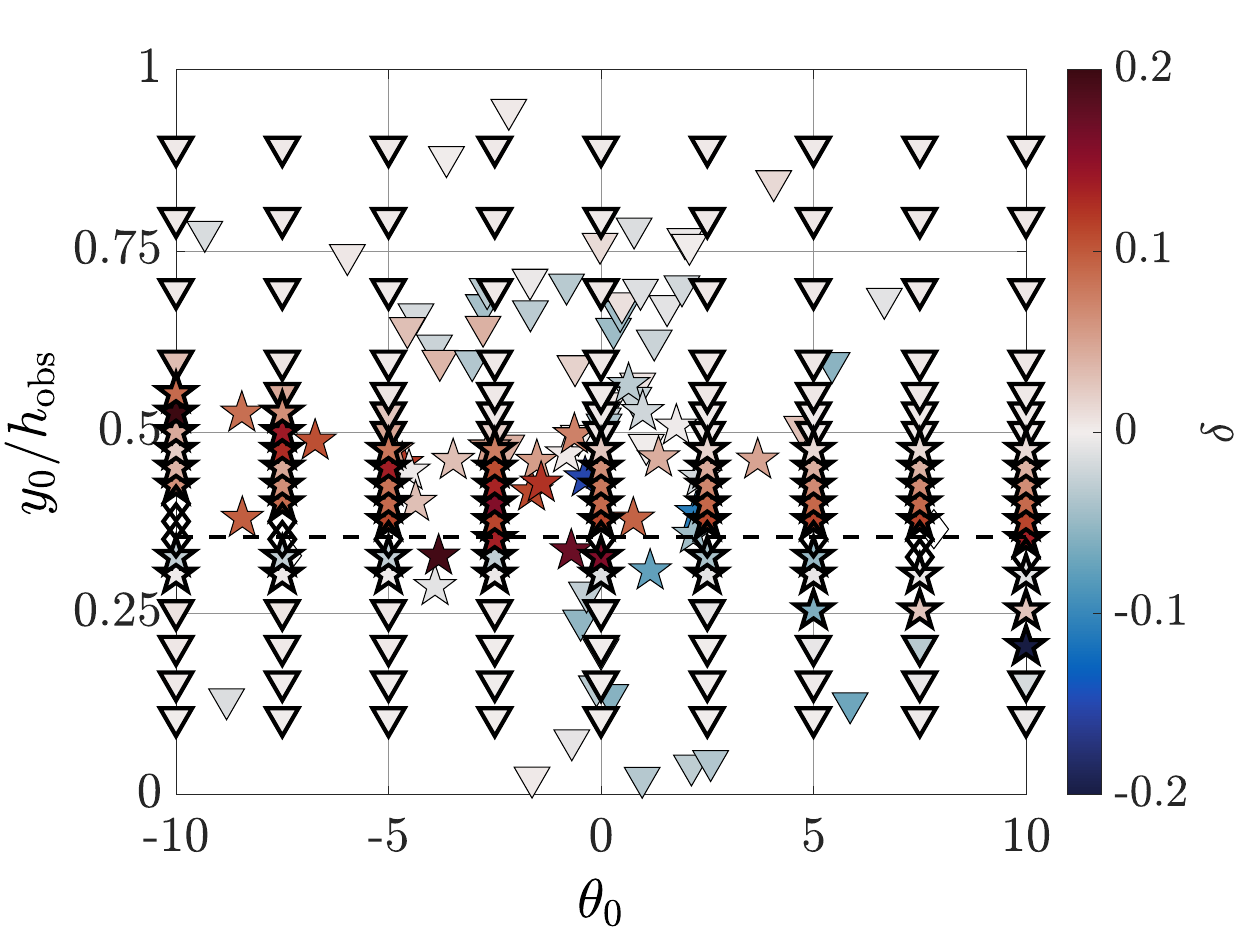} }
    \hfill
     \subfloat[\label{subfig:delta_pdf}]{\includegraphics[width=0.48\textwidth]{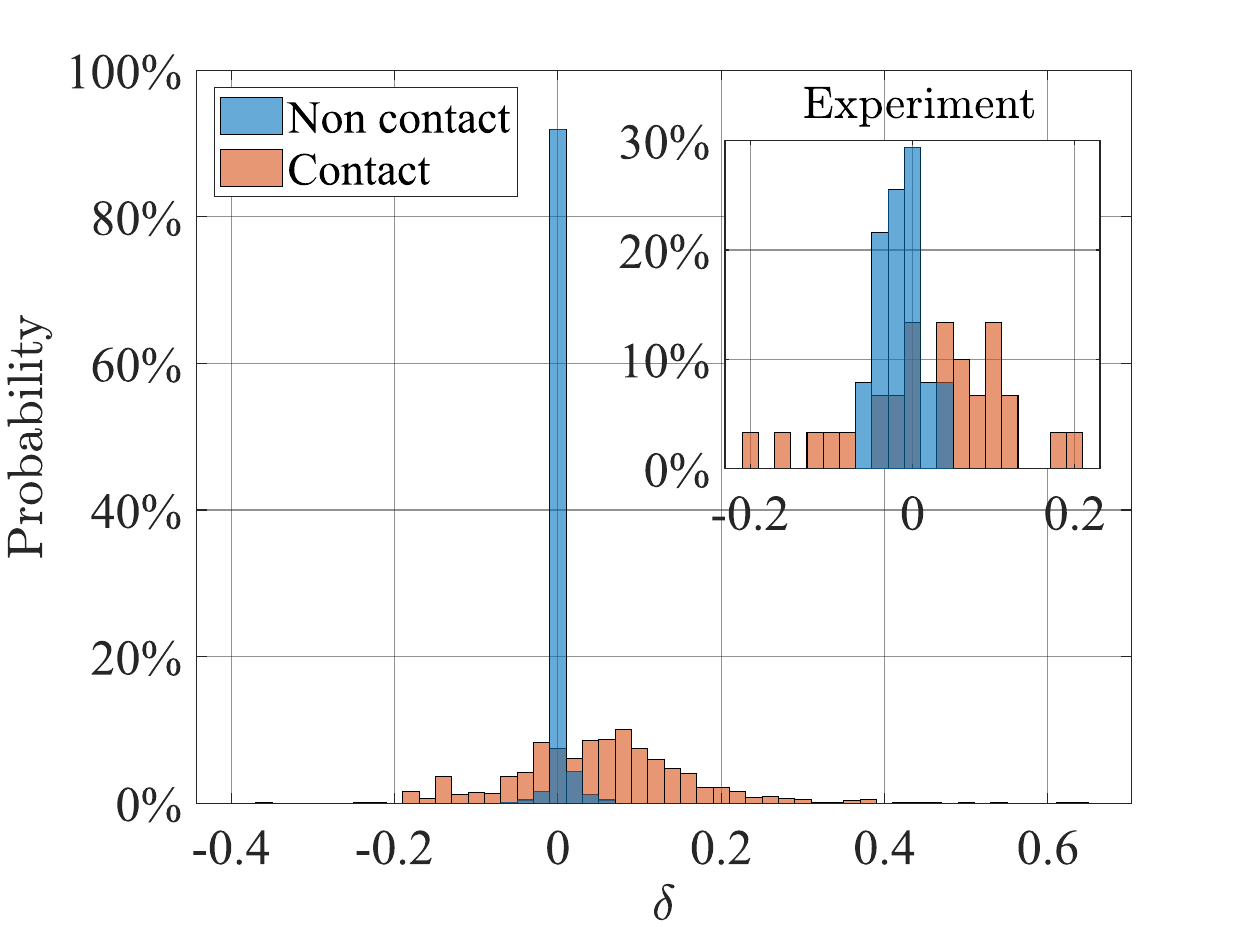}}
    \hfill
    \\[0.5cm]
    \includegraphics[width=0.49\textwidth]{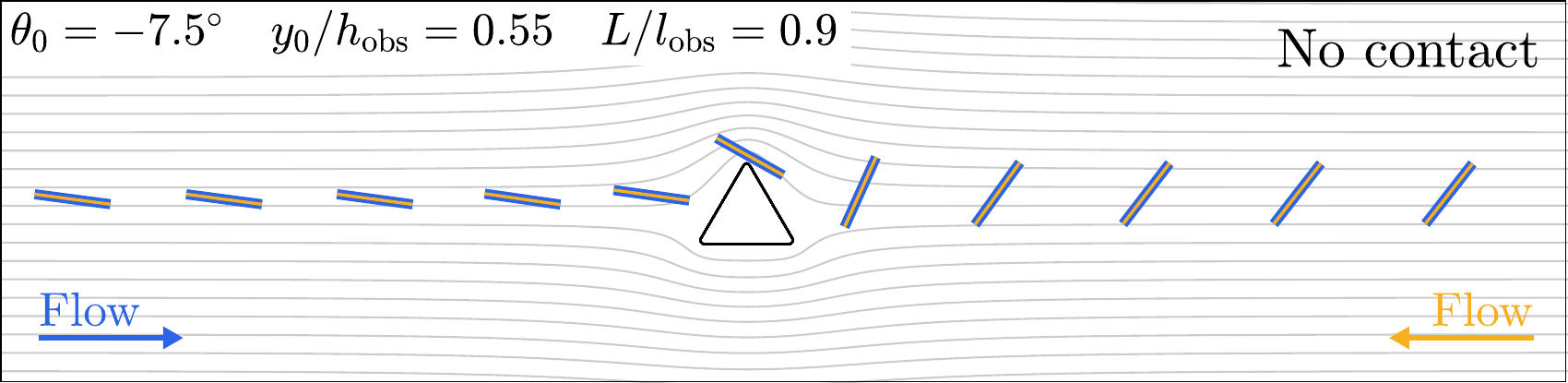}
    \hfill
    \includegraphics[width=0.49\textwidth]{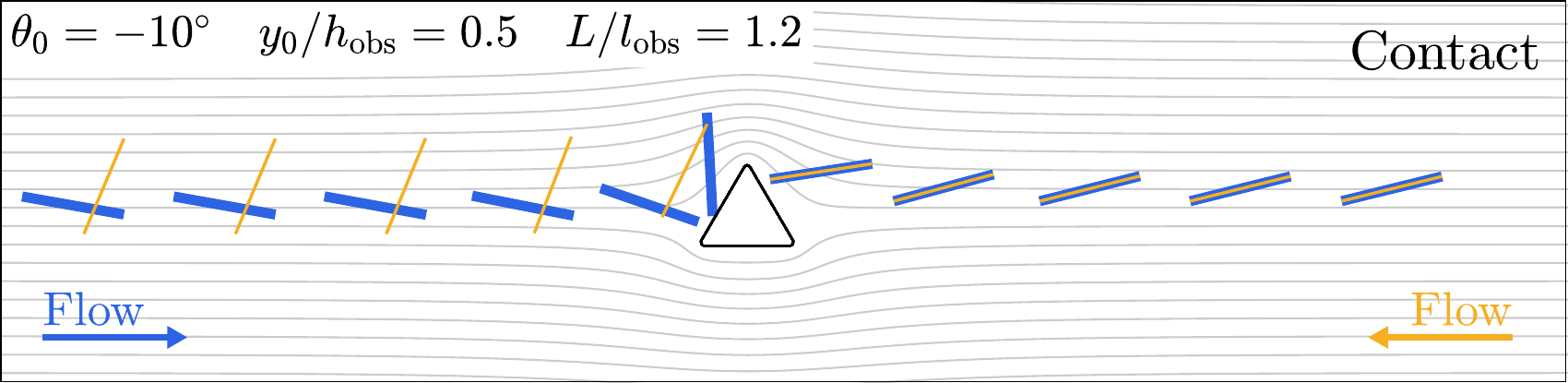}
    \\
    \subfloat[\label{subfig:trajectories_theta0_m7o5_y0_0o55_L_0o9}]{\includegraphics[width=0.49\textwidth]{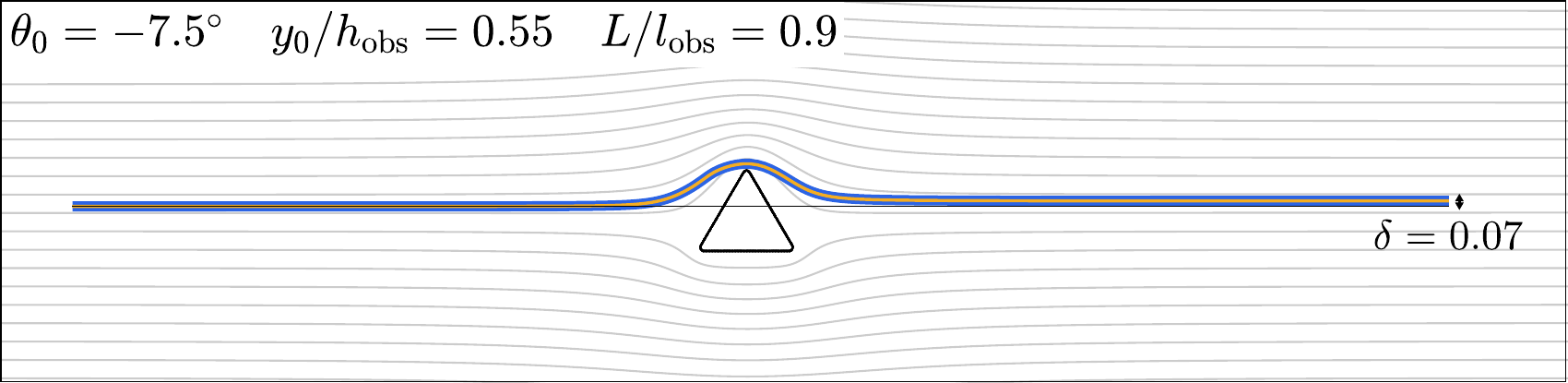}}
    \hfill
    \subfloat[\label{subfig:trajectories_theta0_m10_y0_0o5_L_1o2}]{\includegraphics[width=0.49\textwidth]{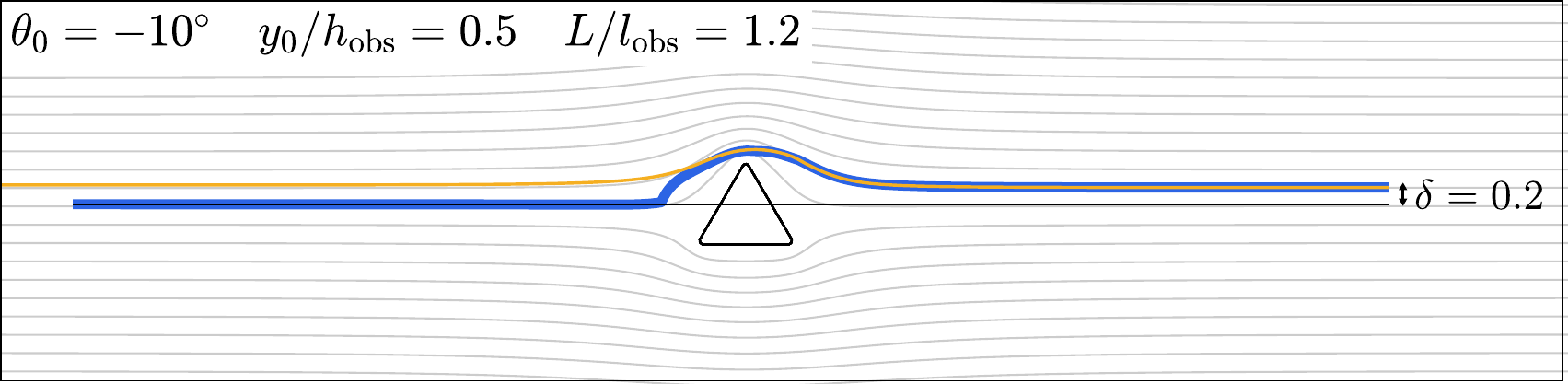}}
    \caption{Influence of contact on the fiber trajectory.
    (a) Phase diagram showing a close link between direct contact and high lateral deviation. \rev{The dashed line represents the position of the flow separatrix at the entrance of the channel.}
    (b) Probability of $\delta$ with and without contact in the simulations. Inset shows the experimental probabilities. Initial conditions are: $-10^{\circ} < \theta_{\rm 0} < 10^{\circ}$, $0 < y_0/h_{\rm obs} < 1$, and $0.5 < L/l_{\rm obs} < 1.5$ \rev{both in the experiments and simulations}.
    (c) Typical example of chronophotographs and trajectories in the absence of contact (blue: flow from left to right, yellow: flow from right to left). Reversibility of the fiber motion is shown by the superimposition of blue and yellow trajectories. Small lateral deviation, $\delta$, is obtained (this case corresponds to the highest observed value of $\delta$ in the absence of contact).
    (d) Typical example of chronophotographs and trajectories in case of contact, showing irreversibility of the fiber motion (blue and yellow trajectories do not superimpose), and larger lateral displacement is observed. 
    }
    \label{fig:effect_of_contact_on_deviation}
\end{figure}

Figure \ref{subfig:deviation_contact_combined_num_exp} shows whether contact has taken place (stars) or not (pointing-down triangles) and the corresponding lateral displacement in our experiments and simulations at $L/l_{\rm obs}=0.8$. When $y_0/h_{\rm obs} < 0.2$ or $y_0/h_{\rm obs} > 0.55$ the fibers do not enter in direct contact with the pillar. In most cases, without contact, fibers nearly go back to their initial lateral position with no or very small deviation ($|\delta| < 0.07$). On the contrary when $0.2 \leq y_0/h_{\rm obs} \leq 0.55$ the fibers touch the pillar and their deviation is non-negligible. %(up to $|\delta|  \approx 0.4$).
Direct contact with the pillar therefore significantly enhances the lateral displacement.

This enhancement is quantified in Fig.~\ref{subfig:delta_pdf} showing the probability distribution of lateral displacements with and without contact. In the absence of contact, both simulations and experiments exhibit a peaked distribution around $\delta = 0$. When contact occurs, the distributions widen significantly: the standard deviation $\sigma_{\delta}$ increases by a factor 13 in the simulations (from $\sigma_{\delta} \approx 0.0085$ to $\approx 0.11$) and by a factor 3.5 in the experiments (from $\sigma_{\delta} \approx 0.026$ to $\approx 0.092$).

%However, the obtained lateral displacements with a single pillar are rather small and lead to a limited sorting efficiency. The sorting efficiency could be significantly improved by considering a pillar array, as done in DLD devices to sort other types of particles such as spherical particles \cite{Huang2004,Joensson2011}, red blood cells \cite{Kabacaoglu2018,Chien2019} and bacteria \cite{Holm2011,Beech2018}.

By reversing the flow  after the fiber has passed the obstacle in the simulations we have tested the reversibility of the trajectories. The trajectories where no contact between the fiber and the obstacle occurs remain reversible (Fig.~\ref{subfig:trajectories_theta0_m7o5_y0_0o55_L_0o9}) as required for flows at vanishing Reynolds numbers whereas trajectories with contact (Fig.~\ref{subfig:trajectories_theta0_m10_y0_0o5_L_1o2}) are not reversible. Contact thus strongly modifies the nature of the trajectories. The good agreement between experimental and simulated trajectories in the case of contact  confirms that contact properties but also the occurrence of contact are correctly captured by the effective approach of the simulations and reasonably well detected by our visual observations.

Altogether, these results confirm that contact strongly enhances lateral displacements. However, we would like to stress that contact is \textit{not necessary} to induce asymmetric fiber trajectories and  cross-stream migration. As shown above and in Fig.~\ref{subfig:trajectories_theta0_m7o5_y0_0o55_L_0o9}, $\delta$ can reach non-zero, yet small, values in the absence of contact, without breaking the reversibility of the Stokes equations. It is well-known from Faxen's laws that a finite object can migrate across streamlines in the presence of a shear gradient if the flow has some curvature in the direction normal to the streamlines \cite{kim2013}. For instance, the trajectory of a sphere transported by a uniform flow around a spherical obstacle is fore-aft symmetric, but deviates from the streamlines as it approaches the obstacle, where the flow is curved, and follows them back as it moves away. The situation is more complex for fibers whose orientation also strongly influences the flow sampled by the object. Changes in orientation allow the fiber to jump streamlines asymmetrically with respect to the obstacle but the trajectory remains reversible. When direct contact occurs between fiber and obstacle, as shown in Fig.~\ref{subfig:trajectories_theta0_m10_y0_0o5_L_1o2}, stronger cross-stream migration is observed, leading to larger deviations, and the trajectory becomes irreversible.

\

\section{Conclusions}
\label{sec:conclusions}

In this work, we have presented a joint experimental and numerical investigation of the interaction between a rigid fiber and a triangular obstacle in confined microchannel flow.
One major output of this study is the identification and classification of four dynamics based on the initial fiber conditions at the channel entry: lateral position $y_0$, orientation $\theta_0$, and length $L$.

When the lateral position is far enough from the flow separatrix, separating streamlines going above and below the obstacle, $y_0/h_{\rm obs} > 0.55$ or $y_0/h_{\rm obs} < 0.2$, the fibers simply follow the streamlines, situations which are referred to as the ``Above" and ``Below" dynamics.
When the fibers are initially close to the separatrix, $0.2 \leq y_0/h_{\rm obs} \leq 0.55$, they approach the obstacle very closely and two more interesting dynamics,  ``Pole-vaulting" and ``Trapping", appear. The primary mechanism of the  dynamics in this region is the competition between the rotation induced by the strong flow disturbances  around the obstacle and the contact force if the fiber directly touches the obstacle.
Because of the low Reynolds number flow, the initial condition of the fiber $(\theta_0,y_0,L)$ determines its configuration in the vicinity of the obstacle $(\theta_{\rm c},y_{\rm c})$ determining the dynamics of the fiber. 
%and yields the phase diagram of Fig.~\ref{subfig:dynamics_3D}.

Another important finding of the present study is the sorting potential of an individual triangular pillar for a rigid fiber. Sorting occurs when fibers with different properties exhibit different cross-stream migration for identical initial conditions. Such migration can have two origins, the interaction of the slender object with the complex disturbance flow or direct fiber/obstacle contact. We show that reversible interactions with the disturbance flow can indeed lead to small lateral deviations. 
%Such deviations can for example occur for finite sized particled in flows with curved streamlines and in the presence of shear gradients as described by Faxen's law.
However in our situation such reversible lateral displacements remain very small and negligible compared to displacements induced by direct fiber/obstacle contact. Such contact leads to irreversible trajectories and strong lateral displacement.

The fact that direct contact is the primary mechanism for cross-stream migration might seem surprising for transport at small Reynolds number. The confined channel geometry we are working with concentrates the flow disturbance very close to the obstacle and enhances velocity gradients there. Fiber trajectories are thus only affected when passing very close to the pillar. On the other hand the concentration of streamlines near the obstacle also promotes fibers to get very close to the obstacle. Together with the finite length of the fiber, which is larger or comparable to the scale of the flow disturbance, this increases the probability of fiber/obstacle contact.

Longer fibers tend to have larger lateral deviations after passing the pillar than shorter fibers when the initial positions are in the range of $0.2 \leq y_0/h_{\rm obs} \leq 0.55$. However, the obtained lateral displacement with a single pillar remains overall rather small with a maximum of 60\% of the obstacle height and leads to a limited sorting efficiency. We have also shown that small variations of the initial condition within this range can have a large influence on the lateral displacement, masking the effect of the fiber length. To obtain fiber sorting, a very precise control of the initial condition is thus necessary, a condition that is difficult to fulfill in the experiments.

To increase the sorting potential one could in the future optimize the shape of an individual pillar to tune different fiber dynamics and trajectories. Optimizing the microchannel for fiber sorting could also be possible by considering more pillars and by arranging their layout. The sorting efficiency of such pillar arrays has been shown before in DLD devices for spherical particles \cite{Huang2004,Joensson2011}, red blood cells \cite{Kabacaoglu2018,Chien2019} and bacteria \cite{Holm2011,Beech2018}. And finally, as fiber/obstacle contact is crucial for large lateral displacements one could for example modify the obstacle roughness to induce more direct contacts. On the other hand the range of initial conditions where trapping occurs increases with increasing fiber/obstacle frictions as would occur for rougher pillars. Trapping could lead to clogging, preventing fiber sorting. Modifying the direct fiber-obstacle interactions would thus have to be done very carefully.

%Since trapping is due to a subtle balance between the hydrodynamic forces around the contact point (here it is left vertex of the triangle), it should be unstable in the absence of friction. However, in both experiments and simulations the trapped fibers never escape, which suggests that even a small amount of friction could play a significant role. A stability analysis of the trapping state  would provide some insight on this interesting question. \Anke{I this still correct? I think we now know that friction is not needed for trapping?}

%%%%%%%%%%%%%%%%%%%%%%%%%%%%%%%%%%%%%%%%%

% References
\begin{acknowledgments}
We thank Camille Duprat for stimulating this collaboration. BD\ and CB\ acknowledge support from the French National Research Agency (ANR), under award ANR-20-CE30-0006. AL and OdR acknowledge funding from the ERC Consolidator Grant PaDyFlow (Agreement  682367). This work has received the support of Institut Pierre-Gilles de Gennes (\'Equipement d’Excellence, ``Investissements d’avenir'', program ANR-10- EQPX-34). ZL acknowledges funding from the Chinese Scholarship Council.
\end{acknowledgments}

\bibliography{Bibliography}

\appendix

\section{\rev{Discussion on fiber-walls and fiber-obstacle hydrodynamic interactions}}
\label{sec:fiber-wall HI}
In this appendix we discuss the role hydrodynamic interactions (HI) between the fiber and the channel walls and obstacle surface, and show that they can be neglected in the model.

\rev{First, it has been shown previously that the small confinement in the vertical  ($2a/H_{\rm ch} \approx 0.1$) and lateral ($L/W_{\rm ch}\approx 0.1$) direction weakly affects the fiber velocity  \cite{Nagel2017}. HI with the channel walls can therefore be neglected.  In addition the fiber remains in the mid-plane at $z = H_{\rm ch}/2$ so that HI with the upper and bottom walls cancel each other by symmetry.}

\begin{figure}[H]
    \centering
     \includegraphics[width=0.99\textwidth]{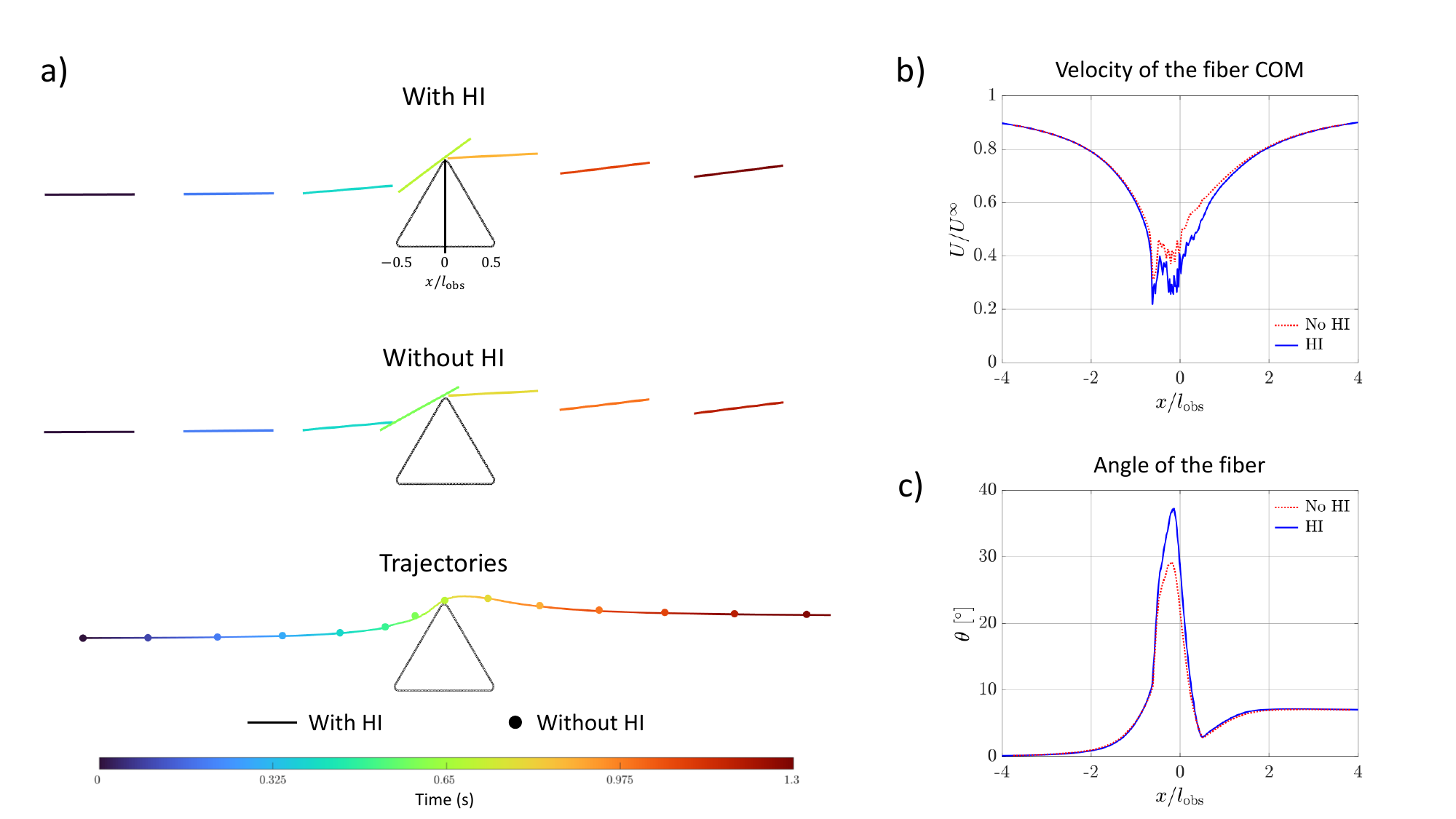}
    \caption{\rev{Effect of fiber-obstacle HI on fiber motion. (a) Chronophotographs and comparison of the trajectories of the fiber COM between the simulations with and without fiber-obstacle HI.  Fiber velocity (b) and orientation (c) as a function of the horizontal distance from the obstacle $x/l_{\rm obs}$. Blue line: with fiber-obstacle HI. Red dashed line: without fiber-obstacle HI. }}
    \label{fig:HI-obstacle}
\end{figure}

\rev{In order to quantify the effect of HI with the obstacle, we have ran two identical simulations (same initial conditions, contact forces, incident flow $U^{\infty}$, etc...): one with fiber-obstacle HI (using a regularized version of the Boundary Element Method (see \cite{balboa2017hydrodynamics,gidituri2023,Makanga2023}) and one without fiber-obstacle HI. In both simulations HI with the channel walls are neglected. We chose an initial condition that favored fiber-obstacle interactions, which is a critical configuration. Figure\ \ref{fig:HI-obstacle}a shows the chronophotographs and the trajectories of the fiber center of mass (COM) for both cases. Even though the fiber without HI is  slowed down less near the obstacle, the trajectories match closely, which is the most important feature to quantify the lateral displacement $\delta$ (see Section V), and the agreement in time is also very good.  Figure\ \ref{fig:HI-obstacle}b-c compares the fiber speed and angle as a function of horizontal distance from the center of the obstacle. We see that the velocity correction due to fiber-obstacle HI slightly slows down the fiber near the obstacle ($\approx 10\%$ slower near contact), and the reorientation is larger. However, the angle is recovered almost exactly in the wake of the pillar. This quantitative comparison shows that neglecting fiber-obstacle HI does not change appreciably the fiber trajectory, as expected from the very good quantitative agreement with experiments.
Furthermore, neglecting fiber-obstacle HI significantly reduces the computing time (simulations are about 100 times faster without HI) which was necessary to perform a large parametric study with 1300 simulations.}

\section{Effect of the channel depth on the flow field}
\label{sec:supplementary_material_flow}
\begin{figure}[H]
\centering
\subfloat[\label{subfig:velocity_field_H40um}]{\includegraphics[width=0.32\textwidth]{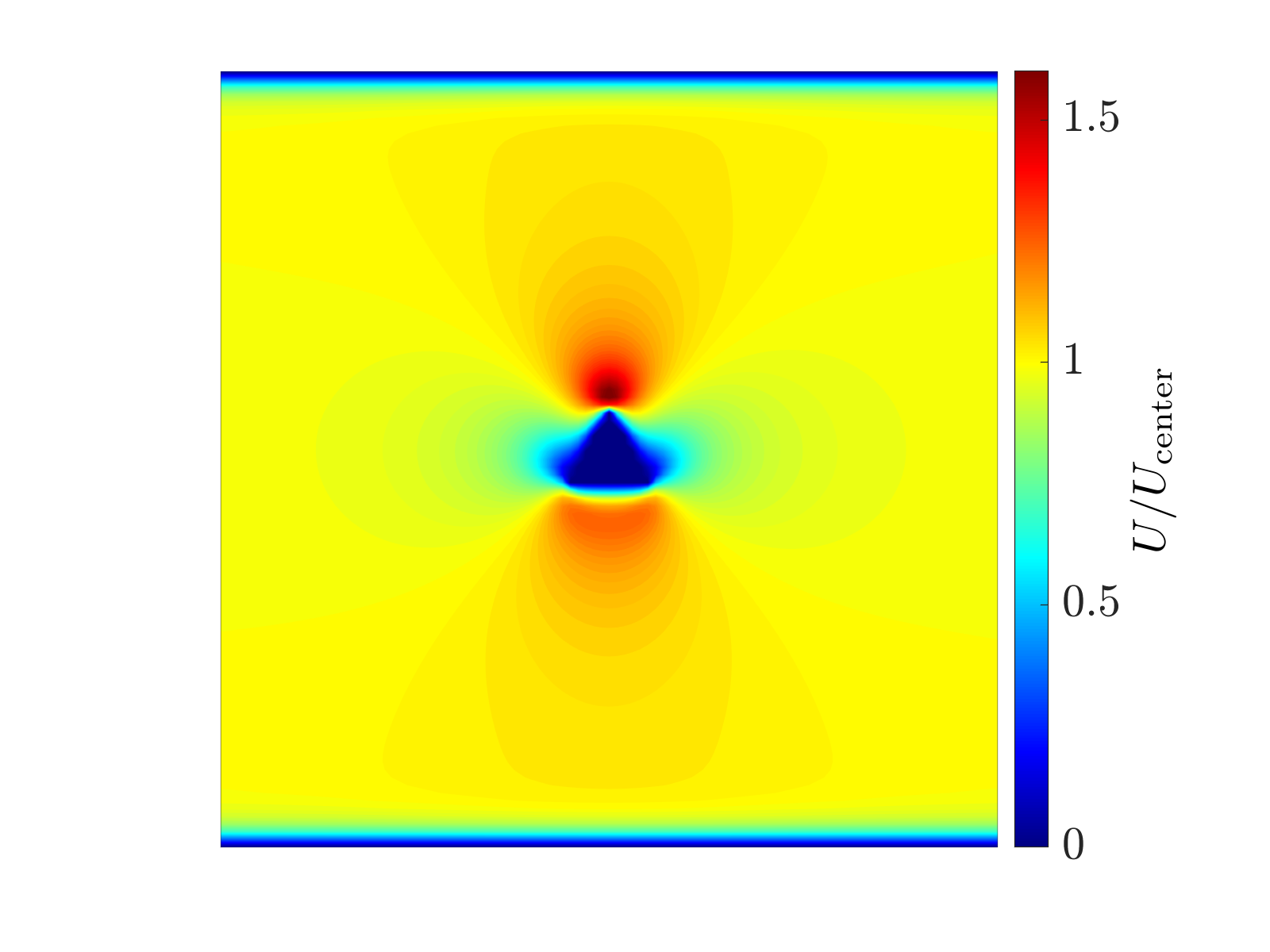}}
    \hfill
    \subfloat[\label{subfig:velocity_field_H80um}]{\includegraphics[width=0.32\textwidth]{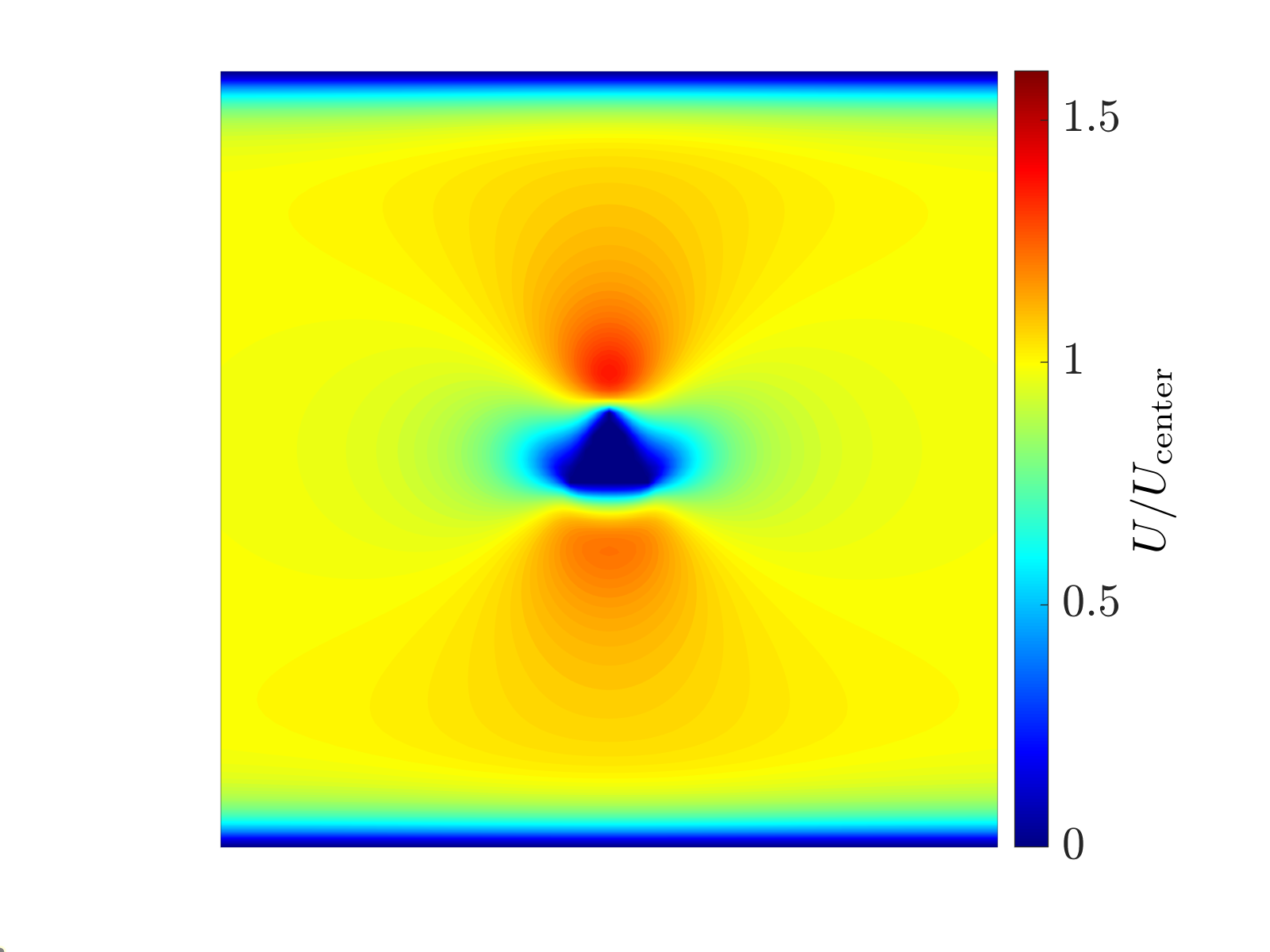}}
    \hfill
    \subfloat[\label{subfig:velocity_field_H120um}]{\includegraphics[width=0.32\textwidth]{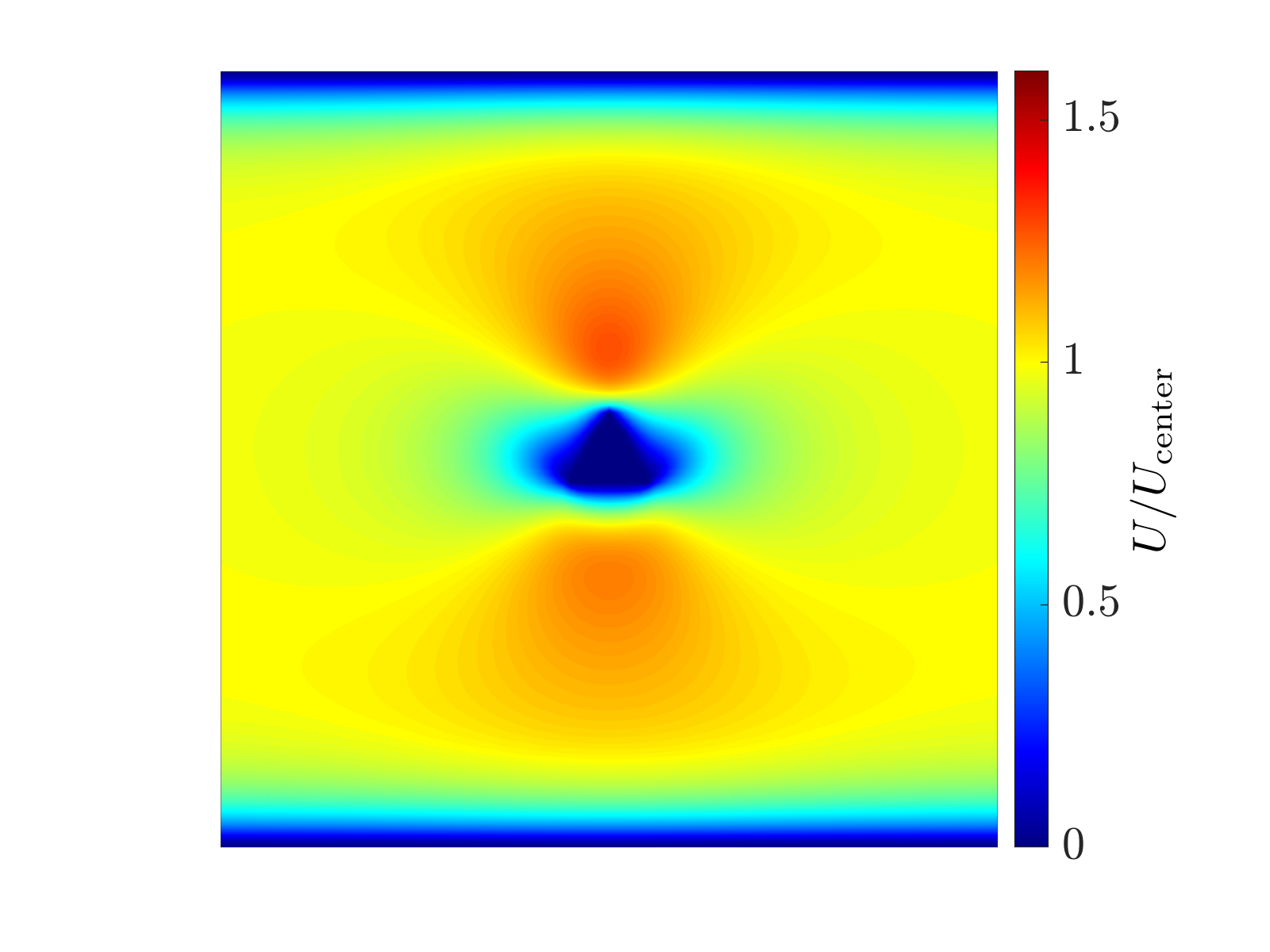}}
    \\
    \subfloat[\label{subfig:velocity_profiles_along_x_different_heights}]{\includegraphics[width=0.4\textwidth]{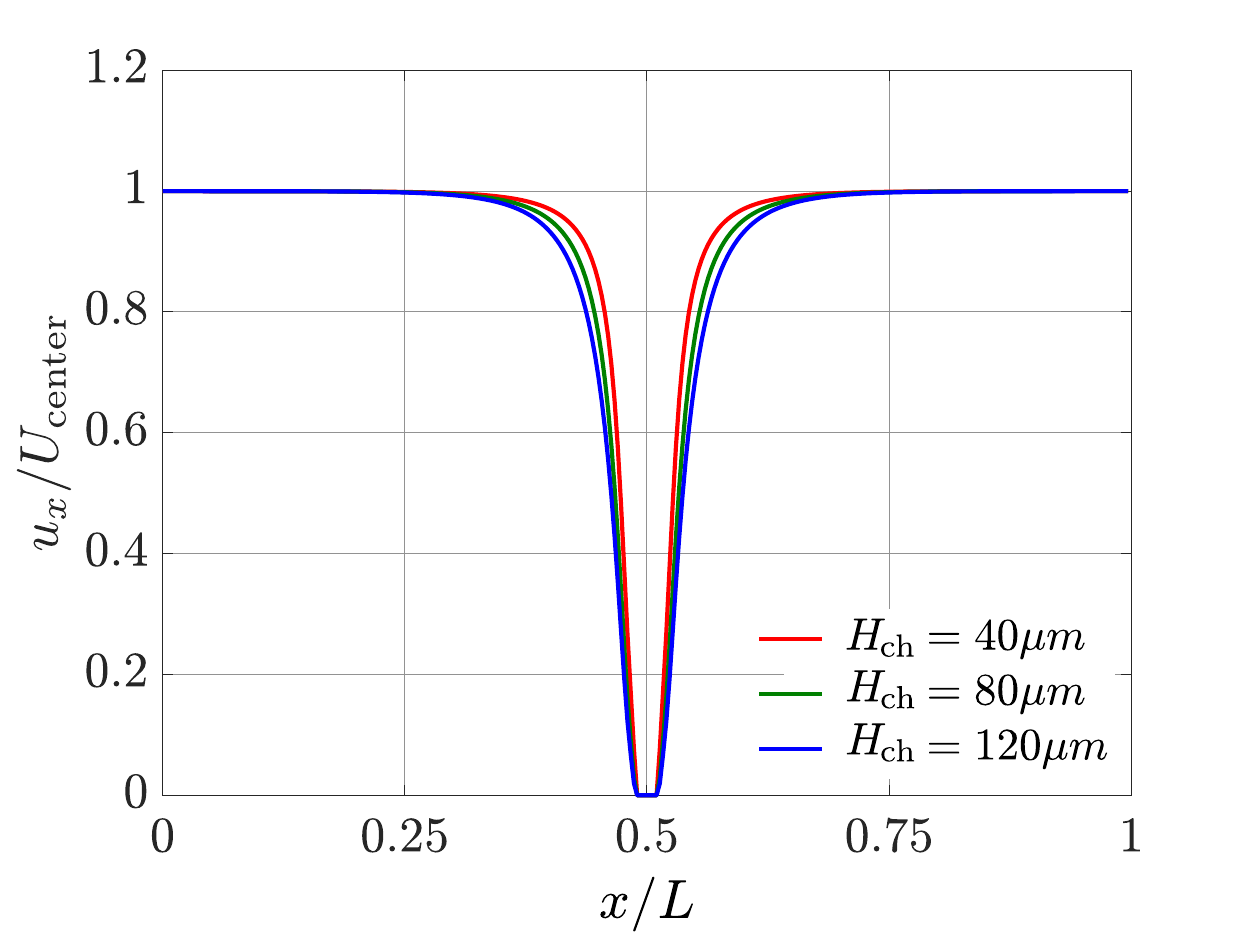}}
    \hspace{2cm}
    \subfloat[\label{subfig:velocity_profiles_along_y_different_heights}]{\includegraphics[width=0.4\textwidth]{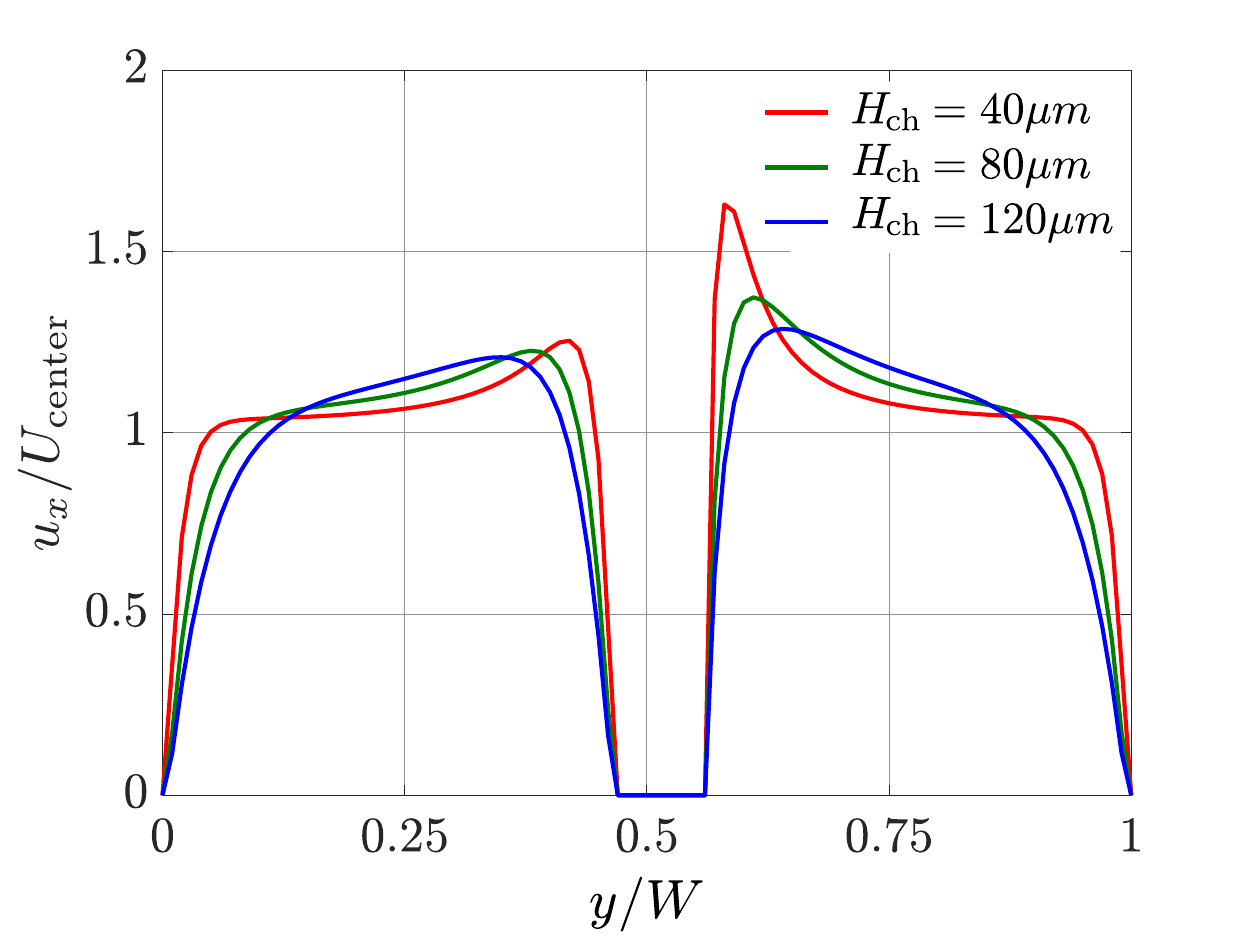}}
    \caption{Velocity fields and profiles around the triangular pillar computed by LBM for different channel heights: (a) $H_{\rm ch} = 40$\,µm,  (b) $H_{\rm ch} = 80$\,µm, (c) $H_{\rm ch} = 120$\,µm. (d) and (e) respectively show the velocity profiles along the $x$ and $y$ axis. \rev{$U_{\rm center}$ is the velocity magnitude at the channel centerline.}}
    \label{fig:effect_channel_height_velocity_fields}
\end{figure}

The dimensions of the channel significantly alter the flow field around the pillar.
Figure \ref{fig:effect_channel_height_velocity_fields} shows the velocity fields and the velocity profiles computed by the lattice Boltzmann method (LBM) in the neighborhood of the pillar for three different channel heights: $H_{\rm ch} =$ 40\,µm, 80\,µm and 120\,µm.
The perturbation of the flow field induced by the presence of the obstacle enlarges with the channel height, resulting in a lower velocity magnitude and lower gradients in the vicinity of the pillar.
Shallower channels are thus expected to promote interactions between the fibers and the pillar.
However, further decreasing the height of the channel below $H_{\rm ch} = 40\,\unit{\um}$ would also make more difficult to focus the fibers in the midplane.
This would result in many fibers flowing and aggregating close to the channel walls, which is highly undesirable.
In this work we chose a channel height $H_{\rm ch} = 40$\,µm, which is shallow enough to have strong interactions between the fibers and the pillar, and deep enough so that the lateral walls do not affect the fibers trajectories.

\section{\rev{Hydrodynamic forces in the ``Trapping" case}}
\label{sec:appendix_hydrodynamic_forces_trapping}
\begin{figure}[H]
    \centering    \subfloat[\label{subfig:hydrodynamic_forces_t_2o5_L_1_y0_0o35}]{\includegraphics[width=0.3\textwidth]{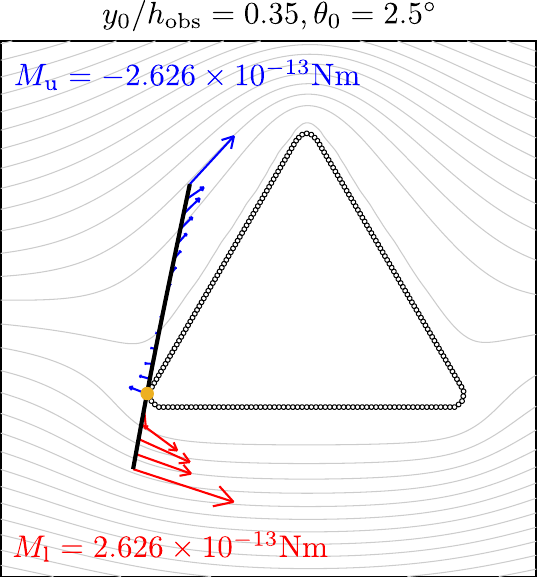}}
    \hfill
    \subfloat[\label{subfig:hydrodynamic_forces_t_5_L_1_y0_0o325}]{\includegraphics[width=0.3\textwidth]{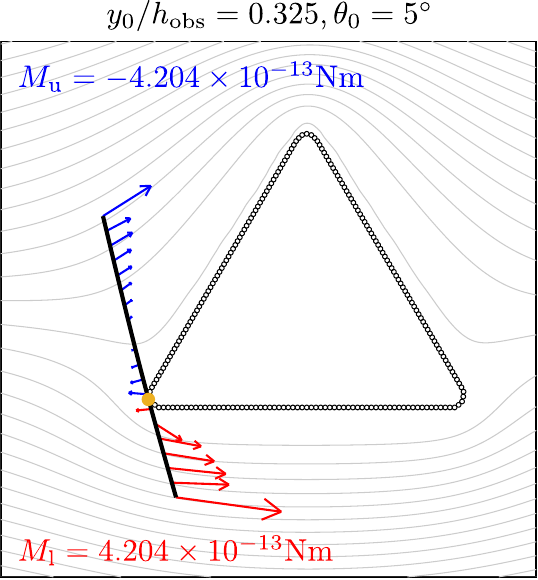}}
    \hfill
\subfloat[\label{subfig:hydrodynamic_forces_t_m10_L_1_y0_0o375}]{\includegraphics[width=0.3\textwidth]{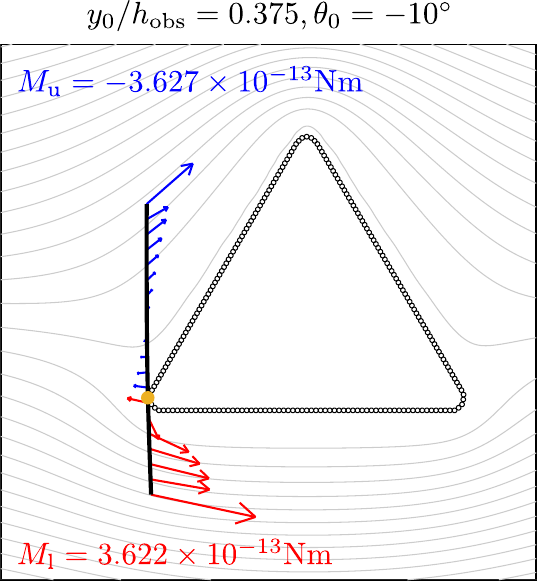}}
    \caption{\rev{Hydrodynamic forces along three permanently trapped fibers. Blue and red arrows respectively show the hydrodynamic forces above and below the contact point (yellow dot). Their moment $M_{\rm u}$ and $M_{\rm l}$ on both sides of the contact point counterbalances, which prevents the rotation of the fiber. The fiber length is $L/l_{\rm obs} = 1$ and the initial conditions are: (a) $y_0/h_{\rm obs} = 0.35$, $\theta_0 = 2.5^\circ$, (b) $y_0/h_{\rm obs} = 0.325$, $\theta_0 = 5^\circ$, (c) $y_0/h_{\rm obs} = 0.375$, $\theta_0 = -10^\circ$.}}
    \label{fig:hydrodynamic_forces_trapping}
  \end{figure}

\rev{
Trapping events result from a balance of the hydrodynamic forces, and more precisely of the moment of the hydrodynamic forces on both sides of the contact point between the fiber and the pillar.
The hydrodynamic forces ${\bf F}^{\rm H}$ and their moments ${\bf M}^{\rm H}$ are computed as
\begin{align}
{\bf F}^{\rm H} = -{\bf M}^{-1}\left({\bf U} - {\bf u}\right), & &
{\bf M}^{\rm H} = \sum_{i=1}^n \left({\bf r}_i - {\bf r}^{\rm CP}\right)\times{\bf F}_i^{\rm H}
\end{align}
with ${\bf M}$ the mobility matrix, ${\bf U}$ the velocity of the fiber beads (${\bf U} = {\bf 0}$ in case of trapping), ${\bf u}$ the velocity of the beads induced by the background flow, ${\bf r}_i$ the position of the $i$th fiber bead and ${\bf r}^{\rm CP}$ the position of the contact point.
They are represented in Fig.~\ref{fig:hydrodynamic_forces_trapping} along three permanently trapped fibers of length $L/l_{\rm obs}=1$ starting at three different initial conditions.
Blue and red arrows respectively show the hydrodynamic forces above and below the contact point (yellow dot), and $M_{\rm u}$ and $M_{\rm l}$ are respectively the moments of the hydrodynamic forces on the upper and lower parts of the fiber with respect to the contact point.
As the velocity magnitude is stronger below the pillar, the equilibrium position of the fiber is asymmetric to verify $M^{\rm H} = M_{\rm l} + M_{\rm u} = 0$, which prevents the rotation of the fiber.
}

\section{Influence of the fiber initial position and length on the contact configuration}
\label{sec:appendix_effect_fiber_initial_position_length_contact}
\begin{figure}[H]
    \centering
    \subfloat[\label{subfig:vectorMap_delta-theta_yc-y0_L1}]{\includegraphics[width=0.49\textwidth]{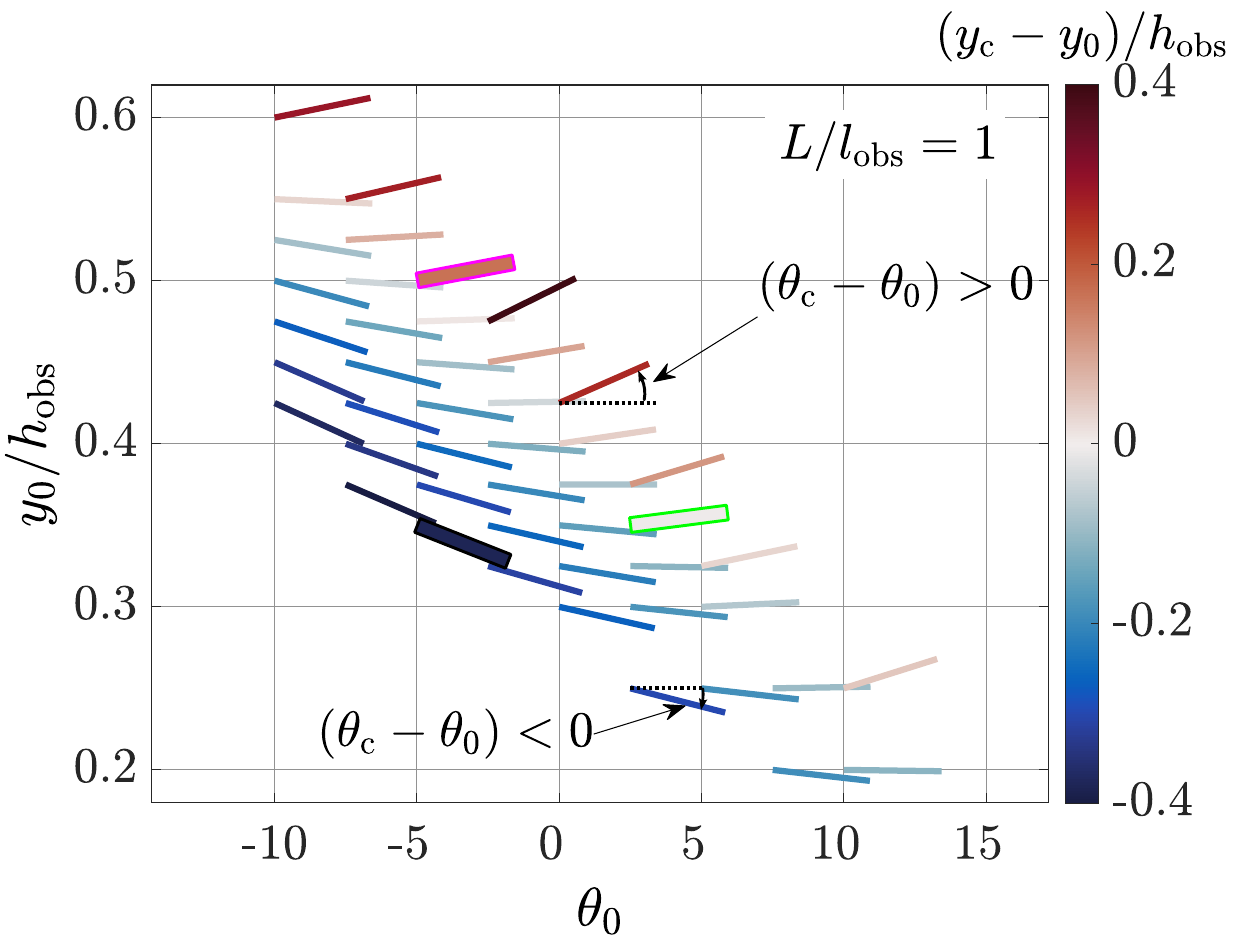}}
    \\
    \subfloat[\label{subfig:effect_L_on_contact}]{\includegraphics[width=0.49\textwidth]{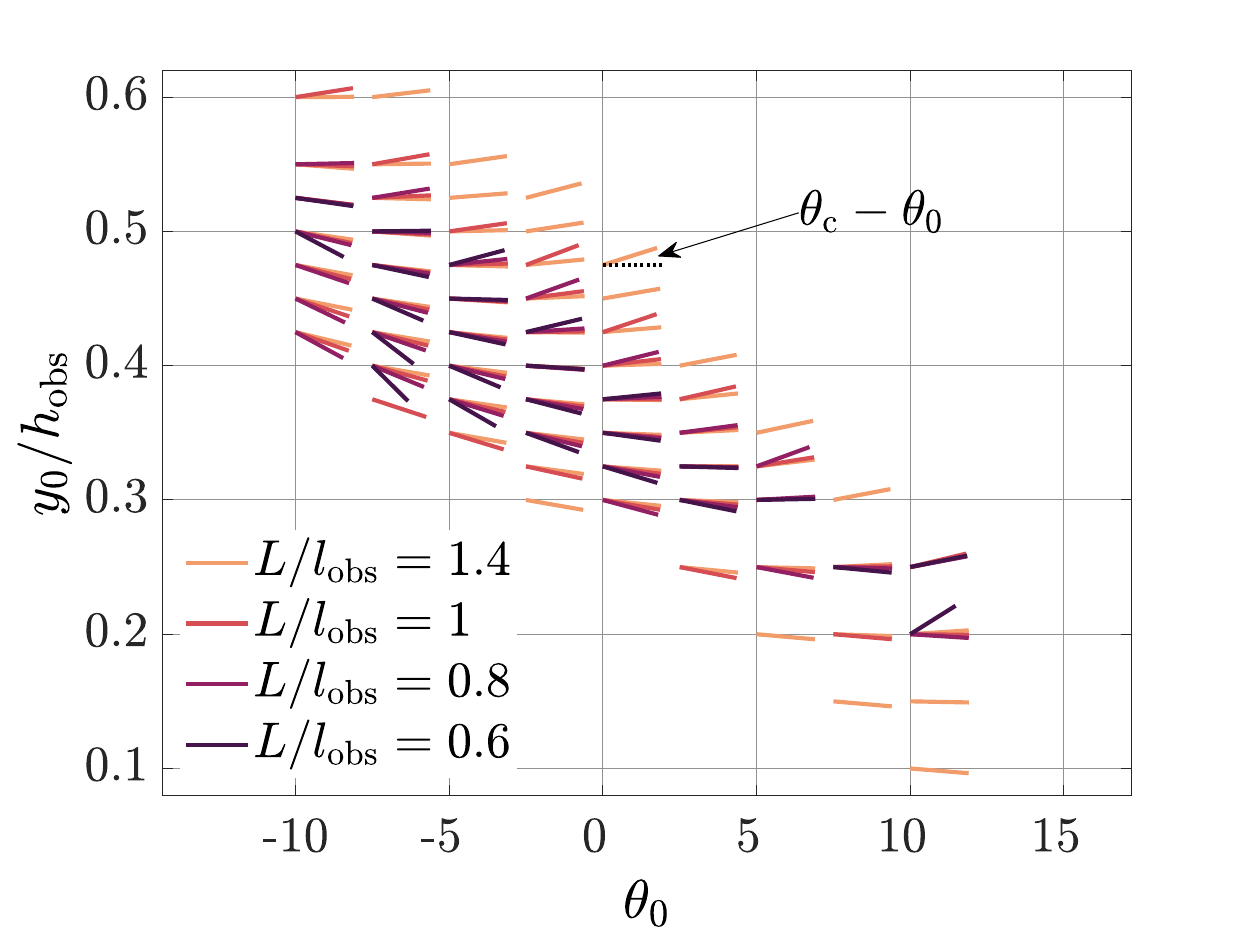}}
    \hfill
    \subfloat[\label{subfig:orientation_x}]{\includegraphics[width=0.49\textwidth]{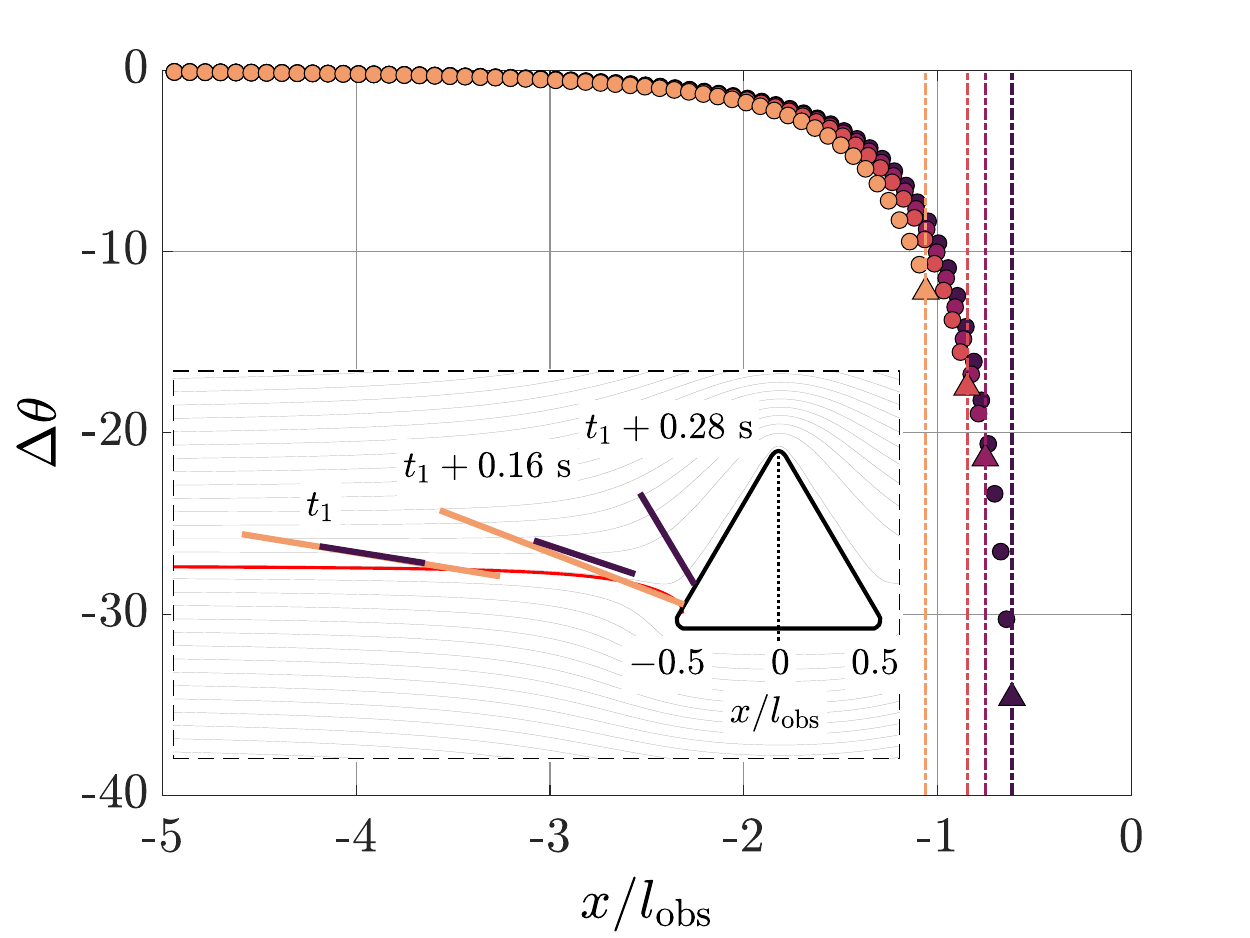}}
    \caption{Influence of the fiber initial position, initial orientation and length on the contact configuration. (a) Angle and lateral position differences between the initial and first contact configurations as a function of ($\theta_0,y_0$) for a fiber length $L/l_{\rm obs}=1$. The colors and angle of the lines respectively indicate $(y_{\rm c}-y_0)/h_{\rm obs}$ and $\theta_{\rm c}-\theta_0$. (b) Influence of the fiber length $L$ and initial condition ($\theta_0,y_0$) on the contact angle $\theta_{\rm c}$. (c) Fiber orientation difference with respect to the initial angle, $\Delta \theta = \theta(x) - \theta_0$, as a function of the $x$-position of the fiber's center of mass for $y_0/h_{\rm obs}=0.425$ and $\theta_0=-7.5^{\circ}$. The triangular symbols and dashed lines are the positions of the first contact with obstacle surface. The inset shows snapshots of two fibers of length $L/l_{\rm obs}=0.6$ and $L/l_{\rm obs}=1.4$ approaching the obstacle. When the longer fiber hits the obstacle at $t = t_1+0.16\,\unit{\second}$, the shorter fiber is still transported and rotated by the flow until it first touches the pillar at $t = t_1+0.28\,\unit{\second}$, leading to very different contact angles between the two fibers.}
    \label{fig:vectorMap_deltatheta_L}
\end{figure}

In this Appendix we carry out a sensitivity analysis of the mapping between the fiber initial position $(\theta_0,y_0)$ and length $L$ on the contact configuration $(\theta_{\rm c},y_{\rm c})$ when the fiber first touches the pillar.
The relation
\begin{equation*}
    (\theta_{\rm c},y_{\rm c}) = f(\theta_0,y_0,L)
\end{equation*}
is complex due to the strong disturbances of the flow field in the vicinity of the pillar and the elongated asymmetrical shape of the fiber.
Figure \ref{fig:vectorMap_deltatheta_L} provides insights of this complex mapping based on data extracted from numerical simulations for which direct fiber/obstacle contact occurs.

Panel \subref{subfig:vectorMap_delta-theta_yc-y0_L1} explores the influence of $\theta_0$ and $y_0$ on $\theta_{\rm c}$ and $y_{\rm c}$ at a given fiber length $L/l_{\rm obs} = 1$.
The color and angle of the lines respectively indicate the difference of lateral position, $(y_{\rm c}-y_0)/h_{\rm obs}$, and orientation, $\theta_{\rm c} - \theta_0$, between the initial and contact configurations.
It shows that both $(y_{\rm c}-y_0)/h_{\rm obs}$ and $\theta_{\rm c}-\theta_0$ increase with the initial position $y_0$ for a given initial angle.
This is due to the curvature of the streamlines in the vicinity of the obstacle.
Streamlines above the flow separatrix are curved upwards to pass above the obstacle, while those below the separatrix are curved downwards to pass below the obstacle.
As the fibers overall follow the streamlines, those that are initially located at a higher lateral position are transported by the upward-curved streamlines resulting in a higher $(y_{\rm c}-y_0)/h_{\rm obs}$.
It is the opposite scenario for the fibers starting at a lower lateral position.
This is illustrated in Fig.~\ref{fig:vectorMaps} (top) that displays snapshots of the fibers framed in black ($\theta_0=-5^\circ$, $y_0=0.35$) and magenta ($\theta_0=-5^\circ$, $y_0=0.5$).
The uppermost fiber follows upward-curved streamlines, it is thus transported upwards ($y_{\rm c} - y_0 > 0$) and rotated counter-clockwise ($\theta_{\rm c} - \theta_0 > 0$) by the flow.
It is the opposite for the black fiber which initially lies on the flow separatrix (that bends downwards).

The difference of initial position $(y_{\rm c}-y_{\rm 0})/h_{\rm obs}$ and orientation $\theta_{\rm c}-\theta_{\rm 0}$ also increases with the initial position $\theta_0$ at a given $y_0$.
This is again illustrated in Fig.~\ref{fig:vectorMaps} (middle) showing snapshots of the same fiber framed in black ($\theta_0=-5^\circ$, $y_0=0.35$), and the one framed in green ($\theta_0=2.5^\circ$, $y_0=0.35$).
Both fibers are initially located on the flow separatrix, but they sample different streamlines as they have different initial orientations.
When the green fiber approaches the obstacle, its head feels streamlines above the separatrix that bend upwards, while the head of the black fiber feels streamlines that bend downwards.
As a result, the green fiber rotates counter-clockwise ($\theta_{\rm c} - \theta_0 > 0$) and has $y_{\rm c} - y_0 > 0$, while the black fiber rotates clockwise ($\theta_{\rm c} - \theta_0 < 0$) and has $y_{\rm c} - y_0 < 0$.

Panels \subref{subfig:effect_L_on_contact} and \subref{subfig:orientation_x} show the additional effect of the fiber length on the contact configuration.
The lighter the color, the longer the fiber.
They reveal that shorter fibers rotate more (either clockwise or counter-clockwise) than longer ones before touching the pillar.
This is due to geometry effects, as illustrated in Fig.~\ref{subfig:orientation_x}  which shows the difference of orientation of the fiber with respect to the initial angle, $\Delta \theta = \theta(x) - \theta_0$, during its transport by the flow.
This figure indicates that longer fibers reach the obstacle earlier than shorter ones which continue rotating until they touch the pillar.
The inset in Fig.~\ref{subfig:orientation_x} gives the trajectories of two fibers of length $L/l_{\rm obs}=0.6$ and $L/l_{\rm obs}=1.4$ starting at the same initial configuration $\theta_0=-7.5^{\circ}$ and $y_0/h_{\rm obs}=0.425$.
When the longer fiber first hits the obstacle, the shorter one is still carried and rotated by the flow, leading to two very different contact angles between both fibers.

Additionally, fibers which start from a higher $y_0$ with $\theta_0 > 0$ or a lower $y_0$ with $\theta_0 < 0$ have the same tendency to follow the direction of the streamlines close to the obstacle.
So, it is easier for them to bypass the obstacle without contact, which explains why the data in panels \subref{subfig:vectorMap_delta-theta_yc-y0_L1} and \subref{subfig:effect_L_on_contact} distributes into a parallelogram shape, with no data in the bottom left and top right corners.

\section{Role of obstacle roughness on the fiber dynamics}
\label{sec:appendix_role_friction}
\subsection{Simulations with a perfectly smooth pillar}
\begin{figure}[H]
    \settoheight{\imageheight}{\includegraphics[height=0.28\textwidth]{dynamics_exp_num_L0o8-eps-converted-to.pdf}}
    \centering
    \hspace{4.1cm}\includegraphics[width=0.45\textwidth]{legend_dynamics_only-eps-converted-to.pdf}
    \\[-0.4cm]
    \subfloat[\label{subfig:pillar_beads_Bezier}]{\tikz\node[minimum height=\imageheight]{\includegraphics[width=0.2\textwidth]{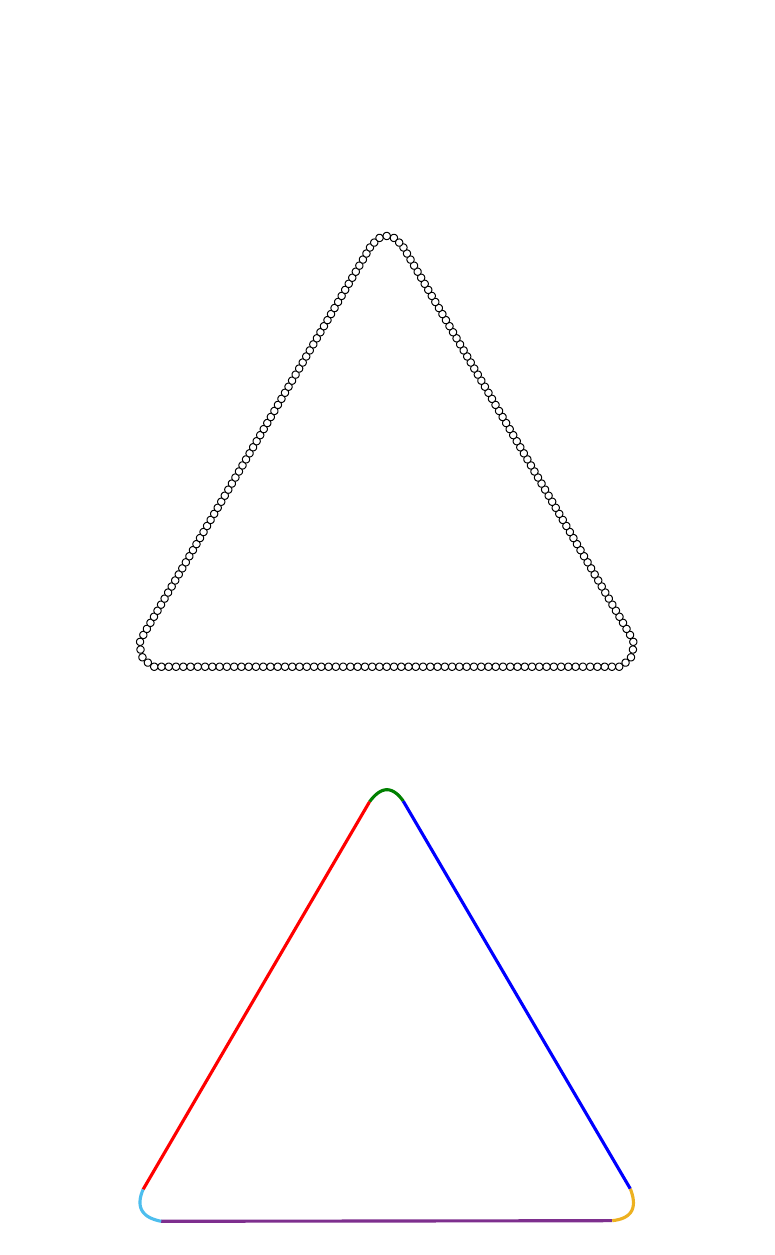}};}
    \hfill
    \subfloat[\label{subfig:phase_diagram_beads}]{\tikz\node[minimum height=\imageheight]{\includegraphics[height=0.28\textwidth]{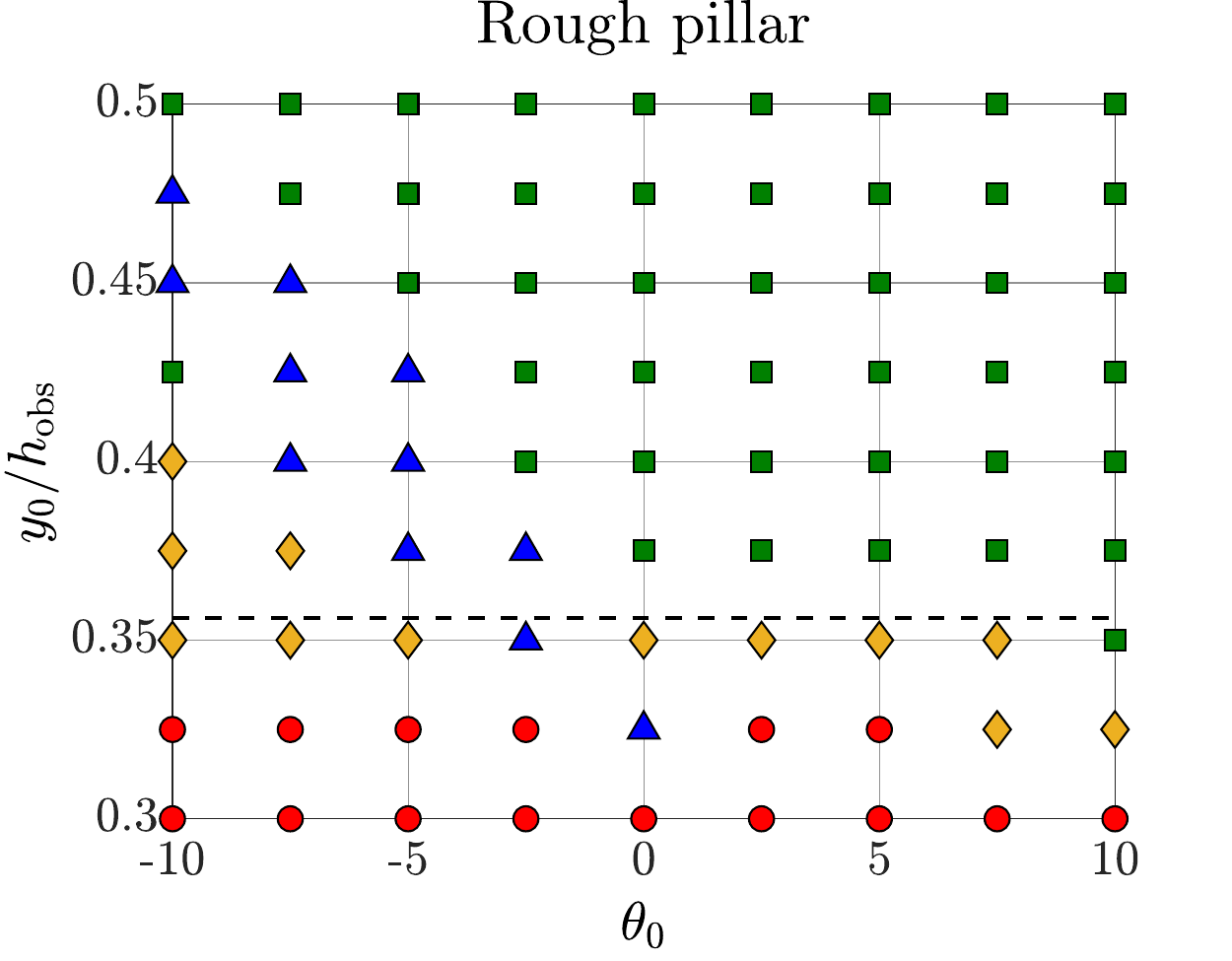}};}
    \hfill
    \subfloat[\label{subfig:phase_diagram_Bezier}]{\tikz\node[minimum height=\imageheight]{\includegraphics[height=0.28\textwidth]{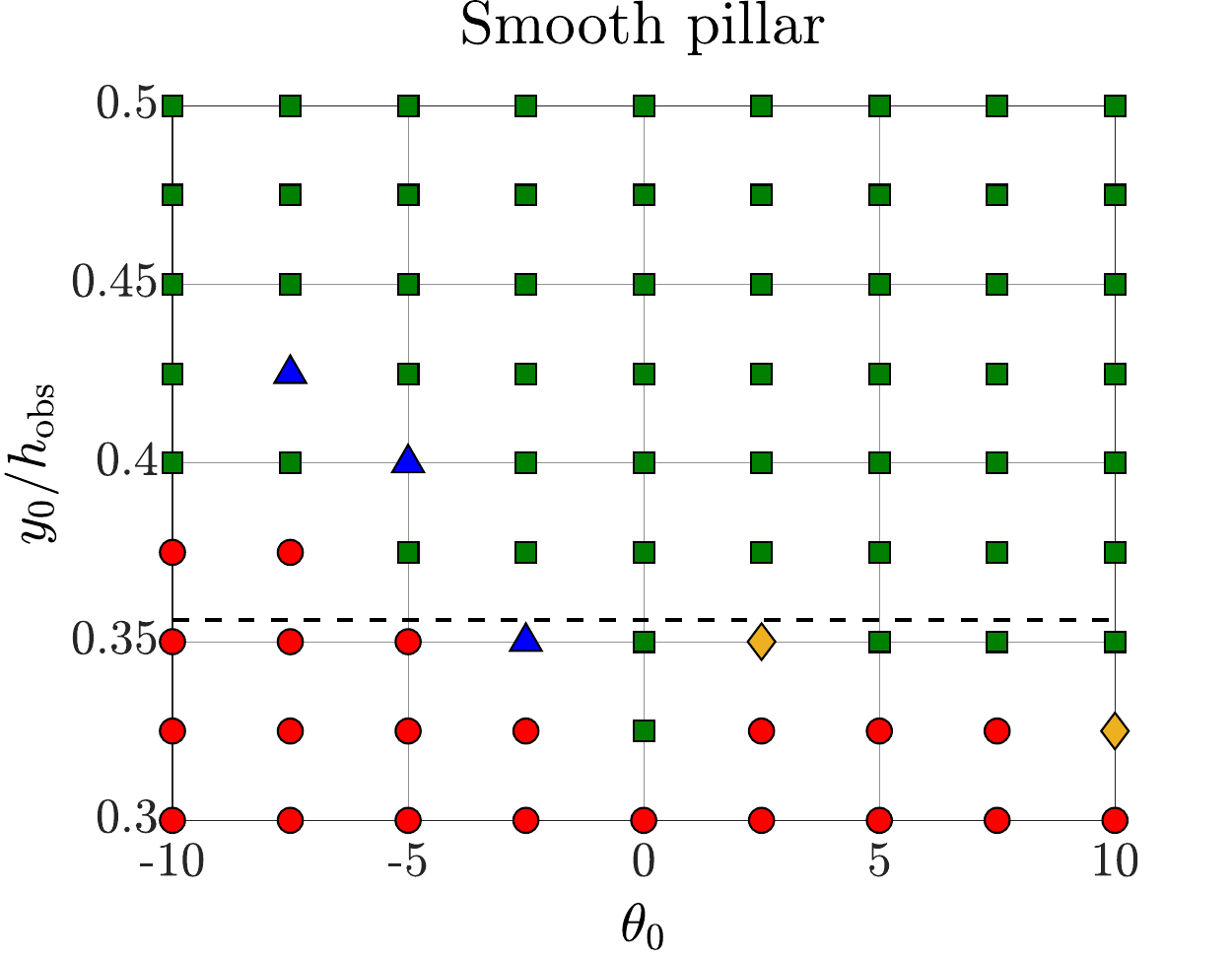}}; }
    \caption{(a) Discretization of the pillar using beads (top) and 6 quadratic Bézier curves (bottom). (b) Phase diagram showing the fiber dynamics computed with the rough pillar discretized using spherical beads. (c) Phase diagram showing the fiber dynamics computed with the perfectly smooth pillar described by Bézier curves. The fiber length in the simulations is $L=0.8l_{\rm obs}$. \rev{Data are only shown in the range $0.3 \leq y_0/h_{\rm obs} \leq 0.5$ for which direct fiber/pillar contact occurs}. \rev{The dashed line in (b) and (c) represents the position of the flow separatrix at the channel entrance.} Circles: Below; Squares: Above; Triangles: Pole-vaulting; Diamonds: Trapping.}
    \label{fig:phase_diagrams_beads_Bezier}
\end{figure}

In all the simulations presented in Secs.~\ref{sec:results_flow_fiber_dynamics} and \ref{sec:results_lateral_displacement}, the fiber and the obstacle are discretized by spherical beads of radius $a=2\,\unit{\um}$ and $a_{\rm obs}=0.6\,\unit{\um}$, respectively.
As discussed in Sec.~\ref{sec:numerical_method}, this discretization using beads leads to a repulsive force which is not strictly normal to the pillar surface, and therefore generates friction.
To investigate the role of obstacle roughness on the fiber dynamics, we also performed simulations with a perfectly smooth pillar described by quadratic Bézier curves \cite{Bezier1972}, and having the same shape as the pillar used in Secs.~\ref{sec:results_flow_fiber_dynamics} and \ref{sec:results_lateral_displacement} (see Fig.~\ref{subfig:pillar_beads_Bezier}).
\rev{This perfectly smooth pillar allows us to model contact with a purely normal repulsive force between the fiber and the pillar, in contrast to the pillar discretized by beads.}
More technical details on the computation of the repulsive force between the fiber and the perfectly smooth pillar are provided in the next subsection.

Figures \ref{subfig:phase_diagram_beads} and \ref{subfig:phase_diagram_Bezier} respectively represent the phase diagrams of the fiber dynamics computed with roughness (pillar discretized using beads) and without roughness (smooth pillar described by Bézier curves).
As can be seen in panel \subref{subfig:phase_diagram_Bezier}, pole-vaulting and trapping events are also observed in the absence of roughness.
However, the number of pole-vaulting and trapping events is much smaller without roughness.
This means friction is not needed for the fiber to pole-vault or to remain trapped on the pillar, but it promotes pole-vaulting and trapping by increasing the range of initial conditions leading to those events.

\subsection{Computation of the repulsive force}
\begin{figure}[H]
\centering
\includegraphics[width=0.3\textwidth]{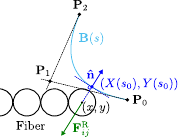}
\caption{Sketch showing the computation of the repulsive force between a fiber bead and a quadratic Bézier curve.}
\label{fig:sketch_Bezier_curve_repulsion}
\end{figure}

The smooth pillar used in the simulations presented in Fig.~\ref{subfig:phase_diagram_Bezier} consists of 6 adjacent quadratic Bézier curves ${\bf B}(s)=(X(s),Y(s))$.
Each of them is defined by 3 control points ${\bf P}_0=(x_0,y_0)$, ${\bf P}_1=(x_1,y_1)$ and ${\bf P}_2=(x_2,y_2)$ (see Fig.~\ref{fig:sketch_Bezier_curve_repulsion}) such as
\begin{equation*}
{\bf B}(s) = (1-s)^2{\bf P}_0 + 2(1-s)s{\bf P}_1 + s^2{\bf P}_2, \quad 0\leq s\leq 1.
\end{equation*}
Rearranging terms gives second-order polynomials for $X(s)$ and $Y(s)$
\begin{equation*}
    \begin{cases}
        X(s) = a_xs^2  + b_xs + c_x \\
        Y(s) = a_ys^2  + b_ys + c_y
    \end{cases}
    \quad 0\leq s\leq 1,
\end{equation*}
with
\begin{align*}
        &a_x = x_0 - 2x_1 + x_2& && &a_y = y_0 - 2y_1 + y_2&\\
        &b_x = 2(x_1 - x_0)& &\text{and}& &b_y = 2(y_1 - y_0)&\\
        &c_x = x_0& && &c_y = y_0.&
\end{align*}

Computing the repulsive force applied on a given fiber bead requires to compute the shortest distance between this bead and each of the 6 Bézier curves.
Let $d_{ij}(s)$ be the distance between the center of mass of the $i$th fiber bead ${\bf r}_i = (x,y)$ and a given point ${\bf X} = (X(s),Y(s)) $ belonging to the $j$th Bézier curve of the pillar.
\begin{equation*}
    d_{ij}(s) = \sqrt{\left(x-X(s)\right)^2 + \left(y-Y(s)\right)^2}.
\end{equation*}
The shortest distance between the $i$th fiber bead and the $j$th Bézier curve is obtained by solving for $d_{ij}'(s)=0$ in the range $0\leq s \leq 1$, with
\begin{equation*}
    d_{ij}'(s) = \frac{X'X + Y'Y}{\sqrt{(x-X)^2 + (y-Y)^2}}.
\end{equation*}
Solving for $d_{ij}'(s)=0$ is equivalent to computing the roots of the third-order polynomial $P(s)$ defined as
\begin{equation*}
     P(s)=X'X+Y'Y=as^3+bs^2+cs+d
\end{equation*}
with
\begin{align*}
&a=2(a_x^2+a_y^2)\\
&b=3(a_x b_x+a_yb_y)\\
&c=2\left[a_x (c_x-x) + a_y (c_y-y) + b_x^2 +b_y^2\right]\\
&d = b_x (c_x-x) + b_y (c_y-y).
\end{align*}
Let $s_0$ be the root of $P(s)$ minimizing $d_{ij}(s)$, and $R_{ij}=d_{ij}(s_0)$.
The repulsive force ${\bf F}_{ij}^{\rm R}$ acting on the $i$th fiber bead due to the $j$th Bézier curve is then computed as
\begin{align*}
    {\bf F}^{\rm R}_{ij} = \begin{cases}
        -\frac{F_{\rm ref}}{a}\left[ \frac{R^2_{\rm ref} - R_{ij}^2}{R^2_{\rm ref} - a^2} \right]^{4}R_{ ij}{\bf \hat{n}} & \text{if}~ R_{ij} < R_{\rm ref} \\
        {\bf 0} & \text{otherwise}
      \end{cases}
\end{align*}
\rev{where $a$ is the bead radius, $R_{\rm ref}$ the cutoff distance at which the repulsive force is activated}, and ${\bf \hat{n}} = {\bf n}/\left|{\bf n}\right|$ with ${\bf n}=(-Y'(s_0),X'(s_0))$
the unit vector normal to the Bézier curve at ${\bf X}_0 = (X(s_0),Y(s_0))$.

\end{document}